\documentclass[a4paper,oneside,Palatino]{article}
\usepackage{latexsym}
\usepackage[dvips]{color}
\usepackage[dvips]{graphics}
\usepackage{amsmath}
\usepackage{amssymb}
\usepackage{latexsym}
\usepackage{cite}
\usepackage[dvips]{color}
\usepackage[dvipdfm]{graphicx} 
\usepackage{appendix}
\usepackage{caption}
\usepackage{amsmath} 
\catcode`@=11

\@addtoreset{equation}{section}
\catcode`@=12
\title{\centerline{\small BHU/HEP/2020 \hfill hep-ph/2002276}\bigskip
\bf Magnetized matter effects on dilaton photon mixing}

\author{Ankur Chaubey,$^{a}$ Manoj K. Jaiswal$^a$ and Avijit K. Ganguly$^a$\thanks{e-mail addresses: avijitk@hotmail.com }  \thanks{corresponding author},\\ 
\normalsize 
$a)$ Institute of Science, Department of Physics, Banaras Hindu University, 
Varanasi- 221005, INDIA.\\
\normalsize 
}

\begin{document}
\maketitle

\begin{abstract}
  Dilatons ($\phi(x)$) are a class of bosonic scalar particles associated with scaling
  symmetry and its compensation 
  (under the violations of the same).  They are capable of  interacting gravitationally with
  other 
  massive bodies.  As they have coupling to two photons ($\gamma$), so they are (also)
  capable of decaying to the two photons. However, the decay time is long and that makes
  them good candidate for dark matter (DM).
  Furthermore due to two photon coupling, they can produce optical signatures in a magnetic
  field.  
  In vacuum  or plain matter they couple to one of the transversely polarized state of the
  photon. But 
  in a magnetized matter, they couple to both the transversely polarized state of photon
  (due to emergence
  of a parity violating part of photon self energy contribution- from a magnetized matter).
  Being spin zero 
  scalar, they could  mix with spin zero longitudinal part of photons but they don't.
  A  part of this work
  is directed  towards understanding this issue of mixing of scalar with various polarizations
  states 
  of photon in a  medium ( magnetized or unmagnetized ) due to the constraints from
  different
  discrete 
  symmetries  e.g.,  Charge conjugation (${\bf C}$), Parity (${\bf P}$) and Time reversal
  (${\bf T}$)  associated with the interaction. 
  Based on these symmetry aided arguments, the structure of the mixing matrix  is found to
  be $3 \times 3$, 
  like in the case of neutrino flavour mixing matrix.Thus  there exists non-zero finite
  probabilities 
of oscillation between different polarization states of photon  to dilaton. 
Our analytical and numerical analysis show  no existence of periodic oscillation length
either 
in temporal or spatial  direction for most general values of the parameters in the theory.
Possible
astrophysical consequences of these results, those can be detected through observations
are discussed.
  \end{abstract}

%############################################################################ 
\section{Introduction}
%############################################################################
\indent
The issue of unification of four fundamental forces of nature 
by the introduction of an additional scalar (fields)$(\phi(x))$ is probably one  
issue that may hold  key to the solution of dark 
energy/matter puzzle. These postulated fields are found to  make 
their  appearance in two kinds of theories,  one is in quantum theories of unification 
and other is in higher dimensional theories of cosmology/gravity.\\
\indent
In theories beyond the standard model (of particle physics), these additional scalar fields  
often appear in theories of unification (\cite{sugra} --\cite{scherk}), like in 5-dimensional Kaluza Klien theory, 
 in super-string theory and also in theories of extended super 
gravity \cite{sugra} etc.
In higher dimensional theories, like  the String theory,  the scalar fields -- termed moduli (fields) (\cite{witten}--\cite{Dhuria}), 
are necessary to  produce a four-dimensional effective theory 
from the original higher dimensional theory,  by compactifing the  extra  dimensions. \\
\indent
On the other hand, scalar fields ( both interacting  and massive ) had also been postulated 
independently, in models of  cosmology  to shed light 
on the aspects of dark matter, they are called chameleon whose presence remains imprinted in the cosmological 
observations of cosmic microwave background radiations  (\cite{chameleon} -- \cite{burrage}).
Also, the compelling observational aspects of dark matter physics has  motivated 
particle-astrophysics community to construct particle physics models with  similar fields 
(\cite{bento} --\cite{essig}) and verify their suitability in explaining the experimental data.
A detailed description of the  possible candidates of dark matter can be found in \cite{essig}. And possible other applications can be
found in (\cite{Kusenko}--\cite{sachadavidson}).\\ 
\indent
It is not so often that the solution of an unresolved issue in one area of physics holds {\emph{the}} key to another  unresolved issue in other area of the same. The issues of 
unification of four forces of nature and the missing mass and energy (dark matter and  dark energy)
problem of cosmology, might turn out  to represent such an event of rarity.\\
\indent
The notable feature that this field $\phi(x)$ exhibits in these theories 
lie in the structure of their interaction with photons. 
This is given by, an interaction  Lagrangian of the form, 
$\frac{1}{M} \phi F^{\mu\nu}F_{\mu\nu}$,  where $M$, is the  energy scale of the physics, involved.
In place we may write, $g_{\phi\gamma\gamma} = \frac{1}{M}$, where $g_{\phi\gamma\gamma}$ is the dimension full coupling constant. 
This interaction term has most of the desirable properties of an ideal Lagrangian except one. 
Although it is local, remains invariant under Lorentz, gauge and {\bf CPT} symmetry transformations,  
however, is non-renormalizable,  compromised for having interaction term of 
{{\em mass dimension}} five. Incidentally, similar interaction terms  are also possible for
loop induced Higgs photon interaction, given by, $g' H F^{\mu\nu}F_{\mu\nu} $, where $g'$
is the effective coupling constant, obtained after integrating out the heavy degrees of freedom (dof), 
$H$ is the Higgs field, and the rest of the pieces have their usual meaning.\\ 
\indent
Many of these fields can have coupling ({\it very weak though}, since $\frac{m}{M} \ll 1$  ) to 
standard model particles, there by, may be detected indirectly, through collider or astrophysical 
observations. Some of
their possible signatures are (i) existence of {\it  fifth force} distinguishable from gravity
(ii) violation of Lorentz invariance\cite{kahniasvili}  and (iii) spectro-polarimetric
signals those discussed in the phenomenal papers like, \cite{Miani}, \cite{Raffelt},\cite{sikivie1} and \cite{sikivie2} among many others of similar quality. A comprehensive list of other 
possible signatures, can be found in (\cite{pospelov} -- \cite{Hooper}).\\
\indent
A notable and important issue related to this kind of interaction, that is,  capable of distinguishing 
similar looking scalar fields from each other (owing their origin to various symmetries )
-- is related  to the  magnitude of their mass ($m$). The same is not fixed by any 
symmetry argument. However the lower limit  to the same ($m$) is fixed  by the 
torsion-balance fifth force experiment, that is estimated to be greater than $10^{-2} eV$ 
\cite{torsion-balance1}--\cite{torsion-balance4}. And the upper limit, on the same 
(i.e. $m$) can go upto few TeV [\cite{acharya}] -- depending
on the type of model (moduli) one is interested in. In cosmology for example, a limit to the mass $m$ and energy scale $M$ is set from the estimate of 
their life time to two photon decay, given by: 
%$\tau_{\phi \to \gamma \gamma}$
 $\tau_{\phi} = \frac{16 \pi M^2}{m^3}$ \cite{kim, fifthforce1,fifthforce2,fifthforce3}, so that the produced photons do not interfere with the 
big bang nucleosynthesis (bbn) constraints \cite{salati}. Although lately there are proposals of new kind
of dark matter termed as {\it fuzzy} dark matter\cite{fuzzy1} aviailable in the literature, for which torsion balance bounds are not applicable. The bounds on their masses are obtained from sachs-wolfe effect \cite{swe1,swe2}. \\
\indent
 The laboratory based experimental search for these particles were initially
suggested in \cite{sikivie1,sikivie2}  and  some of the variants of the same were in  \cite{CASPEr, ADMX,LSW,borexino, borexino2}. Many of these lab based experiments  have provided some bounds  on  axion coupling constant with other fundamental particles like electron, photon etc. These two experiments \cite{borexino, borexino2} happen to be one of those.   However the launch 
of the satellite missions- EUVE \cite{Lieu1996}, ROSAT \cite{Bonamente2002}, BeppoSAX \cite{Bonamente2001}, 
XMM-Newton \cite{Nevalainen2003}, Chandra  \cite{Bonamente2001,Henriksen2003, Bonamente2007} and Suzaku 
\cite{Lehto2010} sensitive to soft X-ray emissions around $5 - 10$ KeV  from the galaxy 
clusters-- have opened up another possibility of their astrophysical confirmation.  
Interestingly enough, ever since their launch, 
evidence of  0.5-1.0 MeV lines \cite{0.5-1KeV}, 3.5 KeV lines \cite{3.55KeV} 
and 511 keV lines (\cite{.511KeVa}--\cite{.511KeVb}) have been reported in the literature. Some of these signals are believed to be due to Dark matter.\\
\indent
Though many of the laboratory and astrophysical EM signals are complementry to 
each other however, for few the astrophysical ones are better
than the laboratory ones. The presence of {\it strong coherent magnetic field} over 
a large length scale, an ambient plasma and abundance of highly energetic
photons make it convenient to look for EM signals from 
astrophysical sources to test many types of interactions.\\
\indent
For the same reason, the EM signatures from astrophysical sources
for $\frac{1}{M} \phi F^{\mu\nu}F_{\mu\nu}$ interaction, is arguably
better than the ones from the laboratory. Although a large volume of 
literature \cite{ConlonMarsh},\cite{Miani}, \cite{Raffelt}, \cite{pankaj},  
\cite{Craig}, \cite{sikivie2021} are already available on many aspects of the relevant issues.\\

\indent
One of the notable aspect of these studies had been the incorporation of 
magnetized vacuum effects  by considering the Euler-Heisenberg Lagrangian. 
This incorporation lifts the degeneracy between the two transversely polarized
photons and makes only one of  the polarized states of  photon mix with ALP
like particle leaving the other one free. This  turns the vacuum dichoric and 
birefringent that inprinciple can be detected if ALPs exist in nature.\\

\indent
In this note however, in this work we would like to point out another aspect, that 
has rarely been considered important in such investigations, that is background dependence of the dynamics of scalar or pseudoscalar photon interaction.\\ 
\indent
 Photons propagating in a magnetized vacuum with $\phi F^{\mu\nu}F_{\mu\nu}$
interaction  have two transverse {\it polarization}
states. One of them $(|\gamma_{\perp}>)$ is orthogonal to $\rm{B}$ and the other one $(|\gamma_{\parallel}>)$ lies on the  ${k}$-$\rm{B}$ plane.
Following the reasoning of \cite{Raffelt} (performed originally for axion photon system), 
the two polarization states of photon, for this case too, would
transform differently under parity ${\bf P}$ and  charge conjugation ${\bf C}$ symmetry transformations. 
 Since under ${\bf CP}$ transformation, the scalar and the ${\bf CP}$ even 
polarization state of photon would remain even, only these two would couple
during propagation; the ${\bf CP}$ odd  polarized state of the photon would propagate freely.\\
\indent
When medium induced corrections 
are considered, the contribution from the in-medium polarization 
tensor $\Pi_{\mu\nu}(k)$ need to be taken into account. The same,
in absence of any parity violating interaction  or ambient external 
 magnetic field $\rm{B}$, would be ${\bf CP}$ symmetric. 
Hence even in an unmagnetized medium, the propagating modes 
of the scalar ($\phi$) photon ($\gamma$), 
system with dim-5 scalar photon interaction, would remain same as in magnetized vacuum
\footnote{ However, that consideration 
of the ambient plasma effects brings changes in the size of the
contribution to the oscillation probability $P_{{\gamma_{||} \to \phi}}$}.\\
\indent
This picture, however changes, with the introduction of  a new parity violating interaction term to the  effective Lagrangian ($L_{eff}$), that
originates from the magnetized medium induced  corrections to the photon self energy tensor (PSET), $\Pi_{\mu\nu}(k,T,\mu,e\rm{B})$ (sometimes called the Faraday term) in the system.
This tensor (PSET) in a magnetized medium, has parts thats even in $e\rm{B}$ and a part that is odd in the same. It  is the part that is  odd in $e\rm{B}$ is also odd in $\mu$ (chemical potential) and  is parity violating. This part was originally evaluated in \cite{GKP}.\\
\indent
 An effective Lagrangian of the form $A^{\mu} (-k)\Pi_{\mu\nu}(k,T,\mu,e\rm{B})A^{\nu}(k)$ constructed with the same would change the mixing dynamics. The leading order magnetic field effects in a magnetized plasma for $m^{2}>e\rm{B}$, can be obtained by retaining  the O($e\rm{B}$) piece from  $A^{\mu} (-k)\Pi_{\mu\nu}(k,T,\mu,e\rm{B})A^{\nu}(k)$, in the  interaction Lagrangian $(L_{eff})$. We perform the same in this study. 
This formalism was developed earlier in  \cite{Raffelt}, \cite{pankaj}, \cite{GJM} to study the  axion-photon interaction dynamics. We have extended this formalism to the case of scalar photon interaction dynamics by incorporating the correction mentioned above.This causes further mixing between the two 
transverse polarization states of the photon; therefore, the scalar and the two transversely polarized states of the photon mix with each other. As a result $|\gamma_{\perp}>$  part of a 
photon beam, unlike vacuum would evolve with propagation due to the presence of  PSET, and longitudinal dof would propagates freely. 
The effect of the same  on  polarimetric signatures of scalar photon mixing  case has been discussed in \cite{CJG}.Though a similar approach, for polarimetric studies 
had been considered  in \cite{GJM} for 
axion photon system also, but the conversion  probabilities of parallely or perpendicularly polarized photons to pseudo-scalar axions remained unexplored.\\
\indent
We complement the same in this work by  calculating these conversion probabilities for dilaton-photon system.With the introduction of PSET, the probability of conversion of perpendicularly polarized photon to dilaton and vice versa turns out to be finite.Till so far  this aspect remained unreported in the literature. We explore the same  here and its consequences.\\
\indent
We demonstrate here in this note that the inclusion of  PSET, causes the $\gamma-\phi$ mixing matrix to be $3 \times 3$ instead of $2 \times 2$,  usually encountered for a similar process taking place in magnetized vacuum, or unmagnetized plasma. 
\footnote{The situation is also different from axion photon system when PSET is considered, where
there is  mixing between all four degrees of freedom (three 
degrees of freedom of  a photon  in medium and single degree of freedom of axion). Thus one has to deal with a 
$4 \times 4$ mixing matrix  to study the evolution of the axion-photon system.}
Now, as a result of scalar photon mixing, the two transverse degrees of freedom of photon, the $\parallel$ and the $\perp$, can now oscillate into and out of scalar (dilaton) mode in addition to the oscillations among themselves i.e., $\parallel \leftrightarrow \perp$. Under favourable circumstances signals of the same may be within
the future detector sensitivity.\\
\indent
The organisation of the document is as follows: in the next section that is in section two, we have elaborated on the 
form  of the action and propagators in flat and curved space time. We have provided the logic behind sticking to 
the description in flat space time  because of pathological problems encountered in curved spacetime results.
Followed  by that the tensorial structure of  the polarization tensor for  photons in an unmagnetized medium is 
discussed. This is followed by the description of the fermion propagator in coordinate space. The parity
violating part of the photon polarization tensor is discussed  and its tensorial structure is discussed next.\\   
\indent
Section three contains a discussion of  the effective Lagrangian.
This is followed by a brief discussion on the discrete symmetry transformation properties  of the  gauge potentials and the 
terms of the  equations of motion of the $\gamma\phi$ system obtained from the effective Lagrangian under consideration. In section four, we  establish the unitary transformation matrix which diagonalises the mixing matrix. 
In section five, using the same unitary matrix obtained in section four, we diagonalize the equations of motion and identify its similarity with Klein Gordon equation in the diagonal form. We take this equation and following the procedure of \cite{Raffelt}, we estimates the conversion probabilities of various modes into each other. In section six  we discuss the physics of appropriate  astrophysical environments where  the mixing  of the photons with scalar can take place.
  We identify some possible signatures of this mixing from the EM signals coming out of these astrophysical environments. The implications of this modified  system is discussed in section even. In section eight we provide possible implications of our work for some of the DM signatures existing in the literatures followed by an appendix where some technical details are elaborated. 

%############################################################################ 
\section{The polarization tensor}

Considering the form of coordinate space dilaton-photon interaction term presented in \cite{kim} the effective action for scalar photon interaction including medium corrections  
[\cite{GKP}-\cite{ONS}], in configuration space can be expressed as,
\begin{eqnarray}
S \!\!&=&\!\!\int d^4x \sqrt{-g}~\!\! \left(-\frac{1}{4} F_{\mu\nu}(x)F^{\mu\nu}(x)
-\frac{g_{\phi\gamma\gamma}}{4}\phi(x)\,F_{\mu\nu}(x) F^{\mu\nu}(x)
-\frac{1}{2}\int d^4x^\prime\,A_\mu (x)\Pi^{\mu\nu}(x, x^{\prime})A_\nu(x^{\prime}) 
\nonumber  \right.  \\  \left.
\right. &&+  \left. \frac{1}{2}\partial_\mu\phi(x) \partial^\mu \phi(x) -\frac{1}{2}m_\phi^2\,\phi^2(x) \right).
\label{act-config}
\end{eqnarray}

\noindent
In eqn. (\ref{act-config}) $g$ is the determinant of the space-time metric that assumes
the value -1 in Minkowski space. The same ( factor $\sqrt{-g}$ ) is 
multiplied by $ d^4x$
to maintain Lorentz invariance of the measure. In the same  (i.e. eqn.  (\ref{act-config}))
$\Pi^{\mu\nu}(x, x^{\prime}) $ stands for the  in-medium polarization tensor in  configuration space. 
The polarization tensor can be expressed, following \cite{Perez}  as:,
\begin{eqnarray}
\Pi^{\mu\nu}(x, x^{\prime}) = e^2 tr \int \sqrt{-g}~d^4y^{\prime} \left( \gamma^{\mu} S_F(x,y^{\prime})
\gamma^{\nu}S_F(y^{\prime},x) \right) \delta^4(x^{\prime} - y^{\prime}),
\label{polten}
\end{eqnarray}
the trace is defined over Dirac matrices. 
The in medium propagator in coordinate space is defined, in terms of vacuum propagator: $ S^{v}_F(x',x'')= <x'\vert   \frac{1}{{\rlap/\partial} -m +i\epsilon } \vert x'' >  $ as, 
\begin{eqnarray}
 S_F(x',x'') =   S^{v}_F(x',x'') -f(p.u) \Big[  S^{v}_F(x',x'')- S^{v *}_F(x',x'')     \Big]  
\label{fermionprop}
\end{eqnarray}
where $f(p.u)$ happens to be the Fermi-distribution function in a frame moving with four velocity 
$u^{\mu}$.\\  
\indent
The curved spacetime Fermion propagator $S_{Fc}$   can also be represented as \cite{shore},  
\begin{eqnarray}
(i {\rlap /{D}} -m ) S_{Fc}(x,x')=\frac{i}{{\sqrt{-g}}} \delta (x,x').
\label{propcur}
\end{eqnarray}
The reason behind using this alternative representation, is the inability of expressing the
asymptotic vacuum states uniquely  curved space time.  Here $D$ stands for the covariant derivative
in curved space-time. It can further be  written in terms of the scalar green function
\begin{eqnarray}
\left( D^2 +i\sigma^{\mu\nu}F_{\mu\nu} -\frac{R}{4}+m^2 \right) G(x,x')=\frac{-i}
{\sqrt{-g}}\delta(x,x')
\label{scalarprop}
\end{eqnarray}
Finally, the Fermion propagator in curved space time in magnetic field
 is follows from eqn. (\ref{scalarprop}) 
\begin{eqnarray}
S_{Fc}(x,x')= \big( i {\rlap /{D}} +m   \big) G(x,x')
\label{fermionprop2}
\end{eqnarray}
where $R$ happens to be the Ricci scalar. One can express the  above expressions in momentum space, 
by taking the Fourier transform using two point transforms \cite{avramidi},\cite{Rempel}:\\
\begin{eqnarray}
f(x)= \int\frac{dk_{{\mu}'}}{(2\pi)^d}g^{(-1/2)}(x')exp(-i k_{\mu'}\sigma^{\mu'}(x, x')) {\tilde{f}}(k;x').
\label{covfouriertransf}
\end{eqnarray}
\noindent
A strong gravitational background is known to introduce some pathological problem, 
in external electromagnetic field. For example the photon velocity estimated for
Euler-Heisenberg Lagrangian system, predicts superluminal velocity of photon \cite{Drummond}.
Therefore the search for particles like dilaton like particles should be restricted
to space where spacetime curvature is negligible or flat. Usually the spacetime curvature
at a distance $r$ from a  body of mass $M$, is given by the Kretschmann scalar \cite{Richard-Conn-Henry}, 
\begin{eqnarray}
R= \frac{48~ M^2}{r^6}.
\end{eqnarray}
For a body with mass close to solar mass or less, the curvature is negligible for most of the 
regions close to its  surface. Moreover our region of interest happens to be close to the
light cylinder of the star, hence we have not considered the curvature of the spacetime
 for our estimates.\\
 
\noindent
Coming back to flat space time picture, $\Pi_{\mu\nu}(k,\mu, T, eB)$ has contribution coming from (a) magnetized vacuum (b) unmagnetized medium and (c) magnetized medium. Contribution from (c) can further be divided in two pieces those, having algebraic structure that can be written as a polynomial, first $(c_1)$  that is  even $eB$ even $\mu$  and second $(c_2)$ that is  a polynomial  odd in  $eB$ and odd in $\mu$. The even $eB$ odd $ \mu $ and odd $eB$ even $ \mu$ parts make vanishing  contribution  to $\Pi_{\mu\nu}$.  In parameter region where $eB<<m^2$ and momentum $k<<m$, contributions from $(a)$ and $(c_1)$  are suppressed compared to $(c_2)$\cite{Urrutia, Felix}.\\
\indent
Though the contribution  from (a), and $(c_1)$ have their special significance in the mixing dynamics for charge symmetric medium ($\mu = 0$) but inclusion of $(c_2)$ makes a paradigm shift in structure of mixing. So to initiate a discussion on effective Lagrngian we start with a discussion on the structure of polarization tensor in an unmagnetized media in momentum space  in sub-section[2.1]. 
%A discussion on the structure of the polarization
%tensor in unmagnetized media in momentum space is given in the following 
%subsection [\ref{polt}].
\subsection{ Polarization Tensor: Structure}
\label{polt}
%############################################################################
\indent
The linear response to EM excitations of a medium 
at finite density, temperature and an  external field
can studied by evaluating the in-medium photon polarization 
tensor $\Pi_{\mu\nu}(k)$ by the techniques of quantum 
statistical field theory \cite{pal-jose,adas,kapusta,Perez}.
This tensor is supposed to have few essential properties \cite{pal-jose}
e.g., it possesses a  symmetry i.e.,
\begin{eqnarray} 
\Pi_{\mu\nu}(k)= \Pi_{\nu\mu}(-k),
\label{bosesymmetry}
\end{eqnarray}
\noindent
called Bose symmetry. It should obey the Ward identity,
\begin{eqnarray}
 k^{\mu}\Pi_{\mu\nu}(k)=0 ,
\label{WI}
\end{eqnarray}
ensuring gauge invariance (also called charge conservation law). The
requirement of unitarity demands that the polarization
tensor should  have the property $\Pi_{\mu\nu}(k) = \Pi^{*}_{\nu\mu} (k)$.
This when combined with Bose symmetry(Eq.({\ref{bosesymmetry}}) yields: 
 \begin{eqnarray} 
\Pi_{\mu\nu}(k)= \Pi^{*}_{\mu\nu}(-k). 
\label{unitarity}
\end{eqnarray}
\noindent
Therefore, in four dimensions, the tensor  $\Pi_{\mu\nu}(k)$ now needs to be constructed from the 
available  4-vectors and tensors with the system; namely the  medium  centre of mass velocity 4-vector  $u^{\mu}$, the photon 4-momentum 
vector $k^{\mu}=({\omega, {\vec{k}}})$, the metric tensor $g^{\mu\nu}$, Levi-Civita tensor $\epsilon_{\mu\nu\rho\lambda}$ and the Lorentz scalar form factors such that, eqns. ((\ref{bosesymmetry})--(\ref{WI})), are satisfied. \\
\indent
The tensorial structure of the in-medium photon self energy tensor in absence of 
any external field is given by:
\begin{eqnarray}
\Pi_{\mu\nu}(k) = \Pi_{T}R_{\mu\nu}+\Pi_{L} Q_{\mu\nu} \\\, 
\mbox{~~~~where,~~~~}
%\,\,&&\,\, 
\left\{
\begin{array}{c} 
  R_{\mu\nu} =\tilde{g}_{\mu\nu} -Q_{\mu\nu},\\
  \,\,\,\,\,\,\,\,\,\,\,\,\tilde{g}_{\mu\nu}= \left( g_{\mu\nu}  -\frac{k_{\mu}k_{\nu}}{k^2} \right), \nonumber \\
\!\!\!\!\!\!\!\!\!\!\!\!\!\! Q_{\mu\nu}= \frac{{\tilde{u}_{\mu}}{\tilde{u_{\nu}}}}{{{\tilde{u}^2}}},
\\ 

%\!\!\!
\!\!\! \!\!\!\! \tilde{u}_{\mu}= \tilde{g}_{\mu\nu} u^{\nu} .\\

\end{array}
\right.
\label{pit1}
\end{eqnarray}
The dispersive part of $\Pi_{\mu\nu}(k)$, satisfies 
(\ref{unitarity}), in a charge $({\bf C})$ symmetric
( $\mu_{\bar{f}} = \mu_{f}$ ) or asymmetric( $\mu_{\bar{f}} \ne \mu_{f}$ ) medium  automatically. This in turn  dictates the functional 
form of the form-factors $\Pi_{L}$ and $\Pi_{T}$ on the scalars made out of $k^{2}$, (k.u) etc.
The scalar form factor $\Pi_{L}(k)$, corresponding to the 
longitudinal degree of freedom, is given by{\cite{Braaten-Segel}}:
\begin{eqnarray}
\Pi_{L}(k)=-\frac{k^2}{|{\vec{k}}|^2} \Pi_{\mu\nu}(k)u^{\mu}u^{\nu}, \\ 
\mbox{~~~~where~~~~}
u^{\mu}u^{\nu} \Pi_{\mu\nu}(k) =\omega^2_{p}
\left( \frac{|{\vec{k}}|^2}{\omega^2} +   3\frac{|{\vec{k}}|^4}{\omega^4}\frac{T}{m} \right)\nonumber. 
\label{PI^p}
%\nonumber \\
\end{eqnarray}

\noindent
Similarly the transverse form factor $\Pi_{T}$ is given by the  expressions:
\begin{eqnarray}
\Pi_{T}(k)&=& R^{\mu\nu}\Pi_{\mu\nu}(k)
\mbox{~~~~and~~~~}
{\rm{R}}^{\mu\nu} \Pi_{\mu\nu}(k) =\omega^2_{p}
\left( 1+  \frac{|{\vec{k}}|^2}{\omega^2}\frac{T}{m} \right). 
\label{displ}
\end{eqnarray}
In the expressions above $\omega_p$ denotes the plasma frequency. In the classical limit,
to leading order in $\frac{T}{m}$, it is given by:
\begin{eqnarray}
\omega_{p}=\sqrt{\frac{4\pi\alpha n_e}{m}\left(1 - \frac{5T}{2m}\right)},
\label{pf}
\end{eqnarray}

\noindent
where $n_e$, is the number density of electrons. A detailed structure of photon polarization 
tensor $\Pi_{\mu\nu}(k)$ in vacuum, in medium and in magnetized medium has been discussed in \cite{palash2020}. For more insight into it, one can consult to this reference.\\
\indent
Recalling the transformation properties of the background  EM field
 $\bar{F}^{\mu\nu}$,  and the 4-vectors $u^{\mu}$  and $k^{\mu}$ under ${\bf C}$ conjugation and  
${\bf P } $ reversal operations \cite{GKP},
it's easy to figure out that under parity transformation 
the two  orthogonal polarization eigen states of a  photon, 
described in the notation of \cite{GJ} by 
( $ \left( \tilde{\bar{F}}_{\mu\nu}{f}^{\mu\nu}(k)\right)  \triangleq |\perp > $ ) and 
($\left( \bar{F}_{\mu\nu}{f}^{\mu\nu}(k) \right)\triangleq |\parallel > $),
 are at odds with each other.\\
\indent
Now in the light of  Eqn. (\ref{pit1}) $\Pi_{\mu\nu}(k)$ is ${\bf C \rm{~and~} P}$
{\it even}, when the respective transformations are considered individually or together. i.e.,   
\begin{eqnarray}
{\bf P}^{-1}\Pi_{\mu\nu}(k){\bf P} = \Pi_{\mu\nu}(k),
\label{P-pi} \\
{\bf C}^{-1}\Pi_{\mu\nu}(k){\bf C} = \Pi_{\mu\nu}(k),   \\
\hskip 2 cm \left({\bf CP}\right)^{-1}\Pi_{\mu\nu}(k) \left({\bf CP} \right) = \Pi_{\mu\nu}(k). 
\end{eqnarray}
\noindent  
Hence, recalling the issue of {\it coupling of degrees of freedom in a 
unmagnetized material medium} that was initiated in the introduction, it is 
easy to realise that in an unmagnetized medium the parity violating state 
( i.e. $|\perp > $ )
would propagate freely but not the parity preserving state 
($ |\parallel > $).
This one would couple to $\phi(k)$, because of  Eqn. (\ref{P-pi})\footnote{ Since the 
${\bf CP}$ invariant background medium can not compensate for the parity odd property 
of  $|\perp >$, so that it can couple to $\phi(k)$.}. 
Thereby, in a material medium, the dynamics of the system remains as the same 
as it was in a  magnetized vacuum. The kinematics, however changes. In a plain material
medium, the magnitude of the oscillation probability undergoes modification 
vis-a-vis the same in a magnetized vacuum.\\
\indent
In  the presence of an  external EM field, the photon polarization tensor in a magnetized media can be expressed in terms of  the rank two basis tensors constructed out of the field strength tensor ${\bar{F}}^{\mu\nu}$, the  Levi-Civita tensor $\epsilon_{\mu\nu\lambda\sigma}$ along-with the other 4-vectors and tensors mentioned before  and the form factors those are Lorentz scalars constructed using these 4-vectors and tensors. We deliberate on this in the next sub section.

\subsection{Photon polarization tensor in a magnetized 
medium:  all orders in (e$B$)  .}

\noindent
Photon polarization tensor in a magnetized media in configuration space would follow from, 
eqn. [\ref{polten}]  where one needs to use the corresponding expressions for in-medium Fermion 
propagators in  an external  magnetic field [\ref{fermionprop}].  The expression for the
same is provided below. The Fermionic propagator in magnetized vacuum is
\cite{elmfors}
, 
\begin{eqnarray}
iS^{v}_{F} (x', x'') =\!\!\!\! &&\!\!\!\!\!\!\!-\frac{i \Phi(x',x'')}{(4\pi)^2} \int^{\infty}_{0}  ds \frac{eB}{sin(eBs)} 
exp\left[-is\left( -\frac{e  \sigma_{\mu\nu}F^{\mu\nu}}{2}  + m^2  -i \epsilon  \right)\right]  \nonumber \\
&& \times exp \left[  -\frac{i}{4} \left( (x' -x'')^{\alpha}\left(eF~coth(eFs) 
                                                      \right)_{\alpha\beta}(x'-x'')^{\beta} 
                    \right) 
           \right] \nonumber \\
&& \times  \left[ \frac{\gamma^{\lambda}}{2} \left(eF~coth(eFs) + eF
                                                      \right)_{\lambda\rho}(x'-x'')^{\rho}  +m {\bf I}
          \right]
\label{magfermpropvac}
\end{eqnarray}

\noindent
In eqn. [\ref{magfermpropvac}] above the phase factor,$\Phi(x',x'')$ is given by,
\begin{eqnarray}
\Phi(x',x'') = exp \left[  ie \int^{x'}_{x''}  dx^{\mu} \left( A_{\mu} + \frac{1}{2} F_{\mu\nu}\left(x' -x'' 
                                                                                       \right)^{\nu}
                                                     \right)   
                  \right], 
\label{fermionphase}
\end{eqnarray}
The symbol $F$  in eqn [\ref{fermionphase}] stands for the field strength tensor $F^{\mu\nu}$ (suitably contracted when 
they appear in combination with functions of the same or Dirac gamma matrices ), lastly quantity ${\bf I}$ in eqn. 
[\ref{magfermpropvac}] stands for 4$\times$4 unit matrix. Finally the  in-medium propagator in magnetized medium can 
be obtained by using the propagator [\ref{magfermpropvac}]  for  $S^{v}_F(x',x'')$ in eqn. [\ref{fermionprop}]. 

%\indent
%The following observation would be  in order at this stage. The same in extreme environments, like in strong gravitational  and electromagnetic
%background can be obtained by replacing the propagator in flat space with the corresponding one following Schwinger-Dewitt formalism. However these efforts are  not free from pathological issues related to causality even when in use with 
%fields of standard Glashow, Salam Weinberg model.   
%
%%
%%
%

%############################################################################
\subsection{Contribution to Photon self energy from magnetized medium:  all odd orders in (e$\rm{B}$)}
%############################################################################
\indent
In a  magnetized medium, magnetic field $( e\rm {B})$ dependent extra contributions appear in the expression for the photon polarization tensor. They are of two types  one of them  is even the other one is odd in powers of the  field strength ${ e\rm{B}}$. The one
odd in  ${ e\rm{B}}$ turns out to be also odd in $\mu$ (the chemical potential), so that
this term remains even under the operation of charge conjugation {\bf C}. 
This apart, this term satisfies the  conditions given by  Eqns. (\ref{bosesymmetry}), (\ref{WI}) and  Eqn. (\ref{unitarity}).  
This odd ${ e\rm{B}}$ odd $\mu$, contribution to the  polarization tensor violates 
{\it parity}. The exact expression of the same is given by,
\begin{eqnarray}
\Pi^{p}_{\mu\nu}(k,\mu,T,e{\rm B} ) 
&=& 4ie^2 \varepsilon_{\mu\nu\alpha_\parallel\beta} k^\beta
\int \frac{d^4p}{(2\pi)^4} \eta_-(p)
\int_{-\infty}^\infty ds \; e^{\Phi(p,s)}
\int_0^\infty ds' \; e^{\Phi(p',s')} \nonumber\\*
&\times & \Bigg[ 
p^{\widetilde\alpha_\parallel} \tan e{\rm{B}}s + 
p'^{\widetilde\alpha_\parallel} \tan e{\rm{B}}s' 
- {\tan e{ \rm{B}}s \; \tan e{ \rm{B}}s' \over \tan e{ \rm{B}}(s+s')} \;
(p+p')^{\widetilde\alpha_\parallel} \Bigg] \,.
\label{Ofinal}
\end{eqnarray}

\noindent
Here $e$ is the coupling constant for the U(1) gauge theory, 
$\eta_-(p) = \eta_F(p) - \eta_F(-p)$, (with $\eta_F(p)$ is the statistical
factor involving Fermions and their antiparticles \cite{GKP}) and  $\rm{B}$, 
as mentioned before is the background magnetic  field. The functions
 ${\Phi(p,s)}$ and  ${\Phi(p',s)}$  are the contributions from 
Schwinger Propagator \cite{Sch}, having loop momentum p and external momentum k, 
Fermion mass m and  parametric integrating variable s and s'.  The symbol $p'$ on r.h.s stand 
for $(p+k)$ the loop momentum 4-vector $p^{\widetilde\alpha_\parallel}$, appearing 
in  (\ref{Ofinal}) happens to be  components of momentum $p$ which 
 take only the values 0 and 3 ( called the $\parallel$ components); but with a difference. When  ${\alpha_\parallel}$ and ${\widetilde\alpha_\parallel}$ appear  together in any term and are summed up, then, for ${\alpha_\parallel}$ equal to zero,  ${\widetilde\alpha_\parallel}$ would take the  value three and for  ${\alpha_\parallel}$ equal to three, ${\widetilde\alpha_\parallel}$ would take the  value zero. In the same equation, the  symbol  $\varepsilon_{\mu\nu\alpha_\parallel\beta}$ is 
the completely antisymmetric Levi Civita tensor, that takes the values 1 
and -1, for even and odd  permutation of the indices and vanishes when any two 
two indices are the same. 
\begin{eqnarray}
\Phi(p,s) &\equiv& is \left( p_\parallel^2 - {\tan
(e{ \rm{B}}s) \over e{\rm{B}}s} \, p_\perp^2 - m^2 \right) - \epsilon
|s| \,, 
\label{Phi}
\end{eqnarray}  
\begin{eqnarray}
\Phi(p',s') &\equiv& is' \left( {p'}_{\parallel}^2 - {\tan
(e{ \rm{B}}s') \over e{\rm{B}}s'} \, {p'}_\perp^2 - m^2 \right) - \epsilon
|s'| \,. 
\label{Phi}
\end{eqnarray}  

\indent
This expression is exact   to all odd orders in ${ e\rm{B}}$, however performing the integrals and arriving at compact form, from this expression is very difficult.
\footnote{One can perform the integration with the following substitution: 
express the four vector $P^{\mu}$, as 
$p^{\mu}=\alpha_{(1)}u^{\mu}+\alpha_{(2)}n_2 \cdot (k.F)^{\mu}+\alpha_{(3)}n_3\cdot{{B}}^{\mu}+
\alpha_{(4)}n_3 \cdot q^{\mu}$, when $q^{\mu}= \epsilon^{\mu\nu\lambda\rho}F_{\lambda\rho}u_{\nu}$;
and $n_{i}$'s are normalization constants, so as to make the basis vectors orthonormal.}
%############################################################################
\subsubsection{ In medium contribution to photon self energy  to O (e$\rm{B}$) }
%############################################################################
\indent
 The  integral in the equation (\ref{Ofinal}) describing parity violating PSET is
 little difficult to evaluate exactly  using analytical techniques. But perturbative evaluation, to leading order in $e\rm{B}$ is possible \cite{GKP}. The perturbative expression   can be expressed in terms of a scalar form factor $\Pi^{p}(k,\mu,T,e{\rm B})$ and a projection operator ${P}_{\mu\nu}~$,  so $\Pi^{p}_{\mu\nu}(k,\mu,T,e{\rm B})$  to order ($ e\rm{B}$) can be expressed in the following form (\cite{GKP}, \cite{GJM}):
\begin{eqnarray}
\Pi^{p}_{\mu\nu}(k)=\Pi^p(k) \left( i\epsilon_{\mu\nu\alpha_{\parallel}\beta}
\frac{k^{\beta}}{|K|}u^{\tilde{\alpha}} \right) = \Pi^p(k)P_{\mu\nu},
\label{fp0}
\end{eqnarray}
where,
\begin{eqnarray}
{P}_{\mu\nu} =  i\epsilon_{\mu\nu\alpha\beta_{\parallel}}\frac{k^{\alpha}}{|K|}u^{\tilde{\beta}_{\parallel}}.
\label{fpo-1}
\end{eqnarray}
%%%%%%
The tensor ${P}_{\mu\nu}$  given by (\ref{fpo-1}) is hermitian  but it is odd under  
parity transformation. The superscripts with ${\parallel}$, means, they can take only  values between $0$ and $3$.
Furthermore in our notation, 
\begin{eqnarray} 
|K|= \left( \sum^{3}_{i=1}k^{2}_{i} \right)^{1/2}.
\label{wavevect-1}
\end{eqnarray}
The limit $ k \to 0$ in  Eqn. (\ref{fp0}) should be taken in such a way that, 
\begin{eqnarray}
\lim_{|k|\to 0} \left( \frac{k^{i}}{|k|} \right) \to 1.
\end{eqnarray}
%%%%%
\noindent
The scalar form factor $\Pi^p(k)$ appearing in Eq. (\ref{fp0}) is given by
%%%%
\begin{eqnarray}
\Pi^p(k) = \frac{\omega \omega_{B} \omega^2_p}{\omega^2-\omega^2_{B}}, \mbox{~~where~~~}
\omega_{B}=\frac{e\rm B}{m}
\label{faraday}
\end{eqnarray}
is the gyration frequency.\\  
%############################################################################
\section{Effective Lagrangian with magnetized medium effects}
%############################################################################

\indent
 To summarise the observations of the last section, we note that, 
the effects of a  medium to the propagation of excitations of interest
can be considered by evaluating the polarization tensor $\Pi_{\mu\nu}(k)$,
following the methods of finite  temperature quantum field theory \cite{GJM}.
This takes into account the  correction arising out of temperature and
density effects due to the interactions amongst the particles fields that constitute the media.\\
\indent
The action in momentum space, as the quantum corrections due to ambient medium and an external magnetic field to
${\cal O} (e\rm{B})$ are taken into account can be obtained upon taking the 
Fourier transform of eqn. (\ref{act-config}).  The same turs out to be,
%%%%%%%%%%%%%%%%%%%%%%%%%%%%%
\begin{small}
\begin{eqnarray}
S &=& \int d^4k \left[\frac{1}{2} 
A^\nu(-k) 
 \left( -k^2 \tilde{g}_{\mu\nu} + \Pi_{\mu\nu}(k,\mu,T) + \Pi^{p}_{\mu\nu}(k,\mu,T,e\rm{B}) \right) A^\mu(k)
\nonumber  \right. \\ \left.
\right. && \left. + ig_{\phi\gamma\gamma}{\phi(-k)}\bar{F}_{\mu\nu}k^\mu A^\nu(k)+\frac{1}{2}\phi(-k)[k^2 - m^2]\phi(k) \right].
\end{eqnarray}
\end{small}

\noindent
Here $\Pi_{\mu\nu}(k,\mu,T)$ is in medium polarization tensor and $\Pi^{p}_{\mu\nu}(k,\mu,T,e\rm{B})$ is the correction due to magnetized medium effects  PEST as explained before. For the sake of compactness we would be denoting $\Pi_{\mu\nu}(k,\mu,T)$ as  $\Pi_{\mu\nu}(k)$  and $\Pi^{p}_{\mu\nu}(k,\mu,T,e\rm{B})$ as  $\Pi^{p}_{\mu\nu}(k)$ in subsequent sections. We can find  the equations of motion 
in momentum space by standard variational principle. The equations of motion for photons is,
\begin{equation}
\left[-k^2\tilde{g}_{\alpha\nu} + \Pi_{\alpha\nu}(k) + \Pi^{p}_{\alpha\nu}(k) \right] A^\nu(k) =-ig_{\phi\gamma\gamma}\bar{F}_{\mu\alpha}k^\mu{\phi(k)}. 
\label{ph}
\end{equation}

\noindent
It can be simplified further in Lorentz gauge to,

\begin{equation}
k^2 A_{\alpha}(k) - \Pi_{\alpha\nu}(k)A^\nu(k) + \Pi^{p}_{\alpha\nu}(k) A^\nu(k) = ig_{\phi\gamma\gamma}\bar{F}_{\mu\alpha}k^\mu{\phi(k)},
\label{ph1} 
\end{equation}
and other equation of motion for $\phi(k)$ is given by

\begin{equation}
(k^2 - m^2)\phi(k) = ig_{\phi\gamma\gamma}\bar{F}_{\mu\alpha}k^\mu A^\alpha(k). 
\label{sc}
\end{equation}

%#######################################################
\subsection{Expanding the Gauge potential $A_{\mu}(k)$ in orthogonal basis vectors}
%############################################################################
\indent
In order to capture the dynamics of the available degrees of freedom in a medium, in this subsection, we need 
to expand the 4-vector potential $A_{\mu}(k)$ in terms of the available 4-vectors at our disposal. The available 
4-vectors as was noted in section (2.1) are $k^{\mu}$, the 4-momentum of the particles and $u^{\mu} \equiv (1, {\bf 0})$ the center of 
mass 4-velocity of the medium and $\epsilon^{\mu\nu\lambda\rho}$. Using these two and the constant external  field strength tensor $\bar{F}^{\mu\nu}$ 
(where the only nonzero component being $\bar{F}^{12} \ne 0$), two other vectors $b^{(1)\mu}$ and  $b^{(2)\mu}$ can be constructed. They are
given by, 
 \begin{equation}
   b^{(1)\nu} = k_\mu\bar{F}^{\mu\nu}
\label{perp-pol-vect} 
\end{equation}  
and 
\begin{equation}
   b^{(2)\nu} = k_\mu\tilde{{\bar F}}^{\mu\nu},
\label{aux-pol-vect}
\end{equation}  
\noindent
where, $\tilde{{\bar F}}^{\mu\nu}=\frac{1}{2}\epsilon^{\mu\nu\alpha\beta}{\bar{F}}_{\alpha\beta}$. 
The four-vector defining the longitudinal degree of freedom is usually given by
\begin{equation}
 \tilde{u}^\nu = \left( g^{\mu\nu} - \frac{k^\mu k^\nu}{k^2}\right)u_\mu
\label{longt-pol-vect}.
\end{equation}  

\noindent
Furthermore, the vector that is orthogonal to the vectors given in  Eqn.({\ref{perp-pol-vect}}) 
and   Eqn.({\ref{longt-pol-vect}}) is:
\begin{equation}
 I^\nu = \left(b^{(2)\nu} - \frac{(\tilde{u}^\mu b^{(2)}_\mu)}{\tilde{u}^2}\tilde{u}^\nu \right);
 \label{para-pol-vec}
 \end{equation} 

\noindent
 It should be noted, that since the vectors given by, eqns.(\ref{perp-pol-vect}), (\ref{aux-pol-vect}) and 
(\ref{longt-pol-vect}) are space like, so they have to be normalised without compromising the hermitian character
of the gauge fields. The explicit form of the normalisation constants are  given by,
\begin{eqnarray}
N_{1} &=& \frac{1}{\sqrt{-b^{(1)}_\mu b^{(1)^\mu}}}  
=\frac{1}{K_{\perp}\rm{B}}, \\
N_{2} &=& \frac{1}{\sqrt{-I_\mu I^\mu}}=\frac{K}{\omega K_{\perp}\rm{B}}    
\mbox{~~~and lastly}, \\
N_{L} &=& \frac{1}{\sqrt{-\tilde{u}_\mu \tilde{u}^\mu}}=\frac{\sqrt{k^{\mu}k_{\mu}}}{|\vec{k}|}.
%\frac{K}{|K|}.
\end{eqnarray}
\noindent 
With these definitions, the gauge potential 
$A^{\mu}(k)$ can be expressed in terms of these basis vectors and associated form factors as \cite{GJ,mohanty-nieves-pal}:

\begin{equation}
 A^{\nu} (k) =N_1 A_{\parallel} (k) b^{(1)\nu} +N_2 A_{\perp} (k) I^{\nu} +  N_L A_L (k) \tilde{u}^\nu 
+ A_{gf}(k) \frac{k^\nu}{k^2}.
 \label{gp}
\end{equation} 

\indent
In order to get rid of the redundant  degree of freedom of the gauge field, we  choose, $ A_{gf}(k)= 0 $. The form factors $A_{\parallel}(k),~ A_{\perp}(k)$ and $A_{L}(k)$, in  Eqn. (\ref{gp}) 
are Lorentz scalars made out of linear or nonlinear combinations of the tensor and the 4-vectors discussed above,
i.e., $ \omega=k.u $, $ |\vec{k}| = \sqrt{\omega^2 - k^2} $, $ k_{\alpha}{\tilde{\bar{F}}^{\alpha\beta}} u_{\beta}$,
$k_{\alpha}{\bar{F}}^{\alpha\beta}{\bar{F}_{\beta\sigma}}k^{\sigma}$ etc. \\ 
\indent
It is important to note that, the magnetic field
${\cal {B}}_{\mu} = \frac{1}{2} \epsilon_{\mu\nu\lambda\rho}u^{\nu}{\bar F}^{\lambda\rho}$ is actually orthogonal
to $ b^{(1)}_{\nu} $, that is to say ${\cal{B}}^{\nu} \cdot b^{(1)}_{\nu}=0 $, implying the direction of 
the polarization vector $b^{(1)}_{\nu}$ is orthogonal to the external magnetic field. Similar consideration 
would show that, the direction of polarization vector $I_{\nu} $ is along the external magnetic field. 
However, to maintain consistency with our previous work \cite{GJ}, we have denoted the associated 
form factors for the corresponding polarization directions, as $ A_{\parallel}(k) $, $ A_{\perp}(k)$.\\  
\indent
In order to gain an insight of the equations of motions, it is instructive to know
the transformation properties of the vectors, tensors and the form factors used for  Eqn.(\ref{gp}), under {\bf{C}}, {\bf{P}} and {\bf{T}} transformations. We  briefly discuss them  next.

\subsection{ Properties of the basis vectors under {\bf{C}}, {\bf{P}} and {\bf{T}} transformations:}
The non-zero components of the four vectors under deliberation are, $b^{(1)}_{\mu}= \left(0,b^{(1)}_{1},b^{(1)}_{2},0\right) $,
$b^{(2)}_{\mu}= \left(b^{(2)}_{0},0,0,b^{(2)}_{3}\right) $ and $I_{\mu}= \left(I_0,I_i \right)$ when $i=1,2,3$. In order to find out the transformation properties of the same, we need to  first identify the {\bf{C}}, {\bf{P}} and {\bf{T}} transformations of the 4-vectors $k^{\mu}$, $u^{\mu}$ and the tensor $\bar{F}^{\mu\nu}$. For the sake of brevity we do not show the momentum dependence of the form factors here. \\
\indent
The transformation properties of time ($k^{0}$) and space components ($k^{i}$, for $i=$1,2,3 ) of wave propagation vector $k^{\mu}=(k^{0}, k^{i})$ under {\bf{C}}, {\bf{P}} and {\bf{T}} transformations are given by:

\begin{eqnarray}
{\bf{C}}k^{0} {\bf{C}}^{-1}&=k^{0}, ~~{\bf{C}}k^{i} {\bf{C}}^{-1}&=+k^{i},\\
\label{c-k}
{\bf{P}}k^{0}{\bf{P}}^{-1} &= k^{0}, ~~{\bf{P}}k^{i}{\bf{P}}^{-1} &= -k^{i},\\ 
\label{p-k} 
{\bf{T}}k^0{\bf{T}}^{-1} &= k^{0},~~ {\bf{T}}k^i{\bf{T}}^{-1} &= -k^{i}.        
\label{t-k}
\end{eqnarray} 

The centre of mass four velocity of the medium, defined as, $u^{\mu}=\frac{dx^{\mu}}{d\tau}$ (when $d\tau$ is the differential proper time interval), have the following transformation properties under time reversal, parity and charge conjugation transformation:

\begin{eqnarray}
{\bf{C}}u^{0} {\bf{C}}^{-1}&=-u^{0}, ~~{\bf{C}}u^{i} {\bf{C}}^{-1}&=-u^{i},
\label{c-u}\\
{\bf{P}}u^{0}{\bf{P}}^{-1} &= +u^{0}, ~~{\bf{P}}u^{i}{\bf{P}}^{-1} &= -u^{i}, 
\label{p-u} \\
{\bf{T}}u^0{\bf{T}}^{-1} &= -u^{0},~~ {\bf{T}}u^i{\bf{T}}^{-1} &=+u^{i}.      
\label{t-u}
\end{eqnarray} 

%
%\begin{eqnarray}
%{\bf{T}}u^0{\bf{T}}^{-1} &=& -u^{0}, \\ 
%\label{t-u}
%{\bf{P}}u^{i}{\bf{P}}^{-1} &=& -u^{i}, \\
%\label{p-u}
%{\bf{C}}u^{\mu} {\bf{C}}^{-1}&=&-u^{\mu}.
%\label{c-u} 
%\end{eqnarray} 
The first property (\ref{c-u}) follows from the observation, that the statistical part of the thermal 
propagator in real time thermal quantum field theory should remain invariant under the operation of
charge conjugation, as explained in \cite{GKP}. The same ($u^{\mu}$) in the rest frame of the medium,
is given by $ u^{\mu}= \left( 1, 0,0,0 \right)$; for the remaining part of this paper we shall 
assume this to be true.\\       
\indent
The background field strength tensor $\bar{F}^{ij}$, transforms in the following way under $\bf{CPT}$
separately as:

\begin{eqnarray}
{\bf{C}}{\bar{F}^{0i}} {\bf{C}}^{-1}&=-{\bar{F}^{0i}}, ~~{\bf{C}} {\bar{F}^{ij}}{\bf{C}}^{-1}&=-{\bar{F}^{ij}},\\
\label{c-f}
{\bf{P}}{\bar{F}^{0i}} {\bf{P}}^{-1} &= -{\bar{F}^{0i}}, ~~{\bf{P}}{\bar{F}^{ij}}{\bf{P}}^{-1} &= +{\bar{F}^{ij}}, \\
\label{p-f} 
{\bf{T}}{\bar{F}^{0i}} {\bf{T}}^{-1} &= -{\bar{F}^{0i}},~~ {\bf{T}}{\bar{F}^{ij}}{\bf{T}}^{-1} &=- {\bar{F}^{ij}}.      
\label{t-f}
\end{eqnarray} 

%
%\begin{eqnarray}
%{\bf{T}}{\bar{F}^{ij}}{\bf{T}}^{-1} &=& -\bar{F}^{ij}, \\ 
%\label{t-f}
%{\bf{P}}{\bar{F}^{ij}}  {\bf{P}}^{-1} &=& +{\bar{F}^{ij}}, \\
%\label{p-f}
%{\bf{C}}{\bar{F}^{ij}} {\bf{C}}^{-1}&=&-{\bar{F}^{ij}}.
%\label{c-f} 
%\end{eqnarray}  

\indent
With these information we can look into the $\bf{C,P,\mbox{ and } T}$ transformation properties
of the basis vectors. To begin with, we start with vector $b^{(1)}_{\mu}$. It has only two nonzero 
components, and their transformation properties are,

\begin{eqnarray}
{\bf{C}}b^{(1)0}{\bf{C}}^{-1}&=-b^{(1)0}, ~~{\bf{C}}b^{(1)i} {\bf{C}}^{-1}&=-b^{(1)i},\\
\label{c-b1}
{\bf{P}}b^{(1)0}{\bf{P}}^{-1} &= +b^{(1)0}, ~~{\bf{P}}b^{(1)i}{\bf{P}}^{-1} &= -b^{(1)i}, \\
\label{p-b1} 
{\bf{T}}b^{(1)0}{\bf{T}}^{-1} &= +b^{(1)0},~~ {\bf{T}}b^{(1)i}{\bf{T}}^{-1} &= +b^{(1)i}.      
\label{t-b1}
\end{eqnarray} 

%
%\begin{eqnarray}
%{\bf{T}}b^{(1)}_{i}{\bf{T}}^{-1} &=& +b^{(1)}_{i}, \\ 
%\label{t-b1i}
%{\bf{P}} b^{(1)}_{i}  {\bf{P}}^{-1} &=& -b^{(1)}_{i}, \\
%\label{p-b1i}
%{\bf{C}} b^{(1)}_{i} {\bf{C}}^{-1}&=&-b^{(1)}_{i} \mbox{ when i$=$1,2}.
%\label{c-b1i} 
%\end{eqnarray}

Similarly, one can write the transformation properties of $b^{(2)}_{\mu}$, that has one 
time like and one space-like nonzero component. The time like component of $b^{(2)}_{\mu}$ under $\bf{C,P,\mbox{ and } T}$ operations transform as,

\begin{eqnarray}
{\bf{C}}b^{(2)0}{\bf{C}}^{-1}&=-b^{(2)0}, ~~{\bf{C}}b^{(2)i} {\bf{C}}^{-1}&=-b^{(2)i}.\\
\label{c-b2}
{\bf{P}}b^{(2)0}{\bf{P}}^{-1} &= +b^{(2)0}, ~~{\bf{P}}b^{(2)i}{\bf{P}}^{-1} &= -b^{(2)i}, \\
\label{p-b2} 
{\bf{T}}b^{(2)0}{\bf{T}}^{-1} &= -b^{(2)0},~~ {\bf{T}}b^{(2)i}{\bf{T}}^{-1} &=- b^{(2)i}.      
\label{t-b2}
\end{eqnarray} 

%
%\begin{eqnarray}
%{\bf{T}}b^{(2)}_{0}{\bf{T}}^{-1} &=& -b^{(2)}_{0}, \\ 
%\label{t-b20}
%{\bf{P}} b^{(2)}_{0}  {\bf{P}}^{-1} &=& +b^{(2)}_{0}, \\
%\label{p-b20}
%{\bf{C}}b^{(2)}_{0} {\bf{C}}^{-1}&=& -b^{(2)}_{0},
%\label{c-b20} 
%\end{eqnarray}  
%and the space-like component of $b^{(2)}_{\mu}$, under the same transformations transform as, 
%\begin{eqnarray}
%{\bf{T}}b^{(2)}_{i}{\bf{T}}^{-1} &=& -b^{i(2)}, \\ 
%\label{t-b2i}
%{\bf{P}} b^{(2)}_{i}  {\bf{P}}^{-1} &=& -b^{(2)}_{i}, \\
%\label{p-b2i}
%{\bf{C}}b^{(2)}_{i} {\bf{C}}^{-1}&=&-b^{(2)}_{i}.
%\label{c-b2i}
%\end{eqnarray}
Now using above equations, it is easy to 
establish that,
\begin{eqnarray}
{\bf{P}} \left( \tilde{u}\cdot b^{(2)}\right) {\bf{P}}^{-1} &=& \left( \tilde{u}\cdot b^{(2)} \right) 
\label{udotb2}.
\end{eqnarray}  
\noindent
Recalling, the four vector $I^\nu$, to be given by,
\begin{equation}
 I^\nu = \left(b^{(2)\nu} - \frac{(\tilde{u}^\mu b^{(2)}_\mu)}{\tilde{u}^2}\tilde{u}^\nu \right);
 \label{para-pol-vec-2}
 \end{equation} 
 the time and space like components of the same are found to be given by,
\begin{equation}
 I^{0} = \left(b^{(2){0}} - \frac{(\tilde{u}^\mu b^{(2)}_\mu)}{\tilde{u}^2}\tilde{u}^{0} \right) 
\mbox{~and~}
%\begin{equation}
 I^{i} = \left(b^{(2){i}} - \frac{(\tilde{u}^\mu b^{(2)}_\mu)}{\tilde{u}^2}\tilde{u}^{i} \right), 
\mbox{~~(for i$=$1,2,3)}.
 \label{para-pol-vec-20}
 \end{equation} 
 \indent
 Using the ${\bf C}$, ${\bf P}$ and  ${\bf T}$ transformation properties of the individual components of $I^{\mu}$,  $I^{0}$ and $I^{i}$
would transform under the same  as,

\begin{eqnarray}
{\bf{C}}I^{0}{\bf{C}}^{-1}&=-I^{0}, ~~{\bf{C}}I^{i} {\bf{C}}^{-1}&=-I^{i}.\\
\label{c-i2}
{\bf{P}}I^{0}{\bf{P}}^{-1} &= +I^{0}, ~~{\bf{P}}I^{i}{\bf{P}}^{-1} &= -I^{i}, \\
\label{p-i2} 
{\bf{T}}I^{0}{\bf{T}}^{-1} &=- I^{0},~~ {\bf{T}}I^{i}{\bf{T}}^{-1} &= +I^{i}.      
\label{t-i2}
\end{eqnarray} 
%
%\begin{eqnarray}
%{\bf{P}}\,I^{0}\,{\bf{P}}^{-1} = I^{0} \mbox{~~~and~~~} {\bf{P}}\,I^{i}\,{\bf{P}}^{-1}= -I^{i}, 
%\mbox{~~( for i$=$1,2,3 )}. 
%\label{I-parity}
%\end{eqnarray}
%\indent
%Similarly under the ${\bf C}$ and ${\bf T}$ transformations, the same yields,
%\begin{eqnarray}
%{\bf{C}}\,I^{0}\,{\bf{C}}^{-1} = -I^{0} \mbox{~~~and~~~} {\bf{C}}\,I^{i}\,{\bf{C}}^{-1}= -I^{i}, 
%\mbox{~~( for i$=$1,2,3 )},  
%\label{I-C}
%\end{eqnarray}
%\begin{eqnarray}
%{\bf{T}}\,I^{0}\,{\bf{T}}^{-1} = -I^{0} \mbox{~~~and~~~} {\bf{T}}\,I^{i}\,{\bf{T}}^{-1}= +I^{i}. 
%\mbox{~~( for i$=$1,2,3 )},  
%\label{I-T}
%\end{eqnarray}
%\indent
Since the time like components of the gauge field under time reversal transformation, remains the same, i.e.,
\begin{eqnarray}
{\bf{T}}A^{0}{\bf{T}}^{-1}  = A^{0},
\end{eqnarray} 
therefore when  the time reversal transformation is imposed on $A^{0}$, following the definition of $A^{0}$ one should get,
\begin{eqnarray}
{\bf T}^{-1} A^{0}{\bf T}^{-1} ={\bf T}( N_{1} I^{0}  A_{\parallel} + N_{L}\tilde{u}^{0} A_{L}){\bf T}^{-1}= ( N_{1} I^{0}  A_{\parallel} + N_{L}\tilde{u}^{0} A_{L})
\end{eqnarray}
that is the r.h.s  remain invariant.\\
\indent
 Now under time reversal transformation $I^0$ and  $\tilde{u}^{0}$, picks up a -ve sign. So to maintain overall neutrality, 
$A_{\parallel}$ and $A_{L}$ must change sign under ${\bf T}$ transformation. Hence,
\begin{eqnarray}
\left({\bf{T}} A_{\parallel}{\bf{T}}^{-1} \right) = -A_{\parallel} \mbox{~and~}\left({\bf{T}} A_{L}{\bf{T}}^{-1} \right) =- A_{L}.
\label{AL}
\end{eqnarray}

\indent 
We recall that the space like component of $A^{\mu}$ can be expressed as,
\begin{eqnarray}
A_{i}=N_1 b^{(1)}_{i} A_{\parallel} + N_{2}I_{i} A_{\perp}+ N_{L} \tilde{u}_{i}A_{L}.
\label{a-space}
\end{eqnarray}
The same, that is   $A^{i}$, 
under {\bf T} transformation pickup a $-ve$ sign i.e.,
\begin{eqnarray}
({\bf T} A_{i}{\bf T}^{-1}) = -A_{i}.
\label{A-space-T}
\end{eqnarray}

\noindent
It then follows  that,  the right hand side of  Eqn. (\ref{a-space}) should respect transformation  Eqn. (\ref{A-space-T}). Since 
 $b^{(1)}_{i}$,  $I_{i}$ and $\tilde{u}_{i}$  remains invariant  under time reversal symmetry transformation therefore
$A_{\parallel}$, $A_{\perp}$ and $A_L$ should change sign under time reversal transformation symmetry. The   transformation rules of the form factors can now be summarised as,
\begin{eqnarray}
{\bf C} A_{\parallel} {\bf C}^{-1} =  +A_{\parallel},\,\, {\bf C} A_{\perp} {\bf C}^{-1} =  +A_{\perp}, \,\,{\bf C} A_{L} {\bf C}^{-1} =  +A_{L}, \\
\label{chargelike}
{\bf P} A_{\parallel} {\bf P}^{-1} =  +A_{\parallel}, \,\,{\bf P} A_{\perp} {\bf P}^{-1} =  + A_{\perp}, \,\,{\bf P} A_{L} {\bf P}^{-1} =  +A_{L}, \\
\label{spacelike}
{\bf T} A_{\parallel} {\bf T}^{-1} =  -A_{\parallel}, \,\,{\bf T} A_{\perp} {\bf T}^{-1} =  -A_{\perp} , \,\,{\bf T} A_{L} {\bf T}^{-1} = -A_{L}.
\label{timelike}
\end{eqnarray}

\noindent
 So the {\bf C, P} and {\bf T} transformations properties  of the vectors,  tensors and the form factors can further be put  
in tabular form  ( see Table [\ref{table:1}]).

\begin{table*}[h]
\centering
\begin{tabular}{|c | c c c c c  c c c c c c c |} 
 \hline
        & $k_{\mu}$     &   $u_{\mu}$ &    $\tilde{u}_{\mu}$ & $b^{(1)}_{\mu}$ & $b^{(2)}_{\mu}$ & $ I_{\mu}$ & $A_{\parallel}$ & $ A_{\perp} $ &  $A_{L}$   & $i$ & $\epsilon_{\mu\nu\rho\sigma}$  & $\bar{F}_{\mu\nu}$ \\ [0.5ex] 
 \hline\hline
 {\bf C}   & ${\bf  +}k_{\mu}$     & ${\bf - }u_{\mu}$ &  $ {\bf - }\tilde{u}_{\mu}$   & $-b^{(1)}_{\mu}$  & $-b^{(2)}_{\mu}$  &   $-I_{\mu}$  &  $+A_{\parallel}$    & $ + A_{\perp}  $ &  $+A_{L}$   & $+i$ & $+\epsilon_{\mu\nu\rho\sigma}$  & ${\bf - }\bar{F}_{\mu\nu}$    \\ 
 {\bf P}   & ${\bf  +}k^{\mu}$     &   ${\bf  +}u^{\mu}$ &  ${\bf  +} ~\tilde{u}^{\mu}$   & $+b^{(1)\mu}$ & $+ b^{(2)\mu}$ &    $+I^{\mu}$  & $+ A_{\parallel}$    & $  +A_{\perp}  $ &  $+ A_{L}$     & $+i$ & $-\epsilon_{\mu\nu\rho\sigma}$ & $ {\bf  +}\bar{F}^{\mu\nu}$   \\
 {\bf T}     & ${\bf  +}k^{\mu}$     &  ${\bf - }u^{\mu}$ &    ${\bf - }\tilde{u}^{\mu}$  & $ -b^{(1)\mu}$ & $ -b^{(2)\mu}$   &   $-I^{\mu}$  & $ - A_{\parallel} $ &  $ - A_{\perp}$ &  $ - A_{L} $  & $-i$ & $-\epsilon_{\mu\nu\rho\sigma}$ & ${\bf - }\bar{F}^{\mu\nu}$ \\ [1ex]
\hline
\end{tabular}
\caption{ Transformation properties for the vectors, tensors and the EM form-factors used to construct  all the vectors, tensors and form factors used in this work to expand $A^{\nu}(k)$, under ${\bf C}$, ${\bf P}$ and ${\bf T}$.}
\label{table:1}
\end{table*}

\indent
It can be shown that, each term appearing in the equations of motion  of this article when subjected to these transformations, transforms identically. We will demonstrate the same later.

%############################################################################
\subsection{Dynamics Of The Degrees Of Freedom}
%############################################################################

\indent
In this section, we discuss the dynamics of the independent degrees of freedom of 
the system. Recalling the fact that in a medium photon acquire an extra degree of 
freedom in addition to the two transverse degrees of freedom, therefore for the photon scalar interacting system, there should be  four degrees of freedom. 
Moreover, out of the three degrees of freedom of the photon, 
the longitudinal degree of freedom has spin zero, and the other two would be having spin one and minus  one
respectively. Therefore though naively one may expect that, the longitudinal mode of photon 
would couple with the scalar degree of freedom, because they both have same spin assignments, 
however we would  demonstrate in this subsection, by analysing the equations of motion, that, 
this naive expectation doesn't hold good.\\
\indent
Now we  begin with   Eqn.(\ref{ph1}), and substitute the expression for the gauge potential 
from   Eqn.(\ref{gp}) in the same. The resulting equation (considering the Faraday 
contribution to be $\Pi^{p}_{\mu\nu}(k) =- \Pi_{p}(k)P_{\mu\nu}$, when  the  projection operator is defined as $P_{\mu\nu} = i\epsilon_{\mu\nu\beta\delta_{\parallel}}\frac{k^{\beta}}
{\mid k \mid}u^{\tilde{\delta}_\parallel}$){\footnote{Where $\tilde{\delta}_{\parallel} $ 
can takes the value 0 or 3. When ${\delta}_{\parallel} = 0 $ then $\tilde{\delta}_{\parallel} 
= 3$ and vice versa}.} turns out to be, 
\begin{eqnarray}
&&(k^2-\Pi_T(k))[ A_{\parallel} (k)N_1 b^{(1)}_{\alpha} + A_{\perp} (k)N_2 I_{\alpha} + A_L (k) N_L \tilde{u}_\alpha] + \Pi_T(k) A_L(k) N_L \tilde{u}_\alpha  \nonumber \\
 &-& \Pi_L(k) A_L(k) N_L \tilde{u}_\alpha   
- i\Pi^{p}(k)\epsilon_{\alpha\nu\beta\delta}
\frac{k^{\beta}}{\mid{k}\mid}u^{\tilde{\delta}_{\parallel}} [A_{\parallel} (k)N_1 b^{(1)\nu} + A_{\perp} (k)N_2 I^{\nu} \nonumber\\&+ &A_L (k) N_L \tilde{u}^\nu]  
= ig_{\phi\gamma\gamma}b^{(1)}_{\alpha}{\phi(k)},
\label{fde}
\end{eqnarray}
%%%
%%%
\noindent
Next as we multiply  equation (\ref{fde}) by the normalised basis vectors, we get the 
equations of motion for different components of the form factors. For example, if we multiply  equation (\ref{fde}) 
by $b^{(1)\alpha}$, we find the following equation,
%%%
%%%
\begin{eqnarray}
&& (k^2-\Pi_T(k)) A_{\parallel} (k)N_1 b^{(1)}_{\alpha}b^{(1)\alpha} - i\Pi^{p}(k)N_{2}
\left[\epsilon_{\alpha\nu\beta\delta_{\parallel}}
\frac{k^{\beta}}{\mid{k}\mid}u^{\tilde{\delta}_{\parallel}} b^{(1)\alpha} I^{\nu} \right]A_{\perp}(k)\nonumber\\ 
& =& ig_{\phi\gamma\gamma}b^{(1)}_{\alpha}b^{(1)\alpha}{\phi(k)}.
\label{fd0}
\end{eqnarray}
Using the expression of the normalization constants, the same becomes
%%%
\begin{eqnarray}
&&(k^2-\Pi_T(k))A_{\parallel} (k)+ i\Pi^{p}(k)N_{1}N_{2}\left[\epsilon_{\alpha\nu\beta\delta_{\parallel}}
\frac{k^{\beta}}{\mid{k}\mid}u^{\tilde{\delta}_{\parallel}} b^{(1)\alpha} I^{\nu}\right] A_{\perp}(k) 
= \frac{ig_{\phi\gamma\gamma}\phi(k)}{N_1}.
\label{fd1} 
\end{eqnarray}
%%%
\noindent
Similarly,  multiplying   Eqn. (\ref{fde}) by $I^{\nu}$ and $\tilde{u}^\alpha$ 
respectively, we get the following two equations:   
%%
%%
%%%%%%%%%%%%%%%%%%%%%%%
\begin{eqnarray}
 &&(k^2-\Pi_T(k))A_{\perp} (k) - i\Pi^{p}(k)N_{1}N_{2}\left[\epsilon_{\alpha\nu\beta\delta_{\parallel}}
\frac{k^{\beta}}{\mid{k}\mid}u^{\tilde{\delta}_{\parallel}} b^{(1)\alpha} I^{\nu}\right ]A_{\parallel}(k)= 0,
\label{fd2} 
\end{eqnarray}
%%%%%%%%%%%%%%%%%%%%%%%%
\begin{equation}
(k^2 -\Pi_L(k))A_L(k) = 0, 
\label{fd3}
\end{equation}   
%%%
%%%
\noindent
those describe the dynamics of the three degrees of freedom of the photon. Lastly, 
the equation of motion for the scalar field turns out to be, 
%%%%%%%%
\begin{equation}
(k^2 - m^2)\phi(k) = -\frac{ig_{\phi\gamma\gamma} A_{\parallel}(k)}{N_1}.
\label{fd4} 
\end{equation}
%%%%%%%%%%%%%%%%%%%%%%%
\noindent
These four equations  (\ref{fd1}), (\ref{fd2}), (\ref{fd3}) and (\ref{fd4}) describe the dynamics of $\gamma\phi$ interaction in a magnetized medium. The correctness of the above  equations of motion can be established
by performing  {\bf PT} transformation  on these equations.\\
\indent
To demonstrate it, let us choose the equation (\ref{fd2}) and operate $({\bf PT})$ from left and $({\bf PT})^{-1}$ from right     sides of the equation. Following the transformation rules of  table [1], under ${\bf PT}$, the first term of 
the equation will pick up a negative sign due to presence of $A_{\perp}(k)$  which is odd under the same. In the second term, the factors $i,{u}^{\tilde{\delta}_{\parallel}}, b^{(1)\alpha}, I^{\nu}$ and $A_{\parallel}(k)$ are  ${\bf PT}$ odd and rest are   ${\bf PT}$ even as shown below:
\begin{eqnarray}
 \begin{matrix}
                   &   ({\bf PT}) A_{\parallel} ({\bf PT})^{-1} = -A_{\parallel},       & ({\bf PT}) A_{\perp} ({\bf PT})^{-1}  = -A_{\perp} , \\                 
                   &  ({\bf PT}) i ({\bf PT})^{-1}  = -i,        & ({\bf PT}) k^{\beta} ({\bf PT})^{-1}  = +k^{\beta}, \\
                   & ({\bf PT}) u^{\tilde{\delta}_{\parallel}}({\bf PT})^{-1}  = -u^{\tilde{\delta}_{\parallel}},       & ({\bf PT}) b^{(1)\mu} ({\bf PT})^{-1}  =  -b^{(1)\mu}, \\
                   & ({\bf PT}) I^{\nu} ({\bf PT})^{-1}  = -I^{\nu},          & ({\bf PT}) \epsilon_{\mu\nu\delta\beta} ({\bf PT})^{-1} =  +\epsilon_{\mu\nu\delta\beta}.
                 \end{matrix}     
\label{pt-i}
\end{eqnarray}

\noindent
{
The transformation properties of form factor $\Pi^p(k)$ under ${\bf C, P}$ and   ${\bf T}$ can be figured out from the expression, 
\begin{eqnarray}
\Pi^{p}(k) = \frac{{(k.u)}{(e\rm{B_{\parallel}}/m_{e})}}{\omega^{2}- (e\rm{B_{\parallel}}/m_{e})^2}\left(  \frac{n_e}{m_e}\right).
\label{Pipabs}
\end{eqnarray}

\noindent
As can be seen from the equation  (\ref{Pipabs}) that, $\Pi^p(k)$ is  invariant under the  ${\bf PT}$ transformation, i.e.,

\begin{eqnarray}
 ({\bf PT}) \Pi^{p}(k)({\bf PT})^{-1} =  \Pi^{p}(k).
\end{eqnarray} 
\noindent
so collectively, the second term of equation (\ref{fd2})  will also pick up a negative sign. Therefore this equation remains invariant under  ${\bf PT}$ transformation. Same can be established for other equations too using similar logic.\\
\indent
The  {\bf PT} symmetric part of  PSET, when included in the effective Lagrangian, can 
in principle compensate for the {$\bf P$} violation of $ |A_{\perp} > $, when  both appear 
as a product, as they do in the equations of motion. Thus resulting product of the two becomes ${\bf P}$ even, 
making mixing between $\phi(k)$ and  $ |\perp > $ possible.\\
\indent
 Now looking at the problem of mixing, we note that the initial mixing between  $\phi(k)$ 
and $ |\parallel > $, due to $\phi F F$ coupling remains intact,  but introduction of the 
PSET term  causes further mixing between the two orthogonal transverse states, i.e., 
$  |\parallel >$  and  $ |\perp > $;  and lastly the combination of the product of 
PSET term and  $ |\perp > $, as noted in the last paragraph, causes mixing between, 
$ |\perp > $ and $\phi(k)$. Thus the system reduces
to a system of three mutually coupled degrees of freedom, that
evolves following their respective equations of motion.\\    
\indent
It may not be quite out of place, to mention here that,  instead of   $\phi(k) F^{\mu\nu}F^{\mu\nu}$, if   
$a(k) {\tilde{F}}^{\mu\nu}F^{\mu\nu}$ interaction is considered  in a magnetized vacuum or an unmagnetized medium,
the role of  scalar $\phi(k)$ gets interchanged with that of 
pseudoscalar   \cite{peccei}; and  consequently, the role of $|\parallel >$ would get interchanged with $|\perp >$ so that the symmetry remains intact. So the prediction from one can be obtained 
from the prediction of the other.\\
\indent
This simple interrelation between the two, as noted already, however gets modified as one includes
the effect of PSET to, axion photon or scalar photon systems.\\

\indent
Introducing, $F =\Pi^{p}(k)N_{1}N_{2}[\epsilon_{\alpha\nu\beta\delta}
\frac{k^{\beta}}{\mid{k}\mid}u^{\tilde{\delta}_{\parallel}} b^{(1)\alpha} I^{\nu}]$ and $G =\frac{g_{\phi\gamma\gamma} A_{\parallel}(k)}{N_1}$ for the sake of brevity,
the coupled set of equations (of motion) can be presented in matrix form, as 
\indent
\begin{eqnarray} 
\left[ 
  \begin{matrix}
 (k^2 - \Pi_T(k))     &       iF        &       0        &      -iG     \cr     
     -iF             & (k^2 - \Pi_T(k))  &      0        &       0    \cr
     0             &   0          & (k^2 - \Pi_L(k))     &       0     \cr
     iG             &   0         &      0         & (k^2 - m_{\phi}^2) 
  \end{matrix}
\right]
\left[
\begin{matrix} 
{A_{\parallel} (k)} \cr
{A_{\perp} (k)} \cr 
{A_L (k)} \cr 
\phi(k)
\end{matrix}
\right]= 0.
\label{mat}
\end{eqnarray}
%%%%
%%%%
\indent
The longitudinal degree of freedom  of photon ($A_{L}(k)$) as can be seen, has decoupled from 
the rest of the degrees of freedom.

%############################################################################
\subsection{Background dependent mixing pattern}
%############################################################################

\indent
In this section we try to understand the background's influence on the mixing dynamics of the  transverse and longitudinal degrees of freedom of photon with scalar or pseudoscalar 
as the interaction term in the Lagrangian  changes  from $g_{\phi\gamma\gamma}\phi FF$ for  scalar-photon system to $g_{a\gamma\gamma}a \tilde{F}F$ for  pseudoscalar-photon system.\\
\indent
As  was noted in the introduction, that dynamics of the degrees of freedom of photon  in terms of the form factors $A_{\parallel}(k)$ and $A_{\perp}(k)$ in a magnetized vacuum with  $g_{\phi\gamma\gamma}\phi FF$ interaction, can be anticipated from the evolution of these two  form factors in  pseudoscalar photon system ($a\gamma\gamma$),  with the  identification of  the role of  $A_{\parallel}(k)$ with  $A_{\perp}(k)$ and $A_{\perp}$ with $A_{\parallel}$ when the interaction Lagrangian is $g_{a\gamma\gamma}a \tilde{F}F$.\\
\indent
So one can conclude that these system seem to have a symmetry that remains invariant under the exchange of the scalar field with pseudoscalar and parallely polarised state $A_{\parallel}$ with the perpendicularly polarised state $A_{\perp}$ of the photon. Even in material medium, this symmetry behaviour remains the same.\\
 \indent
 In presence of a magnetized medium, however, contributions from the  photon self energy tensor (PSET)
$\Pi_{\mu\nu}(k, \mu, T, e\rm{B})$ that is odd in powers of ${ e\rm{B}}$ lifts this apparent 
degeneracy. The discrete symmetries of nature, like charge conjugation ({\bf C}) 
time reversal ({\bf T}) and parity ({\bf P}) transformations play a  major role 
in removing this apparent degeneracy. We would like to come back to this 
issue in future publications.\\
\indent
Since the longitudinal degree of freedom $A_{L}(k)$,  ({\it with dimension-5 scalar-photon interaction} )
in a magnetized medium,  is decoupled from the rest of the degrees of freedom, therefore we 
won't be considering it any further. Therefore, the resulting matrix   Eqn. (\ref{mat}) can be cast in the following form:
\noindent
\begin{eqnarray} 
\left[ k^2 {\bf I}-\left( \begin{matrix} \Pi_T(k) &   iF     &    -iG \\
-iF &\Pi_T (k) & 0 \\
 iG & 0 & m^2_{\phi}  \end{matrix} \right) \right]\left[
\begin{matrix}
{A_{\parallel} (k)} \cr
{A_{\perp} (k)} \cr   
\phi(k)
\end{matrix}
\right]  =
0.
\label{mat1}
\end{eqnarray}
%%%%%
\noindent
Here {\bf I} is $3 \times 3$ identity matrix.
For the sake of compactness, we would further like to denote the matrix, inside bracket, on the left 
hand side in   Eqn.({\ref{mat1}}) as,
 \begin{equation}
 \left( \begin{array}{ccc} \Pi_T(k) &   iF     &    -iG \\
-iF &\Pi_T(k)  & 0 \\
 iG & 0 & m^2_{\phi}  \end{array} \right)={\bf M},
 \label{mix-mat}
\end{equation} 

\noindent
and  would like to discuss about the elements of the  matrix  ${\bf M} $ for the
physical situation under consideration in next few lines.\\
\indent
 In the long wavelength limit, one can take $\Pi_{T}=\omega^{2}_{p} $, where 
$\omega_p$ is the plasma frequency. With this identification, the other two parameters $F$ and $G$, are 
given by $F=\frac{\omega^{2}_{p} e\rm{B}\cos\bar\theta }{\omega m_e}$ and 
$G= - g_{\phi\gamma\gamma}\rm{B} \sin \bar\theta \omega$. The angle $\bar\theta$  
here corresponds to the angle between the photon propagation vector $\vec{k}$ 
and the magnetic field $\rm{B}$. Lastly the parameter $m_e$ is the electron mass.   
Now identifying $\rm{B} \cos \bar\theta = B_{\parallel}$ and  $\rm{B} \sin \bar\theta= B_{\perp}$, 
the equation of motion can further be written as, 

\begin{equation}
\left[\begin{array}{c} k^2 {\bf I} - 
               \left( \begin{array}{ccc}              
       \omega^2_p                           &    i\frac{\omega^{2}_{p} eB_{\parallel} }{(\omega m_e)}     & \,\,   -i g_{\phi\gamma\gamma}B_{\perp}\omega   \\
-i\frac{\omega^{2}_{p} eB_{\parallel} }{(\omega m_e)}    &          \omega^{2}_{p}                        &                      0                     \\
ig_{\phi\gamma\gamma}B_{\perp}\omega        &                       0                         &                    m^2_{\phi}  
   \end{array} \right)
 \end{array} \right]
\left[\begin{array}{c} A_{\parallel}(k) \\ A_{\perp}(k) \\\phi(k) \end{array} \right]=0.
\label{photon-scalar-mixing-matixx}
\end{equation} 

\noindent
Although, in this work, we will assume  angle $\bar{\theta}$, to be $\pi/4$, however 
depending on spatial  geometry of the emission region this angle 
can vary between zero to $2 \pi$.\\  
\indent
When $\bar{\theta}= n\pi$, for any integer $n$,
the dilatons decouples from the system and the
resulting equation describe electrodynamics of 
magnetized medium. This can be used to study
the propagation of EM wave in magnetosphere 
of pulsars, as was done in \cite{Beskin_pulsar}.
In the other limit, when $\bar{\theta} = \pi/2$, the 
situation reduces to that of $\gamma-\phi$ interaction in an
unmagnetized medium. \\
\indent
 To obtain the solutions of  Eqn. ({\ref{photon-scalar-mixing-matixx}}), 
first we need to have the knowledge of the eigen values of the matrix on the l.h.s of 
 Eqn.({\ref{photon-scalar-mixing-matixx}}) inside square bracket. Once that 
is done, one can find the corresponding eigen vectors and from there obtain the unitary  matrix, that can reduce the matrix ${\bf M}$, to diagonal form.\\
\indent
The eigen values of matrix M can be obtained from the following cubic equation,
that follows from the characteristic equation,
\begin{eqnarray}
c_{1}{\bf E}_{i}^{3}+c_{2}{\bf E}_{i}^{2}+c_{3}{\bf E}_{i}+c_{4}=0.
\label{cubic}
\end{eqnarray}

\noindent
The variables $c_{2}, c_{3}$ and $c_{4}$ appearing in (\ref{cubic}) are further related to the elements of the mixing matrix ${\bf M}$, 
through the following relations:
\begin{equation}  
c_{1}  = 1,
{\mbox{~~~~~}}
 c_{2} = -(2\omega^2_p + m^2_{\phi}), 
 \end{equation}
 \begin{eqnarray}
 c_{3}& =& \omega^4_p + 2 \omega^{2}_p m^2_{\phi}- \left( \frac{e\rm{B}_{\parallel}}{m_e}\frac{\omega^2_p}{\omega}\right)^{2} - (g_{\phi\gamma\gamma}\rm{B}_{\perp}\omega)^2, \\
 \nonumber\\
 c_{4} &= &-\omega^4_p m^2_{\phi} + \left( \frac{e\rm{B}_{\parallel}}{m_e}\frac{\omega^2_p}{\omega}\right)^{2}m^{2}_{\phi}+\omega^2_p (g_{\phi\gamma\gamma}\rm{B}_{\perp}\omega)^2. 
\end{eqnarray} 

\indent
More details  about the  roots and their properties are 
provided in the appendix. Here we just state the result. The roots are,
\begin{eqnarray}
     {\bf E}_{1} &=&   2 {\cal{ R}} \cos (\alpha -\pi/3) - \frac{{ c_{2}}}{3}  , \\
     {\bf E}_{2} &=&   2 {\bf \cal{R}} \cos (\alpha +\pi/3) - \frac{{ c_{2}}}{3}  , \\
     {\bf E}_{3} &=& -2{\bf \cal{R}}  \cos  ( \alpha )  -\frac{{ c_{2}}}{3}.  
     \label{exact-roots-compact}
\end{eqnarray} 

\noindent
Where the variables  $\alpha  =\frac{1}{3} \cos^{-1} \left( \frac{{\cal Q}}{{\bf \cal{R}}^3} \right )$ and  ${\bf \cal{R}} = \sqrt{\left(-{\cal {P}}   \right )} {\bf sgn} \left( {\cal {Q}} \right)$, when 
${\cal {P}}= \frac{\left(3c_{3} - {c_{2}}^2 \right ) }{9}$ and ${\cal {Q}}=\left(\frac{{c_{2}}^3}{27} - \frac{{ c_{2}}{ c_{3}}}{6}+ \frac{c_{4}}{2}\right) $. This completes our knowledge of the roots.

%############################################################################    
\section{The Unitary Diagonalizing matrix U}
%############################################################################

\indent
Before we go on to discuss oscillation between $\gamma$ and $\phi$, we need to evaluate 
the phase space evolution of the individual fields. To achieve that, we need to have the 
solutions for each one of them. In order to obtain the same,
 the mixing matrix 
$ {\bf M}$ (\ref{mix-mat}) has to be transformed to its diagonal form $ {\bf M}_D$, by unitary matrices U 
i.e.,  ${\bf U}^{\dag} {\bf M} {\bf U} = {\bf M}_{D} $, having eigenvalues, 
${\bf E}_{1}, {\bf E}_{2} \mbox{~and ~} {\bf E}_3$  as the diagonal elements.
The construction of the unitary  matrix ${\bf U}$ had been outlined in the {Appendix}. 
Therefore we provide the final result here. Introducing the quantities
 $u_i= (\omega^2_p -{\bf E}_i)(m^2_{\phi} - {\bf E}_i)$,
$v_i =i\frac{e{\rm{B}}_{\parallel}}{m_e}\frac{\omega^2_p}{\omega} (m^2_{\phi} - {\bf E}_i)$ and
$w_i =ig_{\phi\gamma\gamma}{\rm{B}}_{\perp}\omega (\omega^2_p -{\bf E}_i)$ related to the the eigenvectors
of matrix ${\rm {\bf M}}$, the unitary matrix $U$ can be written as,
\begin{eqnarray}
{\bf U}=\left ( \begin{matrix}
                   & \frac{u_1}{\sqrt{u^2_1+ v^2_1+w^2_1}}        & \frac{u_2}{\sqrt{u^2_2+ v^2_2+w^2_2}}         &  \frac{u_3}{\sqrt{u^2_3+ v^2_3+w^2_3}}  \\                 
                   & \frac{v_1}{\sqrt{u^2_1+ v^2_1+w^2_1}}        & \frac{v_2}{\sqrt{u^2_2+ v^2_2+w^2_2}}         &  \frac{v_3}{\sqrt{u^2_3+ v^2_3+w^2_3}} \\
                   & \frac{w_1}{\sqrt{u^2_1+ v^2_1+w^2_1}}        & \frac{w_2}{\sqrt{u^2_2+ v^2_2+w^2_2}}          & \frac{w_3}{\sqrt{u^2_3+ v^2_3+w^2_3}} 
                 \end{matrix}
        \right).
\label{orthonormal-vectors}
\end{eqnarray}

\noindent
Introducing, $\rm{\cal{N}}^{(i)}_{vn}=\frac{1}{\sqrt{u^2_i+ v^2_i+w^2_i}}$ as the normalization constant
for the $i$'th eigen vector, we can further express  Eqn. (\ref{orthonormal-vectors})
in the following form:\\
%\begin{cetner}
%%\begin{strip}
\begin{equation}\\
{\bf U} =  
               \left( \begin{array}{ccc} 
(\omega^2_p -{\bf E}_1)(m^2_{\phi} - {\bf E}_1) \rm{{\cal{N}}}^{(1)}_{vn}     &      (\omega^2_p -{\bf E}_2)(m^2_{\phi} - {\bf E}_2)  \rm{{\cal{N}}}^{(2)}_{vn}                                      &     (\omega^2_p -{\bf E}_3)(m^2_{\phi} - {\bf E}_3)\rm{{\cal{N}}}^{(3)}_{vn}  \\
i\frac{e{\rm{B}}_{\parallel}}{m_e}\frac{\omega^2_p}{\omega} (m^2_{\phi} - {\bf E}_1)\rm{{\cal{N}}}^{(1)}_{vn}    &    i\frac{e{\rm{B}}_{\parallel}}{m_e}\frac{\omega^2_p}{\omega} (m^2_{\phi} - {\bf E}_2) \rm{{\cal{N}}}^{(2)}_{vn}       &  i\frac{e{ \rm{B}}_{\parallel}}{m_e}\frac{\omega^2_p}{\omega} (m^2_{\phi} - {\bf E}_3)\rm{{\cal{N}}}^{(3)}_{vn} \\
 ig_{\phi\gamma\gamma}{\rm{B}}_{\perp}\omega(\omega^2_p -{\bf E}_1)\rm{\cal{N}}^{(1)}_{vn}                     &   ig_{\phi\gamma\gamma}{\rm{B}}_{\perp}\omega (\omega^2_p -{\bf E}_2) \rm{\cal{N}}^{(2)}_{vn}                           &   ig_{\phi\gamma\gamma}{\rm{B}}_{\perp}\omega (\omega^2_p -{\bf E}_3)\rm{\cal{N}}^{(3)}_{vn}
                       \end{array} 
             \right).
 \label{u-mat}
\end{equation} 
%%\end{strip}

\noindent
The explicit expression for the  $\rm{\cal{N}}^{(i)}_{vn}$ is given by,
%%\begin{strip}
%\begin{table*}
\begin{equation}\\
\rm{\cal{N}}^{(i)}_{vn} = \frac{1}{\sqrt{(\omega^2_p -{\bf E}_i)^2(m_{\phi}^2 -{\bf E}_i)^2 + \left(\frac{e{\rm{B}}_{\parallel}}{m_e}\frac{\omega^2_p}{\omega}\right)^2 (m_{\phi}^2 -{\bf E}_i)^2 +  \left( g_{\phi\gamma\gamma}{\rm{B}}_{\perp}\omega \right)^2(\omega^2_p -{\bf E}_i)^2}},
\label{nc}
\end{equation} 
%\end{table*}
%%\end{strip}
where ${\bf E}_i$ stands for  the corresponding eigen value. The hermitian conjugate matrix 
${\bf U}^{\dagger}$ would follow from (\ref{u-mat}). The unitarity relations, 
${\bf U}^{\dagger}{\bf U}= {\bf 1}$ has been verified numerically as well as analytically, and  ${\bf U}{\bf U}^{\dagger} = {\bf 1}$,
numerically. Analytical verification of  ${\bf U}{\bf U}^{\dagger} = {\bf 1}$ is cumbersome, so we have taken recourse to numerical verification, and checked that they are satisfied.\\
\indent 
There are few relations those are satisfied by the elements of ${\bf U}$, and use of them makes it convenient to express the 
probability amplitudes  in compact notation. In order to derive them we first, rewrite  Eqn. (\ref{u-mat}) with respective 
identifications of the elements, as follows,
\begin{eqnarray}
{\bf U}=  
               \left( 
\begin{matrix}
{\hat{u}_1} &  {\hat{u}_2} & {\hat{u}_3} \cr
{\hat{v}_1} &  {\hat{v}_2} & {\hat{v}_3} \cr
{\hat{w}_1} &  {\hat{w}_2} & {\hat{w}_3} \cr

\end{matrix}                        
             \right).    
 \end{eqnarray}
 The condition  $ {\bf U}^{\dagger}{\bf U}={\bf{1}}$  now implies, for i, j=1,2,3, 
 \begin{eqnarray}
{\hat{u}_i}{\hat{u}_j}^{*} +{\hat{v}_i}{\hat{v}_j}^{*}+ {\hat{w}_i}{\hat{w}_j}^{*} =
\left\{  \begin{matrix} 
                1, ~i =   j  \cr
                0, ~i\neq j.
               \end{matrix}
\right.
\label{ortho-1}
\end{eqnarray} 
A similar exercise for ${\bf U}{\bf U}^{\dagger}={\bf{1}}$, further establishes the following relations amongst the elements
of ${\bf U}$. The off diagonal terms of  ${\bf U}{\bf U}^{\dagger}$ will give,
\begin{eqnarray} 
\sum^{3}_{i=1}\hat{u}_{i}\hat{v}_{i}^{*} =\sum^{3}_{i=1} {\hat{u}_{i}}^{*}\hat{v}_{i}  =  \sum^{3}_{i=1}\mid \hat{u}_{i} \mid \mid \hat{v}_{i} \mid = 0,  
\label{ortho2a} \\
\sum^{3}_{i=1}\hat{v}_{i}\hat{w}_{i}^{*} = \sum^{3}_{i=1}{\hat{v}_{i}}^{*}\hat{w}_{i} =\sum^{3}_{i=1} \mid \hat{v}_{i} \mid \mid \hat{w}_{i}\mid = 0,  
\label{ortho2b} \\
\sum^{3}_{i=1}\hat{w}_{i}\hat{u}_{i}^{*} = \sum^{3}_{i=1}{\hat{w}_{i}}^{*}\hat{u}_{i} = \sum^{3}_{i=1} \mid \hat{w}_{i} \mid \mid \hat{u}_{i}\mid = 0.
\label{ortho2c} 
\end{eqnarray}
\noindent
and the diagonal entries of the same will yield,
\begin{eqnarray}
\sum^{3}_{i=1}\mid \hat{u}_{i}\mid\mid \hat{u}_{i} \mid = \sum^{3}_{i=1} \mid \hat{v}_{i} \mid\mid \hat{v}_{i} \mid= \sum^{3}_{i=1} \mid \hat{w}_{i} \mid\mid \hat{w}_{i} \mid=  1.
\label{ortho2d}
\end{eqnarray}

\noindent
 The variables  $\hat{u}^{*}_{i}, \hat{v}^{*}_{i}$, and  $\hat{w}^{*}_{i} $ appearing in above expressions, represent the conjugates of the corresponding elements of  ${\bf U} $.

%############################################################################
\section{Conversion Probability}
%############################################################################

\indent
The conversion probability, of a photon of a particular polarization state to the same of a different polarization state  or scalars can be estimated from the evolution (quantum evolution) equation of the corresponding polarized states of the photon and the scalars. One can perform the same by promoting the momentum variables to corresponding operators and components of the vector potential $A^{\nu}(k)$ and $\phi(k)$ to the corresponding quantum states, following \cite{Raffelt,Mirizzi-raffelt1}. In order to follow the evolution of individual quantum states, one needs to decouple them from each other by the following way. One can multiply equation (\ref{photon-scalar-mixing-matixx}) by ${\bf U}^{\dagger}$ from left
\footnote{ Since ${\bf U}^{-1}$ = ${\bf U}^{\dagger}$.} , 
to reduce it to the following form, 
\begin{equation}
\left[\begin{array}{c} k^2 \mathbf{I} - {\bf U}^{\dag}{\bf M} {\bf U} \end{array} \right]
\left[\begin{array}{c} |A'_{\parallel}(k)> \\ |A'_{\perp}(k)> \\ |\phi'(k)> 
\end{array} \right]=0,
\label{decoupling-the-eom}
\end{equation} 
%%%%%%%%
\noindent
when ${\bf U}^{\dag}{\bf M U} $ is a diagonal matrix. In the diagonal representation, the propagating states
are the diagonal states and they allow principle of superposition.
The matrix ${\bf U} $ is given in Eq. (\ref{u-mat}) and $ {\bf U}^{\dag}$ is the hermitian conjugate  of the same. Here we have denoted \footnote{The set of unprimed and primed column vectors, at places,, 
may be defined collectively as: 
$\left[{\bf A(k)} \right] = \left(A_{\parallel}(k),A_{\perp}(k), \phi(k) \right)^{{\bf T}}$ and
$\left[{\bf A(k)}' \right] = \left(A'_{\parallel}(k), 
A'_{\perp}(k), \phi(k) \right)^{\bf T} $, here the superscript {\bf T} stands for transpose.}
\begin{eqnarray}
 \left[\begin{array}{c} |A'_{\parallel}(k)> \\ |A'_{\perp}(k)> \\ |\phi'(k)> 
\end{array} \right]={\bf U}^{\dagger}
\left[\begin{array}{c} |A_{\parallel}(k)> \\ |A_{\perp}(k)> \\|\phi(k)> \end{array} \right].
\label{primed-degrees-of-freedom}
\end{eqnarray}

%\noindent
The primed states corresponds to the propagating states and unprimed ones are the physical states; They are related to each other by the unitary transformation  by ${\bf U}$ introduced earlier.
For a beam of photon, propagating in the z direction, following the principles stated already, one can promote the momentum  $k_{3}$ to the corresponding operator in 
 $z$ space and write (using natural units $\hbar= c=1$) $k^2 \approx 2\omega(\omega -i\partial_z)$. 
 With this manipulations, the resulting equations get transformed from Klien-Gordon to the form used by \cite{Raffelt}. Recalling that ${\bf U}^{\dag}{\bf M} {\bf U} = {\bf M}_{D}$, 
where ${\bf M}_D$ is the diagonal matrix, equation (\ref{primed-degrees-of-freedom}) 
can further be cast in the form,
%%%%%%%%

\begin{equation}
\left[\begin{array}{c}  (\omega -i\partial_z)  \mathbf{I}  - \left[ \begin{array}{ccc} \frac{{\bf E}_1}{2\omega} &   0     &    0 \\
0 &  \frac{{\bf E}_2}{2\omega}  & 0 \\
0 & 0 & \frac{{\bf E}_3}{2\omega}  \end{array} \right]        \end{array} \right]
\left[\begin{array}{c} |{A'}_{\parallel}(z)>\\| {A'}_{\perp}(z)> \\ |{\phi'}(z)> \end{array} \right]=0.
\label{dmat}
\end{equation} 
\noindent
The matrix evolution  Eqn. (\ref{dmat}) is now easy to solve. Introducing the variables, 
$\Omega_\parallel = \left(\omega - \frac{{\bf E}_1}{2\omega}\right) $,
$\Omega_\perp = \left(\omega - \frac{{\bf E}_2}{2\omega}\right)$ 
and 
$\Omega_\phi = \left(\omega - \frac{{\bf E}_3}{2\omega}\right)$,
we can now directly write down the solutions for the states  vector $\left[|{\bf A}(z)> \right]$ ( where ${\bf A}(z) \equiv {\bf A}(\omega, k_{\perp}, z) $) in the following form,
\begin{equation}
  \left[
\begin{matrix}
{|{A}_{\parallel}(z)>} \cr
{|{A}_{\perp}(z)>} \cr   
{|\phi(z)>}
\end{matrix}
\right] = {\bf U}\left[ \begin{array}{ccc} e^{-i\Omega_\parallel z} &   0    &    0 \\
0 & e^{-i\Omega_\perp z}  & 0 \\
0 & 0 & e^{-i\Omega_\phi z}  \end{array} \right] {\bf U}^{\dagger} 
\left[
\begin{matrix}
{|{A}_{\parallel}(0)>} \cr
{|{A}_{\perp}(0)>} \cr   
|{\phi(0)}>
\end{matrix}
\right]
\label{solnmat}.
\end{equation}
\noindent
The elements of column vector $\left[|{\bf A}(0)> \right]$  in (\ref{solnmat}) and  $\left[|{\bf A_{L}}(0)> \right]$, are normalized  
 such that, $<A_{\parallel}(0)|A_{\parallel}(0)> = <A_{\perp}(0)|A_{\perp}(0)>= <\phi(0)|\phi(0)> = <A_{L}(0)|A_{L}(0)> =1$. With the help of  Eqn. (\ref{solnmat}) one can write down the solutions. And they are as follows:
\begin{eqnarray}
|A_{\parallel}(\omega,z)>&=& 
\left( e^{-i\Omega_\parallel z }{\hat u}_1 { \hat u}^{*}_1  +  e^{-i\Omega_\perp z } {\hat u}_2{ \hat u}^{*}_2   
+ e^{-i\Omega_\phi z } {\hat u}_3{ \hat u}^{*}_3 \right)|A_{\parallel}(\omega, 0)>  \nonumber \\
&+& \left(
e^{-i\Omega_\parallel z }{\hat u}_1 { \hat v}^{*}_1  +  e^{-i\Omega_\perp z } {\hat u}_2{ \hat v}^{*}_2   
+ e^{-i\Omega_\phi z } {\hat u}_3{ \hat v}^{*}_3 \right)|A_{\perp}(\omega, 0) > \nonumber  \\
&+&
\left(
e^{-i\Omega_\parallel z }{\hat u}_1 { \hat w}^{*}_1  +  e^{-i\Omega_\perp z } {\hat u}_2{ \hat w}^{*}_2   
+ e^{-i\Omega_\phi z } {\hat u}_3{ \hat w}^{*}_3 \right) |\phi(\omega, 0)>,
\label{soln-a-parallel}
\end{eqnarray}  
\noindent 
the perpendicular $|A_{\perp}(\omega,z) >$ component is given by, 
\begin{eqnarray}
|A_{\perp}(\omega,z)>&=& 
\left( e^{-i\Omega_\parallel z }{\hat v}_1 { \hat u}^{*}_1  +  e^{-i\Omega_\perp z } {\hat v}_2{ \hat u}^{*}_2   
+ e^{-i\Omega_\phi z } {\hat v}_3{ \hat u}^{*}_3 \right)|A_{\parallel}(\omega, 0)>  \nonumber \\
&+& \left(
e^{-i\Omega_\parallel z }{\hat v}_1 { \hat v}^{*}_1  +  e^{-i\Omega_\perp z } {\hat v}_2{ \hat v}^{*}_2   
+ e^{-i\Omega_\phi z } {\hat v}_3{ \hat v}^{*}_3 \right)|A_{\perp}(\omega, 0) > \nonumber  \\
&+&
\left(
e^{-i\Omega_\parallel z }{\hat v}_1 { \hat w}^{*}_1  +  e^{-i\Omega_\perp z } {\hat v}_2{ \hat w}^{*}_2   
+ e^{-i\Omega_\phi z } {\hat v}_3{ \hat w}^{*}_3 \right)| \phi(\omega, 0)>, 
\label{soln-a-perp}
\end{eqnarray}  
\noindent
and lastly the evolution of the state  $|\phi(\omega,z)>$ is given by
\begin{eqnarray}
|\phi(\omega,z)>&=& 
\left( e^{-i\Omega_\parallel z }{\hat w}_1 { \hat u}^{*}_1  +  e^{-i\Omega_\perp z } {\hat w}_2{ \hat u}^{*}_2   
+ e^{-i\Omega_\phi z } {\hat w}_3{ \hat u}^{*}_3 \right)|A_{\parallel}(\omega, 0)>  \nonumber \\
&+& \left(
e^{-i\Omega_\parallel z }{\hat w}_1 { \hat v}^{*}_1  +  e^{-i\Omega_\perp z } {\hat w}_2{ \hat v}^{*}_2   
+ e^{-i\Omega_\phi z } {\hat w}_3{ \hat v}^{*}_3 \right)|A_{\perp}(\omega, 0)>  \nonumber  \\
&+&
\left(
e^{-i\Omega_\parallel z }{\hat w}_1 { \hat w}^{*}_1  +  e^{-i\Omega_\perp z } {\hat w}_2{ \hat w}^{*}_2   
+ e^{-i\Omega_\phi z } {\hat w}_3{ \hat w}^{*}_3 \right) |\phi(\omega, 0)>.
\label{soln-a-parallel}
\end{eqnarray}  
\noindent
The ways to arrive at these results can be found in  \cite{CJG}.
We would like to end this subsection with the following observation, that the states
defined by $|A_{\parallel}(\omega, 0)>$,   $|A_{\perp}(\omega, 0)>$ and  $|\phi(\omega, 0)>$ are pure states. The corresponding  states denoted by  $|A_{\parallel}(\omega, z)>$,   $|A_{\perp}(\omega, z)>$ and  $|\phi(\omega, z)>$ are the  mixed states those evolve from the pure ones through propagation in phase space through mixing. Even if any one of them is  absent at the beginning, it can be generated later  through mixing much like the neutrinos.  

%############################################################################
\subsection{Oscillation probability $P_{\gamma_{\parallel} \to \phi}$}
%############################################################################

\indent
The amplitude for the transition  of a photon of energy $\omega$ in state $|A_{\parallel}(\omega,0)>$ to $|\phi(\omega,z)>$ after traversing a distance z, is given by $<A_{\parallel}(\omega,0)|\phi(\omega,z)>$.
The probability of the same, $ P_{\gamma_{\parallel} \rightarrow \phi}(\omega,z)$ can be estimated from the evolution equations obtained above by using the formula $ P_{\gamma_{\parallel} \rightarrow \phi}(\omega,z)=|<A_{\parallel}(\omega,0)|\phi(\omega,z)>|^2 $. The same same turns out to be,
\begin{eqnarray} 
P_{\gamma_{\parallel} \rightarrow \phi} = 
\left|     
\left(
e^{-i\Omega_\parallel z }{\hat u}_1 { \hat w}^{*}_1  +  e^{-i\Omega_\perp z } {\hat u}_2{ \hat w}^{*}_2   
+ e^{-i\Omega_\phi z } {\hat u}_3{ \hat w}^{*}_3 \right) 
\right|^{2}. 
\label{gamma-para-to-scalar}
\end{eqnarray}
%%%%
After performing some lengthy algebra, one can observe that, the resulting expression for   Eqn. 
(\ref{gamma-para-to-scalar}) contains a sum of three quadratic pieces, i.e.,
$\left[ \left(  |{\hat u}_1||{\hat w}_1| \right)^2 + \left( |{\hat u}_2||{\hat w}_2| \right)^2 + 
\left(|{\hat u}_3||{\hat w}_3|\right)^2 \right]$
plus three other pieces involving the distance parameter $z$. Upon converting, these
quadratic pieces into a square of their sum and rearranging the resultant expression,
  Eqn.(\ref{gamma-para-to-scalar}) reduces to the following form:
\begin{eqnarray} 
P_{\gamma_{\parallel} \rightarrow \phi} &=&      
 \left[ {\hat u}_1||{\hat w}_1| + |{\hat u}_2||{\hat w}_2|+ |{\hat u}_3||{\hat w}_3| \right]^{2} \nonumber  \\
&-& 2|{\hat u}_1||{\hat w}_1||{\hat u}_2||{\hat w}_2|\left[1 - \cos \left( \left(\Omega_\perp - \Omega_\parallel \right)z \right) \right]
\nonumber \\
&-& 2|{\hat u}_1||{\hat w}_1||{\hat u}_3||{\hat w}_3|\left[1 - \cos\left( \left(\Omega_\parallel - \Omega_\phi \right)z \right)   \right]
\nonumber \\
&-& 2|{\hat u}_3||{\hat w}_3||{\hat u}_2||{\hat w}_2| \left[1- \cos\left( \left(\Omega_\phi      -  \Omega_\perp \right)z \right) \right]. 
\label{gamma-para-to-scalar-2}
\end{eqnarray}
\noindent
At this stage, it is convenient to consider defining, $\mathbb{A}=|{\hat u}_1||{\hat w}_1|$, $\mathbb{B}=|{\hat u}_2||{\hat w}_2|$
and $ \mathbb{C}= |{\hat u}_3||{\hat w}_3|$.  
The constraint,$ |{\hat u}_1||{\hat w}_1| + |{\hat u}_2||{\hat w}_2|+ |{\hat u}_3||{\hat w}_3|=0 $, 
that follows from  equation (\ref{ortho2c}), 
can now be recasted as, 
\begin{eqnarray}
 \mathbb{A} +  \mathbb{B}+ \mathbb{C}= 0.
\label{A+B+C=0}
\end{eqnarray}
\noindent
Now we can make use of  Eqn. (\ref{A+B+C=0}) in  (\ref{gamma-para-to-scalar-2}), to verify that the perfect square term 
vanishes owing to the constraint equation (\ref{A+B+C=0}); and the remaining $z$ dependent pieces, can be manipulated  
further  to provide,
\begin{eqnarray} 
P_{\gamma_{\parallel} \rightarrow \phi} &=& 4 \mathbb{A} \left( \mathbb{A}+\mathbb{C} \right)
\sin^2   \left( \frac{\left(\Omega_\perp - \Omega_\parallel \right)z}{2} \right) \nonumber\\    
&+& 4 \mathbb{B} \left( \mathbb{B}+\mathbb{A} \right)
\sin^2 \left(\frac{\left(\Omega_\phi - \Omega_\perp \right)z}{2} \right) \nonumber\\
&+& 4 \mathbb{C} \left( \mathbb{C}+\mathbb{B} \right)
\sin^2\left(\frac{\left(\Omega_\parallel - \Omega_\phi \right)z}{2} \right).
\label{gamma-para-to-scalar-3}
\end{eqnarray}

\noindent
For the sake of completeness, we provide the expressions for the new variables, $\mathbb{A}$, $\mathbb{B}$ and  $\mathbb{C}$, introduced earlier, 
in terms of the parameters of the matrix ${\bf M }$. And they are: 
\begin{eqnarray}
\mathbb{A} = \rm{\cal{N}}^{(1)}_{vn}  \rm{\cal{N}}^{(1)}_{vn}\left( g_{\phi\gamma\gamma}{B}_{\perp}\omega \right) (\omega^2_p- {\bf E}_1) (\omega^2_p- {\bf E}_1)(m^2_\phi- {\bf E}_1),
\label{bfA} \\ 
\mathbb{B} = \rm{\cal{N}}^{(2)}_{vn}  \rm{\cal{N}}^{(2)}_{vn}\left( g_{\phi\gamma\gamma}{B}_{\perp}\omega \right) (\omega^2_p- {\bf E}_2) (\omega^2_p- {\bf E}_2)(m^2_\phi- {\bf E}_2), 
\label{bfB} \\
\mathbb{C} = \rm{\cal{N}}^{(3)}_{vn}  \rm{\cal{N}}^{(3)}_{vn}\left( g_{\phi\gamma\gamma}{B}_{\perp}\omega \right) (\omega^2_p- {\bf E}_3) (\omega^2_p- {\bf E}_3)(m^2_\phi- {\bf E}_3).
\label{bfC}
\label{bold-ABC}
\end{eqnarray}
The fact that equations (\ref{bfA}--\ref{bfC} ) follow  Eqn. (\ref{A+B+C=0}), to a good accuracy, has been verified numerically.  
\subsection{Oscillation probability $P_{\gamma_{\perp} \to \phi}$}
The oscillation probability for the $A_{\perp}$ component of a  photon  to scalar $\phi$
can be derived similarly. Therefore, instead of going through the same set of arguments,
we would provide the final result. Before we go to the final expression, like before,
we introduce the new set of variables,  $\mathbb{L}$,  $\mathbb{M}$ and $\mathbb{N}$; 
defined as,  $\mathbb{L}= | {\hat v}_1||{\hat w}_1|$,  $\mathbb{M} =|{\hat v}_2||{\hat w}_2| $  
and $\mathbb{N}= |{\hat v}_3||{\hat w}_3| $. Their actual form, in terms of the elements of 
the mixing matrix will be provided shortly. The expression for $ P_{\gamma_{\perp} \rightarrow \phi}$ 
is:\\
\begin{eqnarray} 
P_{\gamma_{\perp} \rightarrow \phi} &=& 4 \mathbb{L} \left( \mathbb{L}+\mathbb{N} \right)
\sin^2   \left( \frac{\left(\Omega_\perp - \Omega_\parallel \right)z}{2} \right) \\    
&+& 4 \mathbb{M} \left( \mathbb{M}+\mathbb{L} \right)
\sin^2 \left(\frac{\left(\Omega_\phi - \Omega_\perp \right)z}{2} \right) \\
&+& 4 \mathbb{N} \left( \mathbb{N}+\mathbb{M} \right)
\sin^2\left(\frac{\left(\Omega_\parallel - \Omega_\phi \right)z}{2} \right).
\label{gamma-perp-to-scalar-1}
\end{eqnarray}
The values of the parameters  $\mathbb{L}$, $\mathbb{M}$ and $\mathbb{N}$ are in terms of the parameters of the theory  are given by,
\begin{eqnarray}
\mathbb{L} = \rm{\cal{N}}^{(1)}_{vn} \rm{\cal{N}}^{(1)}_{vn}  \left( g_{\phi\gamma\gamma}{\rm{B}}_{\perp}\omega \right)\left(\frac{e{\rm{B}}_{\parallel}}{m_e}\frac{\omega^2_p}{\omega}\right)(\omega^2_p - {\bf E}_1)(m^2_{\phi}- {\bf E}_1), 
\label{boldL} \\
\mathbb{M} = \rm{\cal{N}}^{(2)}_{vn} \rm{\cal{N}}^{(2)}_{vn} \left( g_{\phi\gamma\gamma}{\rm{B}}_{\perp}\omega \right)\left(\frac{e{\rm{B}}_{\parallel}}{m_e}\frac{\omega^2_p}{\omega}\right)(\omega^2_p - {\bf E}_2)(m^2_{\phi}-  {\bf E}_2),
\label{boldM} \\
\mathbb{M} = \rm{\cal{N}}^{(3)}_{vn} \rm{\cal{N}}^{(3)}_{vn} \left( g_{\phi\gamma\gamma}{\rm{B}}_{\perp}\omega \right)\left(\frac{e{\rm{B}}_{\parallel}}{m_e}\frac{\omega^2_p}{\omega}\right)(\omega^2_p - {\bf E}_3)(m^2_{\phi}-  {\bf E}_3).
\label{boldN} 
\end{eqnarray}
\noindent
As before, following equation (\ref{ortho2b}),  equations (\ref{boldL}--\ref{boldN}) too have to satisfy the constraint  $\mathbb{L}+\mathbb{M}+ \mathbb{N} = 0$. 
We have tested the same numerically to a good accuracy during the  corresponding probability evaluation. 
%%%%%%%%%%%%%%%%%%%%%%%%%%%%%%%%%%%%%%%%%%%%%%%%%%%%%%%%%%%%%%%%%%%%%%%%%%%%%%%%%%%%%%%%%%
\subsection{Oscillation probability $P_{\gamma_{\parallel} \rightarrow \gamma_{\perp}}$}
%%%%%%%%%%%%%%%%%%%%%%%%%%%%%%%%%%%%%%%%%%%%%%%%%%%%%%%%%%%%%%%%%%%%%%%%%%%%%%%%%%%%%%%%%%
\noindent
Lastly, we provide the  conversion probability for parallel component of photon to perpendicular component 
of photon here i.e.,$P_{\gamma_{\parallel} \rightarrow \gamma_{\perp}}$  . Introducing, as before, the quantities, to be defined later below,
$\mathbb{P}= | {\hat u}_1||{\hat v}_1|$,  $\mathbb{Q}= | {\hat u}_2||{\hat v}_2|$ and 
$\mathbb{R}= | {\hat u}_3||{\hat v}_3|$. The corresponding probablity for conversion, 
$P_{\gamma_{\parallel} \rightarrow \gamma_{\perp}}$ turns out to be:
\begin{eqnarray} 
P_{\gamma_{\parallel} \rightarrow \gamma_{\perp}} &=& 4 \mathbb{P} \left( \mathbb{P}+\mathbb{R} \right)
\sin^2   \left( \frac{\left(\Omega_\perp - \Omega_\parallel \right)z}{2} \right) \\    
&+& 4 \mathbb{Q} \left( \mathbb{Q}+\mathbb{P} \right)
\sin^2 \left(\frac{\left(\Omega_\phi - \Omega_\perp \right)z}{2} \right) \\
&+& 4 \mathbb{R} \left( \mathbb{R}+\mathbb{Q} \right)
\sin^2\left(\frac{\left(\Omega_\parallel - \Omega_\phi \right)z}{2} \right),
\label{gamma-parallel-to-gamma-perp}
\end{eqnarray}
\noindent
where,
\begin{eqnarray}
\mathbb{P} =      \rm{\cal{N}}^{(1)}_{vn}   \rm{\cal{N}}^{(1)}_{vn} \left(\frac{e{\rm{B}}_{\parallel}}{m_e}\frac{\omega^2_p}{\omega}\right)(\omega^2_p- {\bf E}_1)(m^2_{\phi}- {\bf E}_1)^{2},
 \label{boldP} \\
\mathbb{Q} =      \rm{\cal{N}}^{(2)}_{vn}   \rm{\cal{N}}^{(2)}_{vn} \left(\frac{e{\rm{B}}_{\parallel}}{m_e}\frac{\omega^2_p}{\omega}\right)(\omega^2_p- {\bf E}_2)(m^2_{\phi}- {\bf E}_2)^{2},
\label{boldQ}  \\
\mathbb{R} =       \rm{\cal{N}}^{(3)}_{vn}  \rm{\cal{N}}^{(3)}_{vn}  \left(\frac{e{\rm{B}}_{\parallel}}{m_e}\frac{\omega^2_p}{\omega}\right)(\omega^2_p- {\bf E}_3)(m^2_{\phi}- {\bf E}_3)^{2}.
\label{boldR}
\end{eqnarray}
\noindent
Like before we have verified that, the condition, $ \mathbb{P} + \mathbb{Q} + \mathbb{R}=0$  is maintained to a good accuracy all along during the course of  the computation.      \\
\indent
 There is  one important observation that follows from the probabilities derived above; that is: The probability $P_{\gamma_{\parallel} \to \gamma_{\perp}}$ is the only probability out of  the three discussed above, that survives in the limit $g_{\phi\gamma\gamma} \to 0$ and $m_{\phi} \to 0$. This  can be related
to "rotation measure" (RM), that's usually encountered in studies of optical activity relating the angle of rotation of the plane of polarization of a beam of plane polarized light, after travelling some distance L. The RM in this case, can be defined  by $\pi/\bar{\lambda}$, when $\bar{\lambda}$ is the minimum distance that a plane polarized light beam needs to travel, to have $P_{\gamma_{\parallel} \to \gamma_{\perp}}(\omega, \bar{\lambda})=1$. The non-linear dependence of $\bar{\lambda}$  on the parameters of the theory indicates that it is difficult to separate the contributions to  $P_{\gamma_{\parallel} \rightarrow \gamma_{\perp}}$ into  parts originating  from  (a) the magnetized plasma and   (b) the one originating from magnetic field induced - dilatonic interactions. Thus  it should be   considered as a total of the  two  contributions mentioned above. \\
\indent
Lastly, the other important aspect of this analysis, is that the other  probabilities i.e., $P_{\gamma_{\parallel} \to {\phi}}$  and $P_{{\phi} \to \gamma_{\parallel}}$, $P_{\gamma_{\perp} \to {\phi}}$, and $P_{{\phi} \to \gamma_{\perp}}$  along with $P_{\gamma_{\parallel} \to \gamma_{\perp}}$  and $P_{\gamma_{\perp} \to \gamma_{\parallel}}$, turn out to be same as they should be even otherwise.

\begin{figure}[h!]
\begin{minipage}[b]{.50\textwidth}  
\includegraphics[width=1\linewidth]{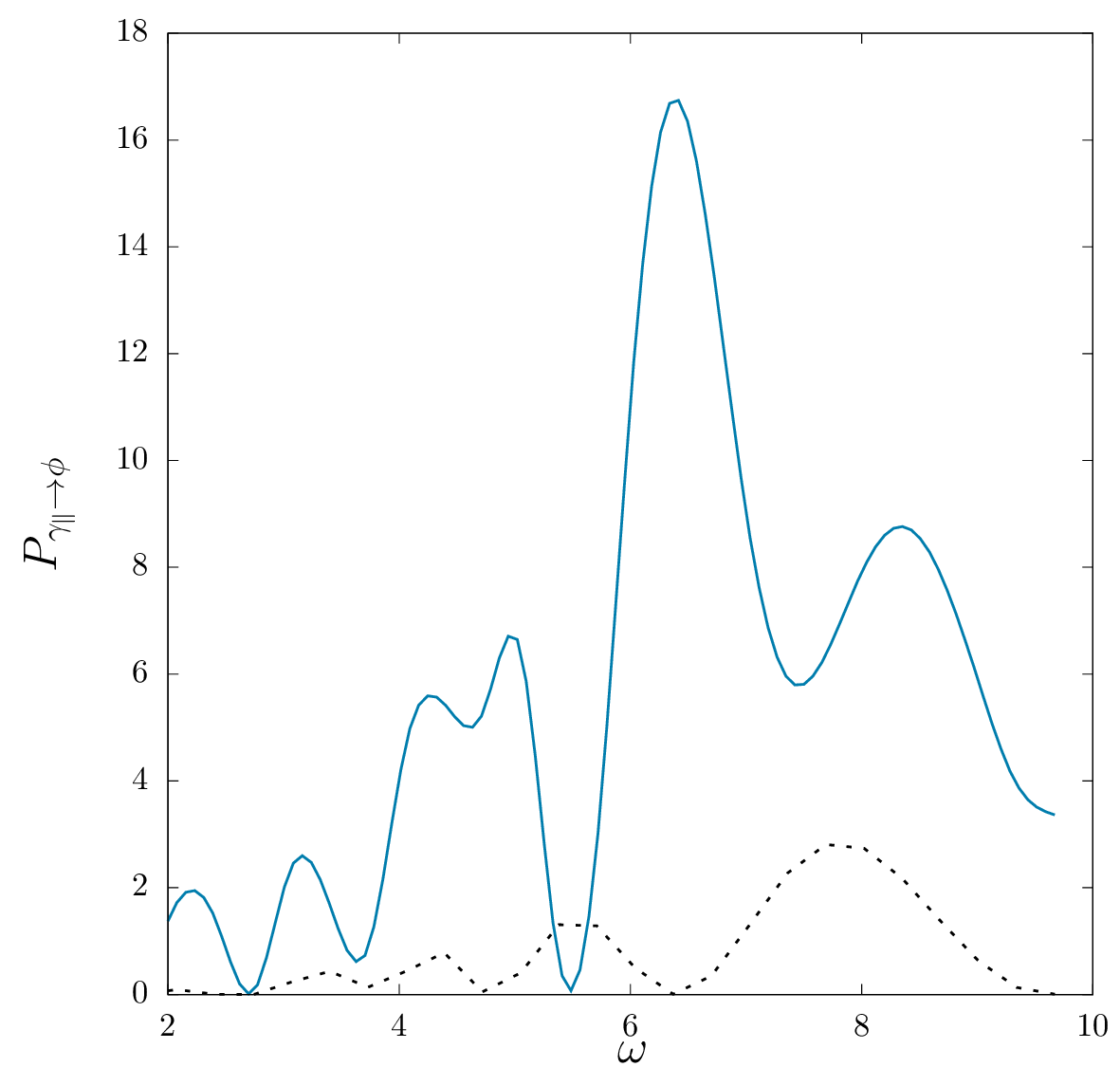}
%\label{f1}
\end{minipage}
\begin{minipage}[b]{.50\textwidth}  
\includegraphics[width=1\linewidth]{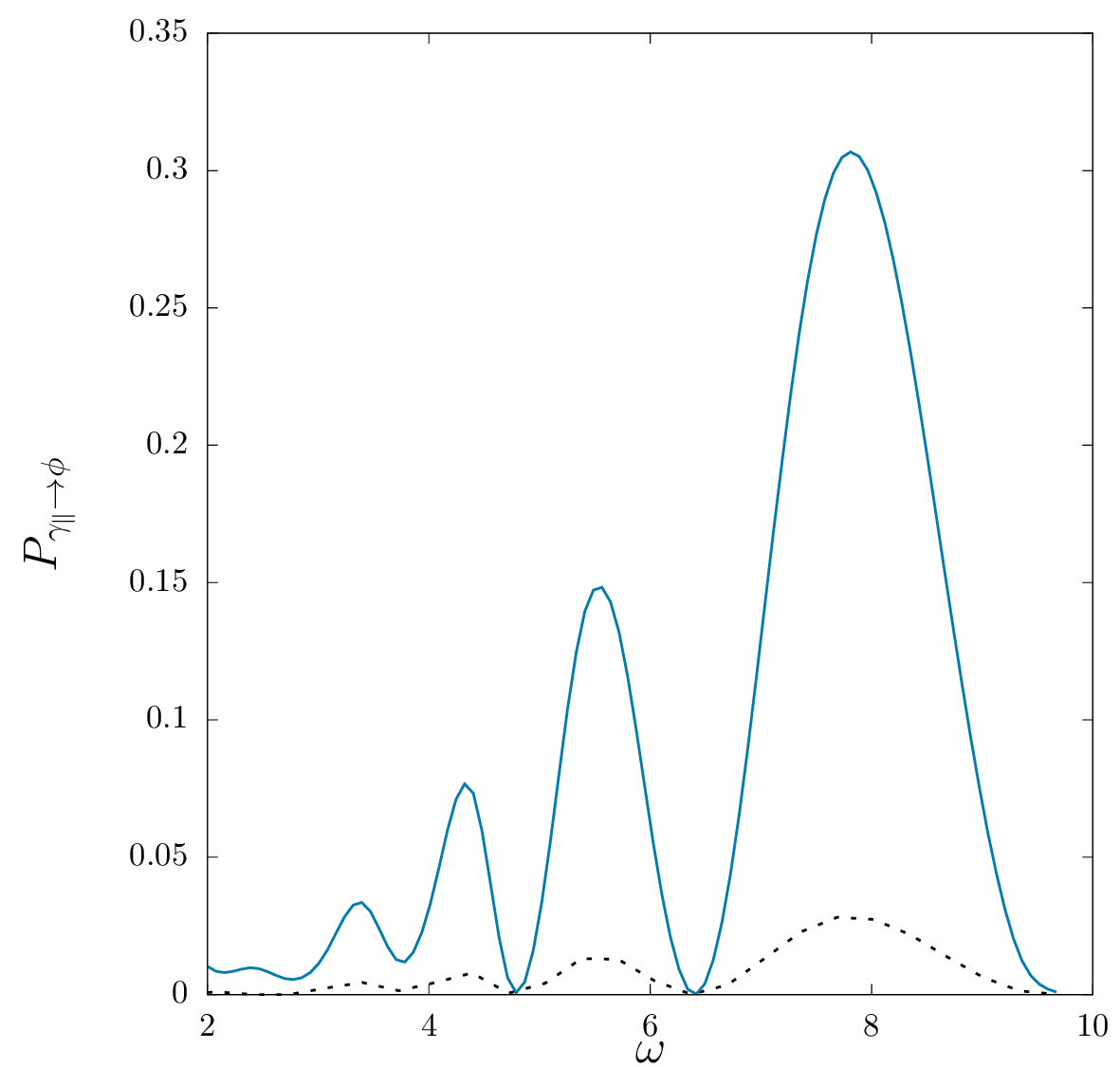}
\end{minipage}
\caption{ Plot of conversion probability of a  parallely polarized photon into a scalar dilaton  in magnetic field $\rm{B} = 10^{12}$ Gauss ({\bf in the  left panel}) and the same in magnetic field $\rm{B} = 10^{11}$ Gauss ({\bf in the right panel}). The solid line is for  conversion probability  in presence of magnetized media and the dashed curve is the same in absence of self energy correction $\Pi^{p}_{\mu\nu}$ from magnetized medium effects.  The abscissa (energy of photon ($\omega$)  in  GeV) is in units of   $10^{-5}$ and the ordinates is plotted  in the units of $10^{-4}$.
Here, the parameters used are: mass of dilaton (scalar) particle $\phi $:  $ (m_{\phi}) = 1.0 \times 10^{-12}$ GeV, coupling constant $ (g_{\phi\gamma\gamma}) = 1.0 \times 10^{-11}$ GeV, photon path length (z) = 1.2 Km, and plasma frequency  $(\omega_{p}) = 1.96 \times 10^{-2}$ eV.}
\label{f001}
\end{figure}

\begin{figure}[h!]
\begin{minipage}[b]{.50\textwidth}  
\includegraphics[width=1\linewidth]{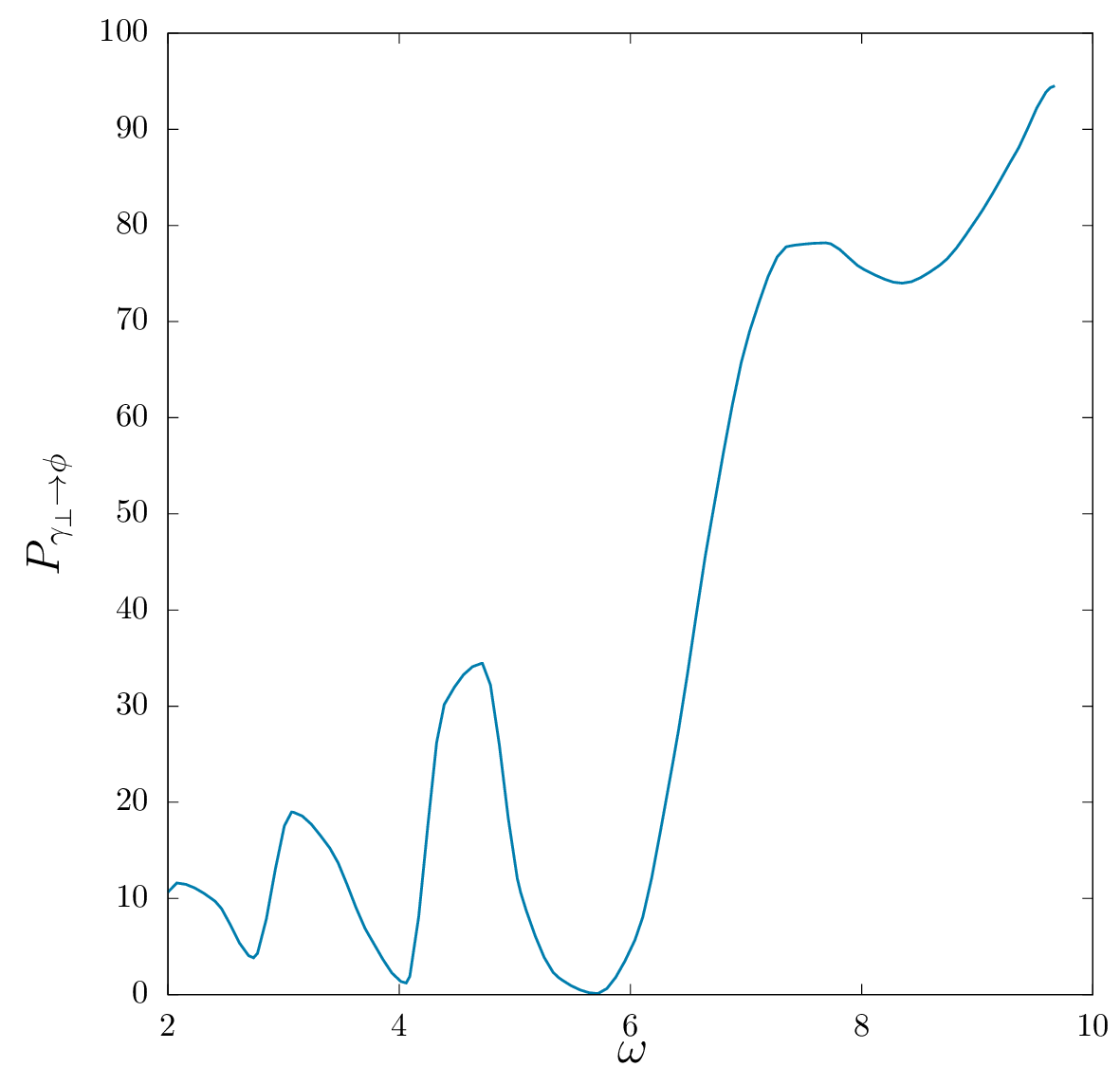}
%\label{f1}
\end{minipage}
\begin{minipage}[b]{.50\textwidth}  
\includegraphics[width=1\linewidth]{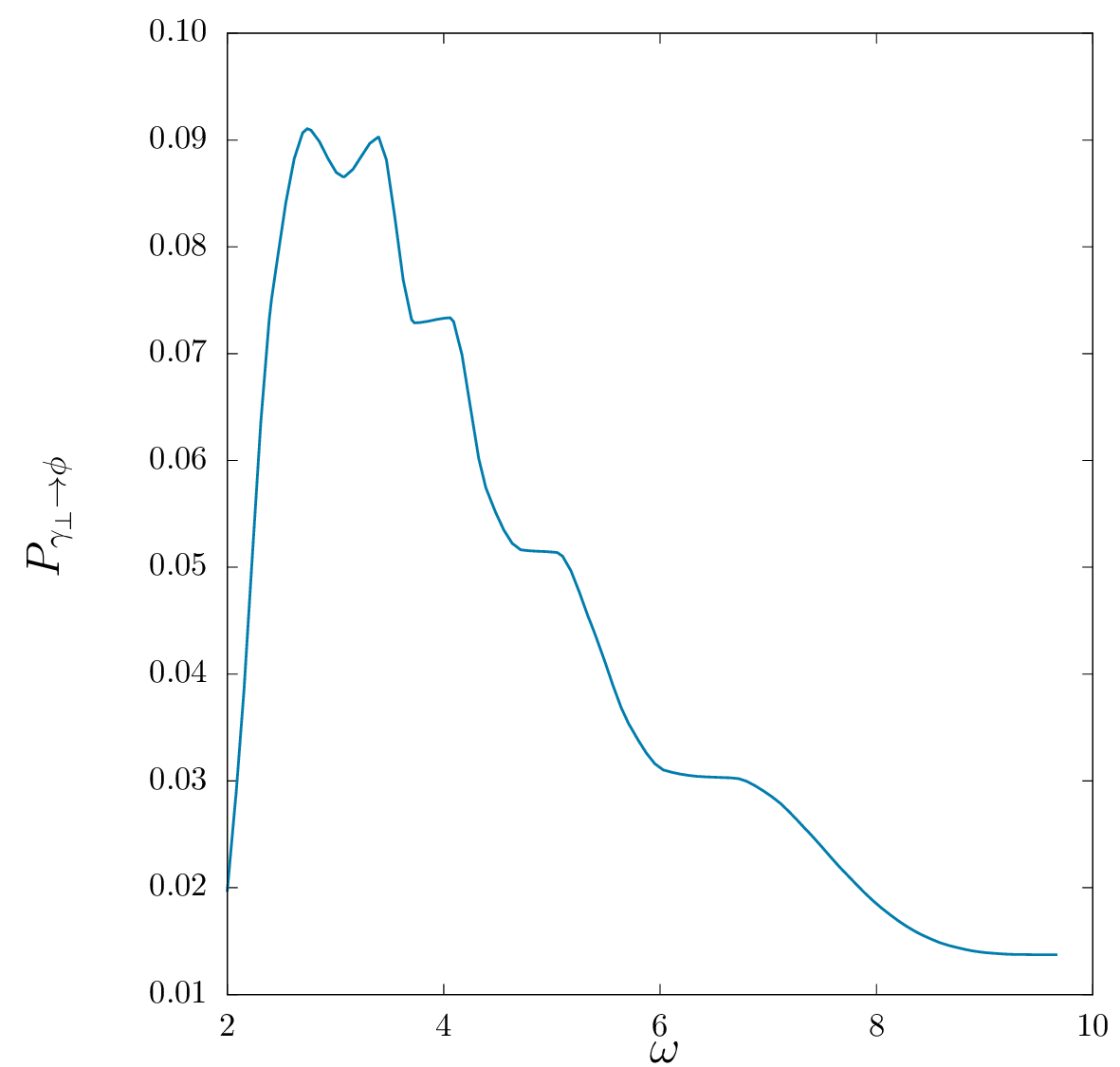}
\end{minipage}
\caption{ Plot of conversion probability of perpendicularly  polarized photon into scalar in magnetic field $\rm{B} = 10^{12}$ Gauss ({\bf in the left   panel}) and the same in magnetic field $\rm{B} = 10^{11}$ Gauss ({\bf in the right  panel}). The abscissa (energy of photon ($\omega$)  in units of $10^{-5}$ GeV)  and the ordinate are plotted in units of   $10^{-5}$.
Here, the parameters used are: mass of dilaton (scalar) particle $\phi $  $ (m_{\phi}) = 1.0 \times 10^{-12}$ GeV, coupling constant $ (g_{\phi\gamma\gamma}) = 1.0 \times 10^{-11}$ GeV, photon path length (z) = 1.2 Km and plasma frequency  $(\omega_{p}) = 1.96 \times 10^{-2}$ eV.}
\label{f002}
\end{figure}

%############################################################################
\subsection{Conversion Probability $P_{\gamma_\parallel \to \phi}$ in an un-magnetized medium.} 
%############################################################################

\indent
In this subsection we provide the photon scalar conversion rate without magnetized medium effects to compare its size with same due to magnetized medium induced effects. The mixing matrix for the same is similar to  that of chamelion-photon mixing in an unmagnetized media, that is 2 $\times$ 2. The probability for the photon scalar transition obtained in \cite{burrage} turns out to be same as that of \cite{Raffelt}. This  probability of transition without any approximation is given by:
\begin{eqnarray} 
P_{\gamma_{\parallel} \rightarrow \phi} = \frac{4{\rm{B_{\perp}}}^2 {\omega}^2}{{(M m_{ef}^2)}^2 +4 {{\rm{B_{\perp}}}}^2 {\omega}^2} \sin^2 \left(\frac{\sqrt{ {{(M m_{ef}^2)}^2 +4 {\rm{B_{\perp}}}^2 {\omega}^2}}}{4 \omega M} z\right).
\label{prgc}
 \end{eqnarray}

\noindent
When in  Eqn. (\ref{prgc}) $ m^2_{ef} = m^2_{\phi} - \omega^2_{p} - \frac{{\rm{B_{\perp}}}^2}{M^2}$ and $m_{\phi}$ is mass of scalar field, as stated before.
 In this case only one degree of freedom of photon (i.e., $A_\parallel$) mix with the scalar, and the other degrees of freedom ($A_\perp$, $A_{L}$) propagate freely.
 It has two special features; One is, in the limit  when the  angle between $\rm{B}$ and $k$ $\to$ zero or $\pi$, the probability in (\ref{prgc}) vanishes, indicating a decoupling of fields. The second one is, in non-interacting limit; i.e., $g_{\phi\gamma\gamma}\to 0$ or $M \to \infty $, $P_{\gamma_{\parallel}\to \phi}\to 0$ for $\omega_{p} \neq m_{\phi}$. But in the same limit if $\omega_{p} = m_{\phi}$, then the oscillation
 length diverges. It is straight forward to verify the same.
In the magnetized plasma case, however the $P_{\gamma_{\parallel}\to \perp}$ and $P_{\gamma_{\perp}\to \parallel}$ remains finite even if the angle between $\rm{B}$ and $k$ $\to$ zero or $\pi$. These checks are useful to prove the consistency of the results.

\newcommand{\R}{\bar{R}}
\newcommand{\B}{\rm{\bf B}_s}
%%%%%%%%%%%%%%%%%%%%%%%%%%%%%%%%%%%%%%%
\section{Astrophysical Applications}

\indent
Presence of very strong magnetic field and an ambient plasma in the environment
of astrophysical compact objects like white dwarves or neutron stars, gamma ray bursters (GRB) etc., provides an  opportunity to look for astrophysical signatures of the kind of particles, we have
been studying in this article. In this work, we have focussed on the  EM
signals from the compact objects that may bear  possible signatures 
of dimension five, $\phi F^{\mu\nu}F_{\mu\nu}$ interaction.\\
\indent
 In this context we have  tried to estimate the ratio of parallel component of the electric flux with the  perpendicular component
. The estimate of this ratio in dipole magnetic field of aligned rotor model is known. We have found the same ratio in presence of the scalar field. So the difference between the two when compared with observations, then one might be able to draw some conclusions about the existence of the field $\phi$.\\
\indent
Following this point of view, we have taken the plasma frequency $\omega_{p}$ to be of the order $10^{-2}$ eV. We have further considered the photon path length to be  $z = 1.2$Km \cite{Lesch}. For these numbers of $\omega_{p}$ and $z$, we have estimated various oscillation probabilities in KeV energy range (20-100)KeV as shown in Fig: [{\ref{f001}] and Fig: [\ref{f002}]. The details  that led to the choice of these parameters  have been provided in supplementary document of this article.

\subsection{Electric field}
In this subsection we provide the expression for the electric field
in a frame of reference where the basis vectors  are orthogonal to the propagation direction $k_{\mu}$. We begin by noting that in the momentum space $k_{\mu}$ the electric field in four-component notation can be written as:
\begin{eqnarray}
{E}_{\mu} = \omega \tilde{A}_{\mu} - k_{\mu}(\tilde{A}.u).
\label{elec}
\end{eqnarray}
 \noindent
It should be noted that in  Eqn. (\ref{elec}), the vector potential $\tilde{A}_{\mu}$ refer to the gauge fields for dynamical   photon (in absence of scalar-photon mixing). Expressing the vector potential using  Eqn.(\ref{gp}) in the basis where $A.k = 0$, the electric field $E_{\mu}$ is, 
\begin{eqnarray}
{E}_{\mu} = \omega \left[ \tilde{A}_{\parallel} (k) \hat{b}^{(1)}_{\mu} + \tilde{A}_{\perp} (k) \hat{I}_{\mu} + \tilde{A}_L (k) \hat{\tilde{u}}_\mu\right] - k_{\mu}(\tilde{A}.u).
\end{eqnarray}
 \noindent
The components of the electric fields vector can be written in terms of the form factors of the vector potentials by contracting it with the corresponding polarization vectors as,
\begin{eqnarray}
 E_{\parallel} =\hat{b}^{(1)}_{\mu}E^{\mu} = \omega \tilde{A}_{\parallel}(k),
 \end{eqnarray}
 \begin{eqnarray}
  E_{\perp} =\hat{I}_{\mu}E^{\mu} = \omega \tilde{A}_{\perp}(k),
  \end{eqnarray}
  \begin{eqnarray}
  E_{L} =\hat{\tilde{u}}_{\mu}E^{\mu} = \tilde{A}_{L}(k),
  \end{eqnarray}
  \begin{eqnarray}
  k^{\mu}\hat{b}^{(1)}_{\mu} = k^{\mu}\hat{I}_{\mu}= k^{\mu}\hat{\tilde{u}}_{\mu}=0. 
  \end{eqnarray}
\noindent
The magnetic field also can be written in terms of these three components. Therefore the total energy stored in the form of EM fields $E_{tot} = \frac{1}{\mu_{0}}({E^{2}}+{B^{2}}$).

\subsection{Intensities of Polarization modes}

\indent
For highly Lorentz boosted electrons, the opening angle for the curvature photons follow the 
relation, $\Gamma \simeq \frac{1}{\theta_e}$, when $\theta_e$, is the opening angle. Therefore 
in this approximation, the amplitudes of the two orthogonal modes of curvature photon in terms of modified Bessel functions $K_{1/3}(\xi)$ and $ K_{2/3}(\xi)$ would be given by \cite{Jackson_book},
\begin{eqnarray}
|\tilde{A}_{\parallel}| & \simeq &\frac{\sqrt{6} \Gamma }{\omega_c} K_{1/3} \left( \xi\right), 
\label{int_para}
\end{eqnarray}
\begin{eqnarray}
|\tilde{A}_{\perp}| & \simeq &\frac{2\sqrt{3} \Gamma }{\omega_c} K_{2/3} \left( \xi \right) . 
\label{int_perp}
\end{eqnarray}
 The argument $(\xi)$ of the modified Bessel functions is defined as , 
\begin{eqnarray}
\xi = \frac{\omega R_{c}}{3c}\left(\frac{1}{\Gamma^{2}} + \theta^{2}_{e}\right)^{3/2}.
\end{eqnarray}

\noindent
Using  Eqn. (\ref{curv-energy}) and the approximation considered above (i.e., $\Gamma \simeq \frac{1}{\theta_e}$ ) the same in terms of $\omega_{c}$ (in the units of $\hbar = c= 1 $) turns out be $\xi  = \frac{\omega}{0.7\omega_{c}}$. The modified Bessel functions when $\omega << \omega_c$, i.e., for small arguments are given by,
\begin{eqnarray} 
K_{2/3}\left(\frac{\omega}{\omega_c} \right) \simeq 2^{-1/3}  \Gamma_{_{E}}\left( \frac{2}{3}\right) \xi^{-2/3} \nonumber   \\
K_{1/3}\left( \frac{\omega}{\omega_c}\right) \simeq 2^{-2/3} \Gamma_{_{E}}\left(\frac{1}{3}\right) \xi^{-1/3}, 
\label{bessfn}
\end{eqnarray}
\noindent
(here $ \Gamma_{_{E}}$ is the Euler gamma function) one can express the square of the  ratio of the amplitude of the two polarization 
components as,
\begin{eqnarray} 
\left| \frac{\tilde{A}_{\parallel}}{\tilde{A}_\perp} \right|^2 \simeq 2^{-5/3} \left[\frac{(\Gamma_{_{E}}(1/3))}{(\Gamma_{_{E}}(2/3))}\right]^2
\xi^{2/3}
\label{amp:ratio}
\end{eqnarray}

\noindent
Since the emitted energy peaks at $\omega_c $, one considers $\omega \simeq 0.7 \omega_c$, 
the ratio of the two amplitudes for this energy comes out  as, $  2^{-5/3} \left[\frac{(\Gamma_{_{E}}(1/3))}{(\Gamma_{_{E}}(2/3))}\right]^2 $.
It should be noted that this ratio is independent of the path traversed by the radiation.
Ideally, for the kind of situation under consideration, this is what one would expect for the ratio 
of the intensities of the two polarization states. However in presence of $\phi F^{\mu\nu}F_{\mu\nu}$ 
interaction, the same will be modified. The square of the modified amplitudes ratio turns out to be,
\begin{eqnarray}
\left|\frac{A_{\parallel}}{A_\perp}\right|^2 \simeq 2^{-5/3} \left[\frac{(\Gamma_{_{E}}(1/3))}{(\Gamma_{_{E}}(2/3))}\right]^2 \xi^{2/3}\times \left[\frac{ 1- P_{\gamma_{\parallel} \to \phi} (\omega,z) }
{ 1- P_{\gamma_{\perp} \to \phi} (\omega,z)  } \right].
\label{amp:ratio-osc}
\end{eqnarray}

 The last term inside the square bracket in  Eqn. (\ref{amp:ratio-osc}), is the modification factor, where we have taken into account the modifications to the intensities in the  $\parallel$ ($\perp$) direction due to oscillation of the  $\parallel$ ($\perp$) mode into $\phi$ after travelling a distance  $z$. It is to be noted that the denominator of the modification factor in  Eqn.(\ref{amp:ratio-osc}), would remain unity, unless the effect of the photon self energy correction in a 
magnetized media to order $e{ \rm{B}}$, is considered. In this work we have retained this piece consistently
in our formalism to find it's contribution to the ratio of the two polarization states.  And in absence of scalar-photon interaction, the ratio would be the same for pure curvature radiation. In presence of 
this interaction ( dimension-five-scalar-photon ), the same would be given by  Eqn.(\ref{amp:ratio-osc}). The contribution of this deviation in intensities can be observed when we take the difference of the two i.e., 
\begin{eqnarray}
\Delta F= \left| \frac{\tilde{A}_{\parallel}}{\tilde{A}_\perp} \right|^2 - \left|\frac{A_{\parallel}}{A_\perp}\right|^2
\label{diff}
\end{eqnarray}
\noindent
The same has been estimated numerically in the right panel of Fig.[\ref{fig3}]. 

\begin{figure}[h!]
\begin{minipage}[b]{.50\textwidth}  
\includegraphics[width=1\linewidth]{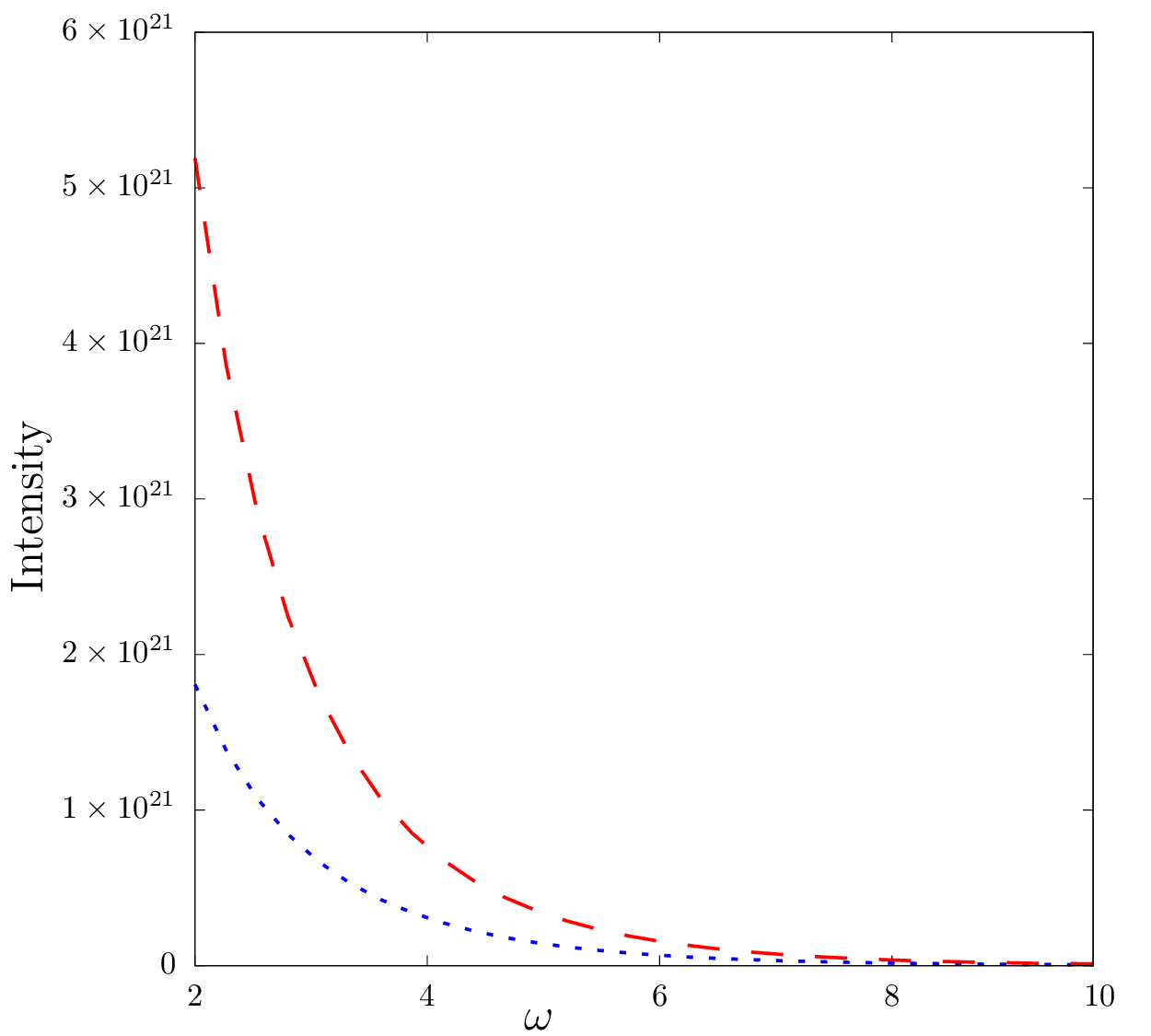}
%\label{f1}
\end{minipage}
\begin{minipage}[b]{.50\textwidth}  
\includegraphics[width=1\linewidth]{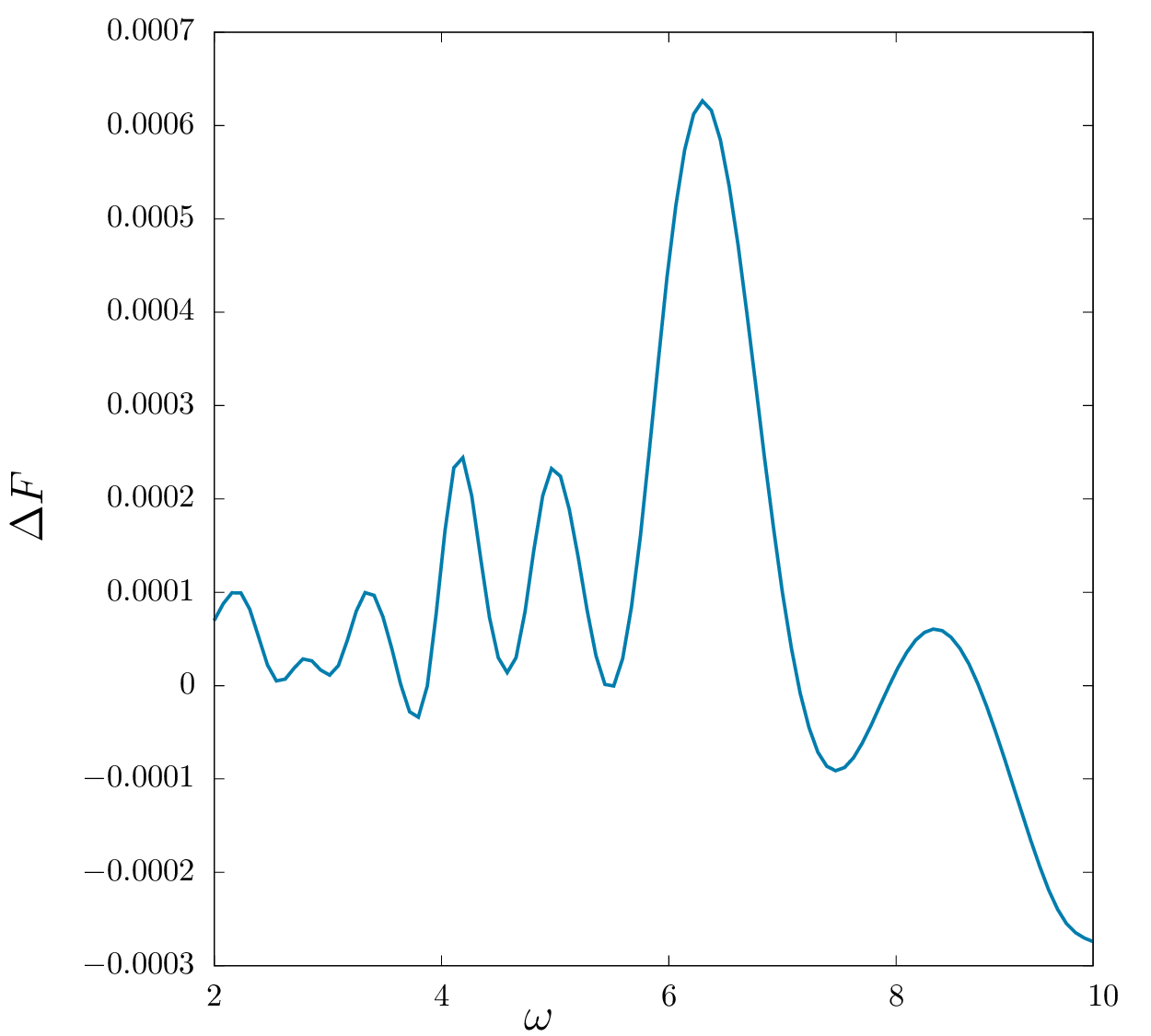}
\end{minipage}
\caption{  {\bf In the left panel} : Intensity of parallely and perpendicularly polarized photons with scalar photon interaction in magnetized media vs energy $(\omega)$. The dot curve represents $|A_{\parallel}|^{2}$ and the dashed curve represents $|A_{\perp}|^{2}$ . {\bf In the right panel }: Difference between ratios of intensity of parallely and perpendicularly  polarized photons(in abscence   and in presence of scalar photon interaction) $\Delta F$. Parameters choosen are;   $\Gamma$ (Lorentz boost factor) $ = 1.0 \times 10^{6}$, $\omega_{c} \sim 5 \times 10^{-5} $ GeV, $\rm{B} \sim 10^{12} $ Gauss. The abscissa (energy of photon ($\omega$)  in  GeV) is in units of   $10^{-5}$ .}
\label{fig3}
\end{figure}
In absence of  PEST in magnetized media ; the result would be same as that of mixing with $2\times2$ mixing matrix.

%############################################################################
\section{Result and discussion}
%############################################################################
\indent
 In this work we have explored the behaviour of the magnetized matter on the probabilities $P_{\gamma_{\parallel,\perp}\to\phi}$ for a electrodynamic system and estimated their magnitudes numerically. 
To perform that we need to  compare the effects of magnetized matter effects with un-magnetized matter  effects in  different magnetic field environments. We have numerically estimated  various probabilities for the  astrophysical parameters given by:  plasma frequency of the stellar environment ($\omega_{p}$) = $1.96 \times 10^{-2}$ eV,  scalar photon coupling constant ($g_{\phi\gamma\gamma}$) = $1.0 \times 10^{-11}$ GeV$^{-1}$, and the photon path length considered here is : $z \sim 1.2$ Km. The results  are plotted in Figs.[\ref{f001}] and [\ref{f002}].  There are three major outcomes of our analysis. Firstly, $ P_{\gamma_{\parallel} \to \phi }$ for magnetized media, turns out to be different from $ P_{\gamma_{\parallel} \to \phi }$ for unmagnetized media, they  admit frequency dependent modification over each others contributions when  the polarized photon has propagated over the same distance. Qualitatively, 
$P_{\gamma_{\parallel} \to \phi }$ conversion   probability was found to be the only non-zero probability  when the  correction of $\Pi^{p}_{\mu\nu}(k)$ is absent. Secondly,  the numerical strength of the probabilities  of conversion $P_{\gamma_{\parallel} \to \phi }$ (or $P_{\gamma_{\perp} \to \phi }$ ),  increases  with increase in strength of the magnetic field.
Since an analytical  estimate of the same is difficult, we have estimated the same numerically and  have 
shown both of them in a same panel (Fig.[\ref{f001}] and Fig.[\ref{f002}]). One interesting feature of  
$P_{\gamma_{\parallel} \to \phi }$ is, it shows  {\it excess enhancement} over $P_{\gamma_{\parallel}\to \phi }$ estimated in an unmagnetized medium at various energies. The third interesting feature is: the intensity of the other orthogonally polarized mode of the 
photon ( denoted as $A_{\perp}$ ), no longer remains the same when the effect of  
magnetized-self-energy  correction to photon  is considered
for studying (photon scalar) oscillation. In absence of magnetized-self-energy correction, the  intensity of  $A_{\perp}$ mode remains constant, 
provided no other absorption or enhancement mechanism is in operation. That is to say, milli-charge  pair production processes in a magnetic field,
($\gamma + \gamma \to e_{m}^{+} + e_{m}^{-}$) that modifies the  $A_{\perp}$ spectra
is considered absent.\\
\indent
The x-ray sources for compact stars can be classified into three categories; one is due to bombardment of pair produced flux on the 
polar cap region of the compact star, the second one is due to conduction heat coming out of the surface of star having surface temperature $T \leq 10^{6} K$. And the last one is due to 
the curvature radiation from the star.\\
\indent 
The beams of  EM radiations from first two sources follow the blackbody radiation pattern and  are unpolarized. However due to the presence of the  magnetic field, the isotropic beams of radiation get  resolved into two components, one of them becomes polarized  parallel   and the other one becomes  polarized perpendicular  to the magnetic field.  Total intensity of each of the components becomes equal to the other. \\
\indent
On the other hand the amplitudes of the two polarized components of the curvature radiation that we have discussed in this paper, denoted by $\tilde{A}_{\parallel}$ and  $\tilde{A}_{\perp}$, are different. As a result, the spectra for their respective polarization, would look different from the usual blackbody  spectrum. The ratios of $\tilde{A}_{\parallel}$ and $\tilde{A}_{\perp}$ would be given by  Eqns. (\ref{int_para}, \ref{int_perp}).  That yields $\frac{\tilde{A}_{\parallel}}{\tilde{A}_{\perp}} \sim {(\frac{\omega}{0.7\omega_{c}})}^{1/3}$ for $\omega_{c}>>\omega$.\\
\indent
 Once the scalar photon interaction is taken into account, and the same ratios are estimated this pattern changes.
 If we take the difference between square of these ratios without and with scalar photon modification i.e.,
 $ \left| \frac{\tilde{A}_{\parallel}}{\tilde{A}_\perp} \right|^2 - \left|\frac{A^{}_{\parallel}}{A^{}_\perp}\right|^2$,
 the resulting plot looks entirely different. The intensities of both the polarization states (i.e., $A_{\parallel}$ and  $A_{\perp} $) under the  circumstances mentioned above plotted separately,  can be found in the left panel of Fig.[\ref{fig3}]. And the differences in their ratios of intensities (without and with scalar photon interaction) can be found in  the right panel of Fig.[\ref{fig3}]. Such an oscillating curve may   bear the possible smoking gun signature  of dilaton-photon interaction from  a compact star. \\

 %############################################################################
 \section{Outlook}
 \indent
 Existence of DM was postulated to explain some astrophysical observations,  such as 
 (i) Galaxy rotation curve
 (ii) observation of x rays from bullet cluster (iii) weak micro lensing effect (iv) structure formation e.t.c.. The candidate particles proposed, to explain the same,  also predicts few additional signatures steaming from   ALP  photon interaction   \cite{Raffelt} like, supernova dimming, modification to distance duality relation, ALPs from supernova, ALP photon conversion in magnetized domain etc., \cite{Mirizzi-raffelt1,Mirrizzi-raffelt2,Tiwari, FERMI_LAT,Soda,Grossman,Bassett,Jackson,ckt1,ckt2} that happens to be some of them. So there has been a surge for the search of ALP signals from the magnetized environments of the compact stars \cite{Pshirkov:2007st} --\cite{ Millar2021}  for their identifications. Therefore to complement this surge in the interest  we have  explored  ALP signal from magnetized media in this work. In course of the investigation
  we find that the formalism employed in this analysis, interpolates between two extreme ends: one, the effective quantum statistical field theory of electrodynamics (QSFED) in a magnetized environment, when the angle between $\rm{B} $ and $k$ is zero or $\pi$; and the other being the interacting theory of ALP with  QSFED at finite density when the angle between $\rm{B} $ and $k$ is $\pi/2$ that follows from  Eqn. (\ref{photon-scalar-mixing-matixx}). These are the checks that can be employed to establish the consistency of our formalism for magnetized media.\\
  \indent
  It has been found that, in addition to $P_{\gamma_{\parallel} \to \phi}$ there are two more additional probabilities of conversion i.e.,  $P_{\gamma_{\perp} \to \phi}$ and $P_{\gamma_{\parallel} \to \gamma_{\perp}}$ possible in a magnetized media according to the number of  possible in-medium polarization states of photon. Our analysis  showed the existence of normal modes corresponding to the three orthogonal directions of propagation of photon.\\
  \indent
 Furthermore, we have noted in the body of the text that   $P_{\gamma_{\parallel} \to \phi}$ in a magnetized medium can be much larger than the contribution of the same in an un magnetized medium in some energy range for suitable values of other parameters.
 Since the produced ALPs would leave the production region (of the star) fast  because of its weak interaction  with the medium, this will lead to anomalous cooling of the star. The indication of presence of ALP in the stellar environment comes from the 
estimate of $\Delta F$ (i.e., $ \left| \frac{\tilde{A}_{\parallel}}{\tilde{A}_\perp} \right|^2 - \left|\frac{A^{}_{\parallel}}{A^{}_\perp}\right|^2$) that becomes zero in absence of ALP and it becomes nonzero when ALP interaction is present. These are the novel signatures that a magnetized compact stars environment can offer for the verification of ALP.\\
\indent
 To conclude, we have shown in this work that, effect of magnetized medium brings non-trivial modifications to the ALP induced signals from compact objects. Though the physics of these objects are complex, however it may be possible to get a tale-tale signature of ALP like objects from the EM signals of the compact stars. Some of these studies would be considered in separate publications.
 
 \section{ Acknowledgements }
 The authors would like to thank Prof. P. B. Pal for going through an earlier version of this article.
 
%############################################################################
\section{Appendix}
\appendix
\section{Technical Details of constructing the unitary matrix}
\label{Technical Details}
%############################################################################

\indent
In order to describe the dynamics of scalar-photon interaction in magnetized media in terms of different dof of mixing, the Hermitian mixing matrix $M_{H}$, obtained from the equations of motion of scalar photon interaction, given by,
\indent
\begin{eqnarray} 
M_{H}=\left[ 
  \begin{matrix}
 \Pi_T(k)   &       iF                &      iG     \cr     
     -iF             & \Pi_T(k)         &       0    \cr
  
    - iG             &   0                &  m_{\phi}^2
  \end{matrix}
\right]
\label{matt}
\end{eqnarray}
\noindent
needs to be diagonalize. Therefore,  to do the same, we write $M_{H}$ as,
\begin{eqnarray} 
M_{H}=\left[ 
  \begin{matrix}
 a    &       ib       &             ic    \cr     
     -ib             & d  &      0         \cr
     -ic             &   0         &      e     
  \end{matrix}
\right]
\label{matt1}
\end{eqnarray}
\noindent
when $a= \Pi_{T}(k)$, $b = F$, $c = G$ and $e = m^{2}_{\phi}$. It can be shown that, an unitary matrix $Q$ and its hermitian  conjugate $Q^{\dagger}$ defined  as, 
\begin{eqnarray} 
Q=\left[ 
  \begin{matrix}
 1   &       0       &             0    \cr     
     0             & i  &      0         \cr
     0            &   0         &      i     
  \end{matrix}
\right],
Q^{\dagger}=\left[ 
  \begin{matrix}
 1    &       0      &             0    \cr     
    0             & -i  &      0         \cr
     0            &   0         &      -i    
  \end{matrix}
\right]
\label{QQdagger}
\end{eqnarray}
\noindent
can transform $M_{H}$ to real symmetric matrix $M_{RS}$ by, $QM_{H}Q^{\dagger} = M_{RS}$, where $M_{RS}$ is, given by,
\begin{eqnarray} 
M_{RS}=\left[ 
  \begin{matrix}
 a    &       b       &         c    \cr     
     b             & d  &      0         \cr
     c             &   0         &      e     
  \end{matrix}
\right]
\label{mat2}
\end{eqnarray}
\noindent
Now, if for a real symmetric matrix $M_{RS}$, there exist a  Unitary matrix ${\bf\tilde{ U}}$ such that  ${\bf \tilde{U}}^{\dagger} M_{RS} {\bf\tilde{ U}}= M_{D}$, where $M_{D}$ is a diagonalised matrix having the eigen  values of matrix $M_{RS}$ as diagonal elements. Now, using this fact, we can subsequently find out the unitary matrix ${\bf {U}}$, that would diagonalize $M_{H}$. The argument goes as follows. Since,
\begin{eqnarray}
{\bf\tilde{ U}}^{\dagger} M_{RS} {\bf\tilde{ U}}&=& M_{D}
\label{trans1}
\end{eqnarray}
\noindent
then, using  $QM_{H}Q^{\dagger} = M_{RS}$ in equation (\ref{trans1}), we get
\begin{eqnarray}
{\bf\tilde{ U}}^{\dagger}Q M_{H} Q^{\dagger}{\bf \tilde{U}}&=& M_{D}\\
{\bf{U}}^{\dagger}M_{H} {\bf{U}}&=& M_{D}
\end{eqnarray}
\noindent
where we have defined $Q^{\dagger}{\bf \tilde{U}} ={\bf{U}}$ and  $(Q^{\dagger}{\bf \tilde{U}})^{\dagger} ={\bf{U}}^{\dagger}$.  The eigen 
 values of the unitarily related matrices $M_{H}$ and $M_{RS}$ remain same. To obtain $M_{D}$, we need unitary  matrix ${\bf{U}}$ that follows from the unitary matrix $\bf{\tilde{U}}$, constructed from the eigen vectors of $M_{RS}$.

%############################################################################
\section{ Eigen vectors of $M_{RS}$}
\subsection{The characteristics equation}
%############################################################################
In this section we denote the eigen values of $M_{RS}$ as ${\bf E}_{i}, i=1,2,3$, then the characterstic equation for the same can be written as,
\begin{eqnarray}
|M_{RS}-{\bf E}_{i}{\bf I}| = 0.
\end{eqnarray}
Where ${\bf I}$ is a 3$\times$3 identity matrix. Hence,
\begin{eqnarray} 
\left|
  \begin{matrix}
 a-{\bf E}_{i}    &       b       &         c    \cr     
     b             & d-{\bf E}_{i}  &      0         \cr
     c             &   0         &      e-{\bf E}_{i}     
  \end{matrix}
\right|= 0
\label{mat3}
\end{eqnarray}\\
 The characteristics equation that follows from the  determinant is,
 \begin{eqnarray}
 (a-{\bf E}_{i})(d-{\bf E}_{i})(e-{\bf E}_{i})-b^{2}(e-{\bf E}_{i})-c^{2}(d-{\bf E}_{i})=0,
 \label{chr-eqn}
 \end{eqnarray}
\noindent
that yields on simplification,
\begin{eqnarray}
{\bf E}_{i}^{3}c_{3}+{\bf E}_{i}^{2}c_{2}+{\bf E}_{i}c_{1} + c_{0}    =0.
\label{cubiceqn}
\end{eqnarray}
\noindent
Where, the coefficients $c_{i}'s$ are defined as :
\begin{eqnarray}
c_{3}&=& 1 \\
c_{2}&=&   -(a+d+e)  \\
c_{1}&=&   (ae+de+ad-b^{2}-c^{2})\\
c_{0}&=& (b^{2}e+c^{2}d-aed)
\end{eqnarray}
\noindent
The nature of the  roots of the cubic equation (\ref{cubiceqn}) must satisfy the following properties,
\begin{eqnarray}
{\bf E}_{1}+{\bf E}_{2}+{\bf E}_{3} &=&-c_{2} \nonumber \\
{\bf E}_{1}{\bf E}_{2}+{\bf E}_{2}{\bf E}_{3}+{\bf E}_{3}{\bf E}_{1}&=& c_{1} \nonumber\\
{\bf E}_{1}{\bf E}_{2}{\bf E}_{3}&=& -c_{0}.
\label{rootsprop}
\end{eqnarray}
%\end{widetext}
\noindent
The roots of  Eqn. (\ref{cubiceqn}) depends on the discriminant, ${\cal D}= {\cal Q}^2 +{\cal P}^3$, where, variables ${\cal P}$
and ${\cal Q}$ are related to the elements of  Eqn. (\ref{cubiceqn}) through the relations,
\begin{eqnarray} 
{\cal P}= \frac{\left(  3 c_{1} - {c_{2}}^2 \right ) }{9} \mbox{~~~and~~~} {\cal Q}=\left(   \frac{{c_{2}}^3}{27} - \frac{{ c_{2}}{ c_{1}}}{6}+ \frac{c_{0}}{2} 
\right). 
\label{disc}
\end{eqnarray}
\noindent
Furthermore, for hermitian matrices, roots are real, we should have ${\cal Q}^2 + {\cal P}^3 \leq 0$. Finally,  following
 \cite{cardano,KOPP}, the roots turns out to be:
\begin{equation}
\begin{array}{cc}
     {\bf E}_{1} =&   {\cal R} \cos \alpha + \sqrt{3} {\cal R} \sin \alpha -{ c_{2}}/3, \\
     {\bf E}_{2} =&   {\cal R} \cos \alpha - \sqrt{3} {\cal R }\sin \alpha -{ c_{2}}/3, \\
     {\bf E}_{3} =& \hskip -1.5cm -2{\cal R} \cos \alpha  -{ c_{2}}/3, \\
\end{array} 
\mbox{~~With~} \\
  \left\{ \begin{array}{c}
                 \alpha  =\frac{1}{3} \cos^{-1} \left( \frac{{\cal {Q}}}{{\cal R}^3} \right )\\
                \,\,\,\,\, {\cal R} = \sqrt{\left(-{\cal P} \  \right )} {\bf sgn} \left( {\cal{Q}} \right) 
     \end{array}
  \right.
\label{exact-roots}
\end{equation}
For having real roots, one should have ${\cal P} < 0$, and $ \mid \frac{\cal{Q}}{{\cal{R}}^3} \mid \le 1 $. 
Although for analytic evaluation of the roots, this condition may not pose any problem, 
however during  numerical evaluation of the same, maintaining this ratio  may become difficult if the magnitudes of elements of the  matrix ${\bf M}$ are close to the precision available for the machine in use.\\
\indent
 One can scale  the matrix  ${\bf M}$  by a suitable numerical factor to avoid this difficulty.
%############################################################################
\subsection{Eigen function and Unitary matrix}
%############################################################################

\indent
The normalized eigen vector  ${\bf \tilde{V}}_{i}$  of the real symmetric matrix $M_{RS}$, can be represented as,
\begin{eqnarray} 
{\bf\tilde{V}}_{i}=N_{i}\left[ 
  \begin{matrix}
 \tilde{u}_{i}   \cr     
     \tilde{v }_{i}  \cr
     \tilde{w}_{i}    
  \end{matrix}
\right],
\label{eign}
\end{eqnarray}\\

\noindent
where $N_{i}$ is the normalization constant.
Therefore, using eigen value equation, i.e., $(M_{RS}-{\bf E}_{i}{\bf I}){\bf\tilde{V}}_{i}=0$. Hence,
\begin{eqnarray} 
\left[ 
  \begin{matrix}
 a    &       b       &         c    \cr     
     b             & d  &      0         \cr
     c             &   0         &      e     
  \end{matrix}
\right]\left[ 
  \begin{matrix}
 \tilde{u}_{i}   \cr     
     \tilde{v }_{i}  \cr
     \tilde{w}_{i}    
  \end{matrix}
\right]={\bf E}_{i} \left[ 
  \begin{matrix}
 \tilde{u}_{i}   \cr     
     \tilde{v }_{i}  \cr
     \tilde{w}_{i}    
  \end{matrix}
\right].
\label{eignfuneqn}
\end{eqnarray}
\noindent
We get,
\begin{eqnarray}
(a-{\bf E}_{i})\tilde{u}_{i}+b\tilde{v}_{i}+c\tilde{w}_{i}&=&0
\label{solns1}\\
b\tilde{u}_{i}+(d-{\bf E}_{i})\tilde{v}_{i}&=&0
\label{solns2}\\
c\tilde{u}_{i}+(e-{\bf E}_{i})\tilde{w}_{i}&=&0.
\label{solns3}
\end{eqnarray}
\noindent
 Using the method discussed in \cite{mybook}, we can obtain analytical expressions of the elements of eigen vector ${\bf\tilde{V}}_{i}$. They are,
\begin{eqnarray} 
{\bf\tilde{V}}_{i}=N_{i}\left[ 
  \begin{matrix}
 \tilde{u}_{i}   \cr     
     \tilde{v }_{i}  \cr
     \tilde{w}_{i}    
  \end{matrix}
\right]
=N_{i}\left[ 
  \begin{matrix}
 (d-{\bf E}_{i})(e-{\bf E}_{i})  \cr     
     -b(e-{\bf E}_{i}) \cr
     -c(d-{\bf E}_{i})   
  \end{matrix}
\right].
\label{Vect}
\end{eqnarray}

\noindent
Here, $N_{i}=\frac{1}{\sqrt{((d-{\bf E}_{i})(e-{\bf E}_{i}))^{2}+(b(e-{\bf E}_{i}))^{2}+(c(d-{\bf E}_{i}))^{2}  }}$, is a normalisation constant. We can now construct the unitary matrix $\bf {\tilde {U}}$, using (\ref{Vect}), as:
\begin{eqnarray} 
{\bf \tilde U}=\left[ 
  \begin{matrix}
 N_{1}\tilde{u}_{1}    &        N_{2} \tilde{u}_{2}       &      N_{3}     \tilde{u}_{3}     \cr     
    N_{1}   \tilde{v}_{1}              &  N_{2} \tilde{v}_{2}   &       N_{3} \tilde{v}_{3}         \cr
     N_{1}  \tilde{w}_{1}              &  N_{2}  \tilde{w}_{2}   &  N_{3}      \tilde{w}_{3}      
  \end{matrix}
\right]
\label{Unit1}
\end{eqnarray}

\subsection{Orthonormality check of ${\bf \tilde{V}}$}
\indent
One can easily check the vectors are normalised i.e., ${\bf   \tilde{V}_{i}. \tilde{V}_{j}}=1$ when $i,j = 1,2,3$ and i$=$j. Furthermore, the verification of the orthogonal properties of  vectors, provided below:
\begin{eqnarray}
{\bf   \tilde{V}_{1}. \tilde{V}_{2}}&=& N_{1}N_{2}(\tilde{u}_{1}\tilde{u}_{2} + \tilde{v}_{1} \tilde{v}_{2}+\tilde{w}_{1} \tilde{w}_{2} )\nonumber\\
                                                 &=&N_{1}N_{2}((d-{\bf E}_{1})(e-{\bf E}_{1})(d-{\bf E}_{2})(e-{\bf E}_{2})+ b^{2}(e-{\bf E}_{1})(e-{\bf E}_{2})\nonumber\\
 \label{v1v2}                                                &+& c^{2}(d-{\bf E}_{1})(d-{\bf E}_{2}))
 \end{eqnarray}  
\noindent
Now, 
\begin{eqnarray}
(e-{\bf E}_{1})(e-{\bf E}_{2})= e^{2} -e({\bf E}_{1}+{\bf E}_{2}) + {\bf E}_{1}{\bf E}_{2}.
\end{eqnarray}

\noindent
 We will convert the above equation having two functions ${\bf E}_{1}$ and ${\bf E}_{2}$ in to single function ${\bf E}_{3}$ by using the properties of roots of cubic equation.
In order to do that, we will try the following substitutions, 
\begin{eqnarray}
{\bf E}_{1} +{\bf E}_{2}& =& [{\bf E}_{1} +{\bf E}_{2} +{\bf E}_{3}]-{\bf E}_{3}\\
{\bf E}_{1}{\bf E}_{2} &= &[{\bf E}_{1}{\bf E}_{2} + {\bf E}_{2}{\bf E}_{3} + {\bf E}_{3}{\bf E}_{1}] -{\bf E}_{3}({\bf E}_{2} + {\bf E}_{1}).
\end{eqnarray}

\noindent
Following the identities of equations (\ref{rootsprop}) and make substitutions in   Eqn. (\ref{v1v2}), we obtain,

 \begin{eqnarray}
{\bf   \tilde{V}_{1}. \tilde{V}_{2}}= N_{1}N_{2}(a-{\bf E}_{3})[(a-{\bf E}_{3})(d-{\bf E}_{3})(e-{\bf E}_{3})\nonumber\\
-b^{2}(e-{\bf E}_{3})-c^{2}(d-{\bf E}_{3})]
\label{v1dotv2}.
 \end{eqnarray}
\noindent
The terms inside the square brackets on the r.h.s of   Eqn.(\ref{v1dotv2}) happens to be zero due to equation (\ref{chr-eqn}) that follows from the characteristics equation. Hence,
\begin{eqnarray}
 {\bf \tilde{V}_{1}. \tilde{V}_{2}}=0.
\end{eqnarray}

\noindent
Following the same procedure one can verify that  ${\bf   \tilde{V}_{1}. \tilde{V_{2}}}={\bf   \tilde{V_{2}}. \tilde{V_{3}}}={\bf   \tilde{V_{3}}. \tilde{V_{1}}}=0$.

\subsection{Analytical check of $ {\bf \tilde {U}^{\dagger}} M_{RS} {\bf \tilde U} = M_{D}$}

\indent
We have obtained the unitary matrix $ {\bf \tilde {U}}$ , which suppose to diagonalise the matrix $M_{RS}$. To check this claim analytically, we first start unitary transformation of $M_{RS}$ by  ${\bf \tilde {U}}$, given as follows:
\begin{eqnarray}
 {\bf \tilde {U}^{\dagger}} M_{RS} {\bf \tilde U} = M_{D}
\end{eqnarray}
\noindent
that implies,
\begin{eqnarray}
\left[ 
  \begin{matrix}
   N_{1}\tilde{u}_{1}    &   N_{1}\tilde{v}_{1}       &       N_{1}\tilde{w}_{1}     \cr     
   N_{2}\tilde{u}_{2}    &   N_{2}\tilde{v}_{2}       &       N_{2}\tilde{w}_{2}         \cr
   N_{3}\tilde{u}_{3}    &   N_{3}\tilde{v}_{3}       &       N_{3}\tilde{w}_{3}      
  \end{matrix}
\right]
\left[ 
  \begin{matrix}
 a    &       b       &         c    \cr     
     b             & d  &      0         \cr
     c             &   0         &      e     
  \end{matrix}
\right]
\left[ 
  \begin{matrix}
 N_{1}\tilde{u}_{1}    &        N_{2} \tilde{u}_{2}       &      N_{3}     \tilde{u}_{3}     \cr     
    N_{1}   \tilde{v}_{1}              &  N_{2} \tilde{v}_{2}   &       N_{3} \tilde{v}_{3}         \cr
     N_{1}  \tilde{w}_{1}              &  N_{2}  \tilde{w}_{2}   &  N_{3}      \tilde{w}_{3}      
  \end{matrix}
\right]=M_{D}
\label{UDMU1}
\end{eqnarray}

\begin{eqnarray}
\hskip -1.0 cm\left[ 
  \begin{matrix}
   N_{1}(\tilde{u}_{1} a +\tilde{v}_{1} b  +  \tilde{w}_{1} c)  & N_{1}(\tilde{u}_{1} b +\tilde{v}_{1} d)& N_{1}(\tilde{u}_{1} c +\tilde{w}_{1} e)  \cr     
   N_{2}(\tilde{u}_{2} a +\tilde{v}_{2} b  + \tilde{w}_{2} c)  & N_{2}(\tilde{u}_{2} b +\tilde{v}_{2} d)& N_{2}(\tilde{u}_{2} c +\tilde{w}_{2} e)      \cr
  N_{3}(\tilde{u}_{3} a +\tilde{v}_{3} b  + \tilde{w}_{3} c)  & N_{3}(\tilde{u}_{3} b +\tilde{v}_{3} d)& N_{3}(\tilde{u}_{3} c +\tilde{w}_{3} e) 
  \end{matrix}
\right]\\
\times\left[ 
  \begin{matrix}
 N_{1}\tilde{u}_{1}    &        N_{2} \tilde{u}_{2}       &      N_{3}     \tilde{u}_{3}     \cr     
    N_{1}   \tilde{v}_{1}              &  N_{2} \tilde{v}_{2}   &       N_{3} \tilde{v}_{3}         \cr
     N_{1}  \tilde{w}_{1}              &  N_{2}  \tilde{w}_{2}   &  N_{3}      \tilde{w}_{3}      
  \end{matrix}
\right]=M_{D}
\label{UDMU2}
\end{eqnarray}

\noindent
Using Eqns.  (\ref{solns1}), (\ref{solns2}) and (\ref{solns3}) in (\ref{UDMU2}), we get, 

\begin{eqnarray}
\hskip -1.0cm \left[ 
  \begin{matrix}
   N_{1}\tilde{u}_{1}{\bf E}_{1}  & N_{1}\tilde{v}_{1}{\bf E}_{1} & N_{1}\tilde{w}_{1}{\bf E}_{1}  \cr     
    N_{2}\tilde{u}_{2}{\bf E}_{2}  & N_{2}\tilde{v}_{2}{\bf E}_{2} & N_{2}\tilde{w}_{2}{\bf E}_{2}    \cr
  N_{3}\tilde{u}_{3}{\bf E}_{3}  & N_{3}\tilde{v}_{3}{\bf E}_{3} & N_{3}\tilde{w}_{3}{\bf E}_{3}
  \end{matrix}
\right]
\left[ 
  \begin{matrix}
 N_{1}\tilde{u}_{1}    &        N_{2} \tilde{u}_{2}       &      N_{3}     \tilde{u}_{3}     \cr     
    N_{1}   \tilde{v}_{1}              &  N_{2} \tilde{v}_{2}   &       N_{3} \tilde{v}_{3}         \cr
     N_{1}  \tilde{w}_{1}              &  N_{2}  \tilde{w}_{2}   &  N_{3}      \tilde{w}_{3}      
  \end{matrix}
\right]\nonumber\\
=\left[ 
  \begin{matrix}
{\bf \tilde{V}_{1}.\tilde{V}_{1}} {\bf E}_{1}    &        {\bf\tilde{V}_{1}.\tilde{V}_{2}} {\bf E}_{1}     &    {\bf  \tilde{V}_{1}.\tilde{V}_{3}} {\bf E}_{1}    \cr     
  {\bf \tilde{V}_{2}.\tilde{V}_{1}} {\bf E}_{2}            &   {\bf \tilde{V}_{2}.\tilde{V}_{2}} {\bf E}_{2}  &      {\bf \tilde{V}_{2}.\tilde{V}_{3}} {\bf E}_{2}       \cr
    {\bf \tilde{V}_{3}.\tilde{V}_{1}} {\bf E}_{3}          &  {\bf \tilde{V}_{3}.\tilde{V}_{2}} {\bf E}_{3}   &  {\bf \tilde{V}_{3}.\tilde{V}_{3}} {\bf E}_{3}      
  \end{matrix}
\right] \nonumber \\
=M_{D}
\label{UDMU3}
\end{eqnarray}
\noindent
Now, if we use the orthonormality of the eigen vectors, we get, 
\begin{eqnarray} 
M_{D}=\left[ 
  \begin{matrix}
 {\bf E}_{1}    &     0      &      0   \cr     
  0             &  {\bf E}_{2}   &       0       \cr
 0             &  0  & {\bf E}_{3}      
  \end{matrix}
\right]
\label{Diagonal}
\end{eqnarray}
 
\section{Data availability}
The data used for plotting Fig. [1] and  Fig [2]  in this article we be shared on request to the corresponding author.

\section{Supplemental section}

\section*{Stellar environment and radiation process}

\indent
 In the paragraphs below we  
outline the essentials of emission mechanism of compact objects common to most of the models
developed for that purpose. There are various models those have been successful to some extent to describe the emission 
mechanism from compact objects, like neutron stars or white dwarfs. Some of the most used 
ones  are polar-cap{\cite{rs75}},  slotgap \cite{as79}, outergap \cite{dh96} models etc..\\ 
\indent
 We will consider  here a simple picture of the enegetic-emission-physics 
from a typical compact object. The specific details can be found in {\cite{beskin-book, meszaros} and the references provided there in. The basic picture according to these models is, the  electric field $\rm{\bf{E}_{||}}$ is produced due to the rotating dipolar magnetic field $\rm{B}$ of the  compact object and is directed parallel to the ambient dipolar field.
%
% the radiation is emitted from the charged particles, those are accelerated by the an electric field $\rm{\bf{E}_{||}}$--produced due to the rotating dipolar magnetic field
% $\rm{B}$ of the  compact object and is directed parallel to the ambient dipolar field. 
 Due to the action of electric field ( $E_{\parallel}$),  charged particles ($e{^+}$ or $e^{-}$) are pulled out of the 
surface of the  compact star. Their number density is $n_{GJ}$ (where $n_{GJ}= \frac{\Omega B}{2\pi c e}$) \cite{usov88}. The radiation is emitted from the charged particles, those are accelerated by the  electric field $\rm{\bf{E}_{||}}$.

 For a typical pulsar of period $0.5$ sec., and magnetic field $\rm{B} \sim 10^{12}$ Gauss, the Goldreich Juliean number density $n_{GJ}$ turns  out to be $3.6 \times 10^{11}/$cm$^{3}$. These charged particles on their way  along the curved magnetic field, radiates EM radiation that is called  curvature radiation. The total  energy released in the process  is given by, 

\begin{eqnarray} 
\dot{E}_{rot} = \frac{2}{3}\frac{\Omega{^4}\bar{R}^{6}\rm{B}^{2}_{s}}{e^{3}} .
\label{e1}
\end{eqnarray}
In the  expression (\ref{e1}) $\bar{R}$ is the radius, $\rm{B}_s$ is the surface magnetic field and  $\Omega$ is the rotational frequency of the compact star. The total observed energy or total luminosity coming from  these objects 
 have been observed to  lie  between $10^{25} - 10^{35}$ erg/sec. That emitted energy can be as high 
as $10^{7}$ MeV for a pulsar having magnetic field $\rm{B} \sim 10^{12}$ Gauss and  rotational  period $P = 1 sec$ \cite{usov93}. The  radiated photons of this  energy 
 may create secondary plasma of $e^{+} $ and $e^{-}$ pairs through the process, $\gamma + \gamma \to e^{+} + e^{-}$. These secondary particles alter the number density of charge particles in the pulsar magnetosphere.\\
\indent
The energy of the emitted photons due to  curvature radiations of the charge particles in the magnetosphere of the compact object can be expressed in terms of the  instantaneous
Lorentz factor ($\Gamma$) of the radiating charged particles. It is given by 
$\omega= \frac{3}{2}\frac{\Gamma^3}{R_c} $, when $R_c$ is the radius of curvature
of  the dipolar magnetic field lines.\\
\indent
 Evolution of $\Gamma$, taking radiation reaction into account and energy gain due to the electric field 
is described by  {\cite{beskin-book}},
\begin{eqnarray}
m \, \frac{d \Gamma}{dx} =e\rm{\bf{E}_{||}} - \frac{2 e^2 \Gamma^{4}}{3R^2_c}.
\label{gama-evoln}
\end{eqnarray}
When  the distance from the centre of the star is given by: $r=x+\R$ ( $\R$ is the radius of the compact object and x is the distance from the surface of the star). This equation can be derived from the Lorentz-Abhram-Diraction equation discussed  in \cite{Jackson_book}.  When the energy-gain becomes equal to energy loss in  Eqn.( \ref{gama-evoln}), a quasi-steady state is reached, that gives the estimate
of $\Gamma$ in terms of electric field. This is given by, 
\begin{eqnarray}
\Gamma=\left(\frac{3\rm{\bf{E}_{||}} R^{2}_c }{2 e}  \right)^{\frac{1}{4}}.
\label{boost}
\end{eqnarray}

\indent
The  electric field, $\rm{\bf{E}_{||}}$ at a position r, 
($r > \R $) from the centre of the {\it{ compact object}} is given by \cite{usov93} and \cite{Arons} in the space charge limited flow in pulsar emission model:
\begin{eqnarray}
\rm{\bf{E}_{||}} \sim \frac{1}{8\sqrt{3}}\left( \Omega \R \right)^{\frac{5}{2}} \rm{B}
\sqrt{\frac{2\R}{r}}.
\label{eparallel0}
\end{eqnarray}
\noindent
In  Eqn. (\ref{eparallel0}), ${\rm{B}}$ is the surface magnetic field of the compact object. Since $\rm{\bf{E}_{||}}$ is position dependent, it introduce position dependence to $\Gamma$,
there by to the emitted radiation, $\omega$. Once the emission region for the 
radiation of energy $\omega$ is determined, the distance it would travel in 
the magnetosphere of the object can be estimated, from the knowledge of
$R_{LC}$ (when $R_{LC}=1/\Omega$, is light cylinder radius).  \\
\indent
Though in principle one can solve  Eqn. (\ref{gama-evoln}), to find out position dependence 
of the Lorentz factor, but solving it analytically is difficult. However it is possible to solve  Eqn. (\ref{gama-evoln}) numerically. A numerical solution is provided in fig.[\ref{fig4}]. The upward turning point ($P$) corresponds to the quasi static limit in  Eqn. (\ref{gama-evoln}). Similar behaviour for the primary particles has also been reported in \cite{Diego}. 

 \begin{figure}[h!]
\begin{minipage}[h]{1.0\textwidth}  
\includegraphics[width=1\linewidth]{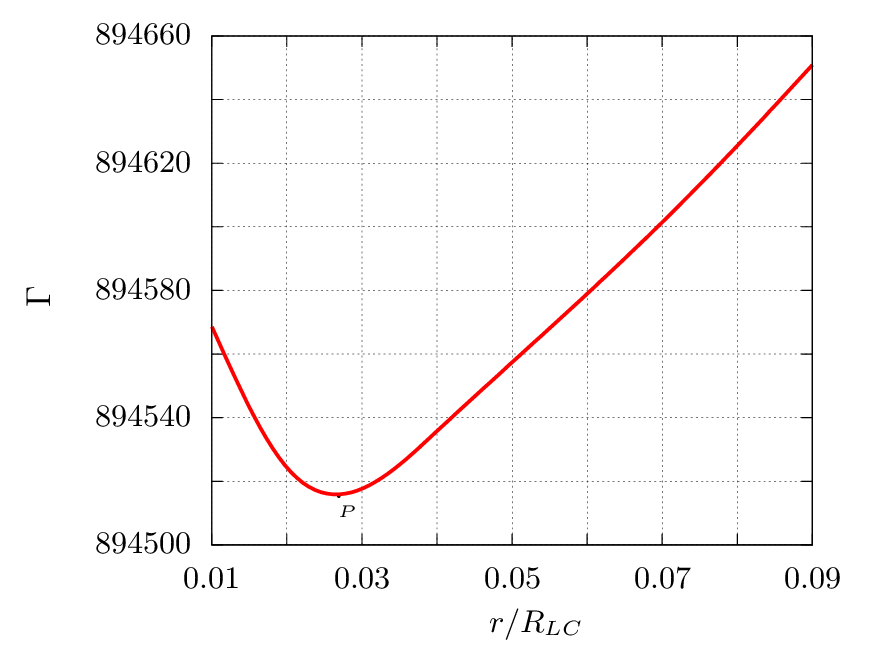}
\end{minipage}

\caption{Evolution of $\Gamma$ (Lorentz factor) of primary charge particle in the pulsar atmosphere with radiation reaction. The lowest point ($P$) in the curve represents the situation when energy gain by the charge particle is equal to energy loss due to radiation reaction i.e.,  when (\ref{gama-evoln}) equal to zero. }
\label{fig4}
\end{figure}

Since the peak emitted energy $\omega_c $, of the curvature photon is give by:
\begin{eqnarray}
\omega_c=\frac{3}{2}\frac{\Gamma^3}{R_c},
\label{curv-energy}
\end{eqnarray}

The equations (\ref{boost}),(\ref{eparallel0}) and (\ref{curv-energy}) can be used  to estimate 
the emission point of photons of a particular energy \footnote{Here we have considered $R_c= \sqrt{r/\Omega}$, 
which is a good approximation for dipole magnetic fields, close to the light cylinder.}. The relation
between, angular speed $\Omega$, the point of emission$\frac{r}{{\rm R}_{LC}}$, radius ${\rm\bar{ R}}$, 
magnetic field ${\rm B}$ and energy of the emitted photon $\omega$, can be expressed in dimension-less quantities as : 

\begin{eqnarray}
\Omega=\frac{1}{{\bar R}}\left[ \frac{\left({\bar{K}}\frac{{\rm e\rm{B}}}{m^2_e}\right)^3}{(m_e {\bar R})^2}
\sqrt{\frac{r}{R_{LC}}} \left( \frac{\omega}{m_e} \right)^4 \right]^{\frac{1}{7}}.
\label{Omega-r-reln}
\end{eqnarray}

In equation (\ref{Omega-r-reln}), we have used a numerical parameter $\bar{K}$,
defined as $\bar{K}=  4 \pi \alpha \frac{160}{\sqrt{6}}\left(\frac{2}{3}\right)^{4/3} $. 
One can estimate the path-length $z = (R_{LC} - r)$ for photons produced at point $r$ with energy $\omega$ from  Eqn.  (\ref{Omega-r-reln}) for various values of neutron star parameters like radius, magnetic field and rotational time period. The estimation of the density of the plasma is provided below.\\
\indent
The net number density of plasma produced in the magnetosphere can be estimated by dividing the  luminosity 
of star by their estimated energy \cite{beskin-book}.  The combined effect of pair produced plasma and the plasma ejected  from the surface of star  can create a final plasma density, more than $n_{GJ}$.  Following this point of view, we have taken the plasma frequency $\omega_{p}$ to be of the order $10^{-2}$ eV. We have further considered the photon path length to be  $z = 1.2$Km \cite{Lesch}. For these numbers of $\omega_{p}$ and $z$, we have estimated various oscillation probabilities in KeV energy range (20-100)KeV as shown in Fig: [{\ref{f001}] and Fig: [\ref{f002}].\\
 \indent
The basic motivation for this choice of parameters, was to find out  the imprints of $\phi F^{\mu\nu}F_{\mu\nu}$ interaction on the non-thermal spectro-polarimetric signals from the star magnetosphere is realised in nature. In presence of such 
interaction, some of these energetic photons may eventually oscillate into a scalars and go out
of the magnetosphere of the compact object, and get detected by oscillating back into photon.    
Although the same could be pointed out, with the results available in the literature 
\cite{ConlonMarsh} and the references there in; however, in this work,  we have  
considered the effect of magnetized medium to the oscillation probability. Our numerical 
estimates show that, emitted photon energy $\omega >>\omega_p $, therefore no self absorption in the medium.\\ 
\indent
However the main process of obstruction, to the  propagation of primary photons,
comes in the form of  absorption. There are few processes those contribute to 
this phenomenon. They are,  (i)$ \gamma + {\rm B}  \to  e^{+} e^{-}$, 
(ii) $\gamma +\gamma \to e^{+} e^{-} $, (iii) scattering with ambient 
electrons in the medium, (iv) synchrotron self absorption etc. .\\
\indent
For  sub-threshold photon energies $(i.e., \omega < 2 m_e)$ (as is the kinematics 
considered here) the  process (i) is forbidden. In order to have the second process 
viable, for photons with $\sim 100 $ KeV energy, the minimum energy required for the
second photon $\omega_{th} \ge m^2_e\omega $ which is also above pair production 
threshold. It is unlikely that  at $r > 0.99\rm{R}_{LC}$ the primary electrons will
generate  photons with energy  $ \omega = 10m_e$ or more. Therefore the second process
also seems unlikely.\\ 
\indent
For stellar, objects of mass $M_s$,
temperature T, the scale hight ( defined as $h=\frac{k_B T {\bar R}^2}{m_p G M_s} $) is the 
quantity that determines the density of the atmosphere of the steller object. Although,
proton mass $m_p$ is used in estimating the same, however if one invokes local charge 
neutrality the scale hight of electrons would also be similar. For a compact object with
about one solar mass and temperature around $10^{5 }$ Kelvin the ambient atmosphere close to the
light cylinder would be extremely thin to contribute to degradation of photon energy through
compton scattering.\\
\indent
Lastly, the synchrotron 
self-absorption coefficient, follows a power law behaviour $\alpha_{\omega} \propto \frac{1}{\omega^s}$, 
where the index $s > 2$. Hence the same may also be neglected for considering towards
absorption. Therefore for the kind of physical situation we have considered so far, the main 
mechanism towards photon depletion would essentially be due to conversion to scalars.\\


\begin{thebibliography}{99}
\bibitem{sugra} V. A. Kostelecky, R. Lehnert, and M. J. Perry,
 %Spacetime-varying couplings and Lorentz violation,
  Phys. Rev. D{\bf 68}, 123511 (2003).
%%%
\bibitem{witten} S. Gukov, C. Vafa, and E. Witten,
 %CFT's from Calabi-Yau fourfolds', 
  Nucl. Phys. B {\bf 584}, 69 (2000).
%%%
\bibitem{asen} A. Sen,  
%Strong weak coupling duality in four-dimensional string theory, 
Int. J. Mod. Phys. A {\bf 9}, 3707 (1994).
%%%
\bibitem{cicoli} L. Anguelova, V. Calo, and M. Cicoli, 
%LARGE Volume String Compactifications at Finite Temperature, 
JCAP {\bf 0910}, 025 (2009).  
%%%
\bibitem{Conlon-Quevedo} J. P. Conlon and F. Quevedo, 
%Astrophysical and cosmological implications of Large Volume String compactifications, 
JCAP {\bf 08}, 019 (2007). 
%%%
\bibitem{Cacciari} M. Cacciari, 
%A Note On the fate of the Landau Yang theorem in non-Abelian gauge theories, 
arXiv.1509.07853v1.
%%%
\bibitem{Kusenko} A. Kusenko,  K. Schmitz, and T. T. Yanagida, 
%Leptogenesis via Axion Oscillation after inflation, 
Phys. Rev.  Lett. {\bf  115}, 011302 (2015).
%%%
\bibitem{Asaka} T. Asaka , J. Hashiba, M. Kawasaki, and T. Yanagida, 
%Spectrum of background x-rays from moduli dark matter, 
Phys. Rev. D {\bf 58}, 023507 (1998).
%%%
\bibitem{Higakikamaga} T. Higaki, K. Kamaga,  and F. Takahashi, 
%Higgs, Moduli Problem, Baryogenesis and Large Volume Compactifications, 
JHEP {\bf 09}, 043 (2012). 
%%%(Note: This work and the one below are set up with similar dynamics.)
%%%
\bibitem{ConlonMarsh} J. P. Conlon, and M. C. David  ~Marsh,
%Excess Astrophysical Photons from a 0.1-1 keV  Cosmic Axion Background,
 Phys. Rev. Lett.  {\bf 111}, 151301 (2013).
%%%
\bibitem{Nakayama} K. Nakayama, F. Takahashi and T. T. Yanagida, 
%The 3.5 keV X-ray line signal from decaying moduli with low cutoff scale, 
Phys. Lett. B. {\bf 735}, 338 (2014).
%%%
\bibitem{Dhuria} M. Dhuria,  C. Hati, and U. Sarkar, 
%Moduli induced cogenesis of baryon asymmetry and dark matter, 
Phys. Lett. B {\bf 756}, 376 (2016). 
%%%\emph{ArXiv: 1508.04144} (Note: This paper discusses other moduli physics related issues.)
%%%
\bibitem{Schmutzer} E. Schmutzer, \emph{ Unified Field Theories of More Than 4 Dimensions}, edited by E. Schmutzer and V. De Sabbata (World Scientific, Singapore, 1983), p. 81. 
%%%
\bibitem{duff} M. ~J.~Duff, B.~E.~W.~Nilsson, and C.~N.~Pope,  
%Kaluza-Klein Supergravity, 
Phys. Rep. {\bf 130}, 01-142 (1986).
%%%
\bibitem{scherk} J. Scherk, \emph{ Supergravity}, edited by P. van Nieuwenhuizen and D. Z. Freedman (North-Holland, Amsterdam, 1979), p. 43. 
%%%
\bibitem{chameleon} J. Khoury, A.  Weltman, 
%Chameleon fields: awaiting surprises for tests of gravity in space,
 Phys. Rev. Lett. {\bf 93}, 171104 (2004).
%
\bibitem{ban1}Narayan Banerjee,  Sudipta Das, and  Koyel Ganguly,
 %Chameleon field and the late time acceleration of the Universe, 
 Pramana {\bf 74}, 3 (2010).

\bibitem{ban2} Nandan Roy, Narayan Banerjee, 
%Dynamical systems study of chameleon scalar field,
Annals of Physics {\bf 356}, 452-466 ( 2015).

\bibitem{guha} Amit Das,  Farook Rahaman,   B. K. Guha,  and  Saibal Ray, 
% Compact stars in f(R,T) gravity, 
The European Physical Journal C {\bf 76}, 654 (2016).
%Fuzzy Dark Matter from Infrared Confining Dynamics
\bibitem{kom} E.Komatsu et.al., 
%Five year Wilkinson microwave anisotropy probe observations: Cosmological interpretation,
 Ap. J. Suppl. {\bf 180}, 330 (2009). 
%
\bibitem{dunk} J. Dunkley et.al., 
%Five year Wilkinson microwave anisotropy probe observations: Liklihoods and parameters from the WMAP data, 
Ap.J. Suppl. {\bf 180}, 306 (2009); and references therein. %\\
%
\bibitem{brax} P. Brax, C. van de Bruck, Anne-Christine Davis, J. Khoury, and  A. Weltman,
 %Detecting dark energy in orbit: The cosmological chameleon, 
 Phys. Rev. D {\bf 70},  123518 (2004).
%
%\bibitem{jk1} J.khoury and A.Weltman, \emph{Chameleon fields: Awating surprises for tests of gravity in space} \emph{ Phys. Rev. Lett.} {\bf 93} 171104 (2004).\\
%
\bibitem{jk2} J. Khoury, and A. Weltman, 
%Chameleon cosmology, 
Phys. Rev. D {\bf 69}, 044026 (2004).
%
\bibitem{brax2} P. Brax, C. van de Bruck, and  Anne-Christine Davis,
 %Compatability of the Chameleon-field model with fifth-force experiments, cosmology, and PVLAS and CAST results, 
  Phys. Rev. Lett. {\bf 99}, 121103 (2007).
%
\bibitem{davis} A. C. Davis, C. A. O. Schelpe, and D. J. Shaw, 
%Effect of a chameleon scalar field on the cosmic microwave background,
 Phys. Rev. D {\bf 80}, 064016 (2009); and references therein.
%
\bibitem{bergmann} P. G. Bergmann, 
%Comments on the scalar-tensor theory, 
Int. J. Theor. Phys. {\bf 1}, 25 (1968).
%
\bibitem{gasperini} M. Gasperini,
 %Constraints on unified theories from the experimental tests of the equivalence principle, 
 Gen. Relativ. Gravit. {\bf16}, 1031 (1984).
%
\bibitem{burrage} C. Burrage, A. C. Davis, and D. J. Shaw, 
%Detecting chameleons: The astronomical polarization produced by chameleon like scalar fields, 
Phys. Rev. D {\bf 79}, 044028 (2009). 
%
\bibitem{bento}  C. Bento, O.  Bertolami, R. Rosenfeld,  and L. Teodoro, 
%Self - Interacting dark matter and the Higgs Boson,
 Phys. Rev. D {\bf 62}, 041302(R) (2000). 
%
\bibitem{davoudiasl} H. Davoudiasl, and C. Zhang,   
%750 GeV messenger of dark conformal symmetry breaking,
 Phys. Rev. D  {\bf 93}, 055006 (2016).
%
\bibitem{Gondolo} P. Gondolo, and L. Visinelli, 
 %Axion Cold Dark Matter in View Of BICEP2 results, 
 Phys. Rev.  Lett.  {\bf  113}, 011802 (2014).
%}
\bibitem{MArsh} D. J. E. Marsh, D. Grin, R. Hlozek, and P. G. Ferreira,  
 %Tensor Interpretation of BICEP2 results Severely Constrains Axion Dark Matter,  
 Phys. Rev.  Lett. {\bf  113}, 011801 (2014).
%
\bibitem{Higaki} T. Higaki, K. S. Jeong,  and F. Takahashi,
 %Solving the tension between high scale inflation and axion isocurvature perturbations, 
  Phys. Lett. B  {\bf  734}, 21 (2014).
%
\bibitem{essig}R. Essig,  
%Dark Sectors and New, Light, Weakly-Coupled Particles,
  arXiv:1311.0029.
%
\bibitem{sachadavidson} S. Davidson and S. Sarkar,
 %Thermalization after Inflation,
  JHEP  {\bf 11}, 012 (2000).
%
\bibitem{kahniasvili} V. Baukh, A. Zhuk, and T. Kahniashvili, 
% Extra dimensions and Lorentz invariance violation, 
  Phys. Rev.  D {\bf 76}, 027502 (2007).
%
\bibitem{Miani} L. Maiani, R. Petronzio, and E. Zavattini, 
%Effects of nearly massless, spin-zero particles on light propagation in a magnetic field, 
Phys. Lett. B {\bf 175}, 359 (1986).
%
\bibitem{Raffelt} G. Raffelt, and L. Stodolsky, 
%Mixing of the photon with low-mass particles,
 Phys. Rev. D {\bf 37}, 1237-1249 (1988).
%
\bibitem{sikivie1} P.~Sikivie, 
%Experimental Tests of the Invisible Axion, 
Phys. Rev. Lett.  {\bf 51}, 1415 (1983) [Erratum-ibid.\  {\bf 52}, 695 (1984)].
%
\bibitem{sikivie2} P.~Sikivie, 
%Detection Rates For 'invisible' Axion Searches, 
Phys. Rev. D {\bf 32}, 2988 (1985), [Erratum-ibid.\ D {\bf 36}, 974 (1987)].
%
\bibitem{pospelov} V.~Flambaum, S.~Lambert, and M.~Pospelov, 
%Scalar Tensor Theories with pseudoscalar couplings, 
Phys. Rev. D {\bf 80},    105021 (2009).
%
\bibitem{Fischbach} E. Fischbach, D. Sudarsky, A. Szafer, C. Talmadge, and S. H. Aronson,  
%Reanalysis of the Eotvos experiment, 
Phys. Rev. Lett. {\bf 56}, 03 (1986).
%
\bibitem{Capparelli} L. M. Cappparelli et. al., 
%Axion-Like particle searches with sub-THz photons,
 arXiv: 1510.06892v1.
%%%%%%%%%%%%%%%%%%%%%%%%%%%%%%%%%%%%%%%%%%%%%%  Done.
%
\bibitem{chaichian1} I.~Brevik and M. M. Chaichian  {\em
Electric current and heat production by a neutral carrier: an effect of 
the axion } Euro. Phys. J {\bf C82 } 3, 202, (2022).\\
%
\bibitem{chaichian2} I.~Brevik and M.~M.Chaichian ,{\emph  Axion electrodynamics: Energy-momentum tensor, and possibilities for experimental tests}  arXiv. 220704807.\\
%
\bibitem{Chen} F. Chen, J. M. Cline, A. Fradette, A. R. Frey, and C. Rabideau,
 %Exciting dark matter in the Galactic Center, 
 Phys. Rev. D  {\bf 81}, 043523 (2010). 
%(Note: This work has marginal relevance with the scenario explored in 
%this paper but this is of the type those explored  origin of gamma ray signals form galactic center through Dark 
%Matter annihilation.)
%
\bibitem{Bartels} R. Bartels,  S. Krishnamurthy, and C. Weniger,  
%Strong Support for the millisecond pulsar origin of Galactic center GeV  excess, 
arXiv:1506.0504v2. 
%(Note: The excess can also be arrtibuted to the  $ \phi -\gamma $ oscillation in the GeV 
%region.  GeV Photon depletion through  $\gamma + \rm{B} \to e^{+}+ e^{-} $  as well as  $\gamma + \gamma \to e^{+} 
%+ e^{-} $ can be evaded through an efficient oscillation channel.)
%
\bibitem{Boyarsky} A. Boyarsky,  J. Franse, D. Iakubovskyi, and O. Ruchayskiy, 
 %Checking the Dark matter Origin of a 3.53 keV Line with the Milky Way Center,  
 Phys. Rev.  Lett.  {\bf  115}, 161301 (2015).
%
\bibitem{Hooper} D. Hooper, I. Cholis, T. Linden, J. M. Siegal-Gaskins, and T. R. Slatyer,
 %Millisecond pulsars cannot account for the inner Galaxy's GeV excess, 
  Phys. Rev. D {\bf  88}, 083009 (2013). 
%(Note: The basic argument presented in this paper is that the sub GeV spectral shape is ``much too soft'' to accomodate this 
%signal. However, this particular observation of the paper further motivate the study of emission model of pulsars with $\phi- \gamma$
%interaction, with photon-self-energy, to order $eB$.)
%
\bibitem{torsion-balance1} E. Fishbach  et. al., Reanalysis of the Eoumltvos experiment, Phys.Rev Lett. {\bf 56}, 3 (1986).
%
\bibitem{torsion-balance1(a)}  S. C. Holding, F. D. Stacey, and G. J. Tuck, 
%Gravity in minesmdashAn investigation of Newton?s law, 
Phys. Rev. D {\bf  33}, 3487 (1986). 
%
\bibitem{torsion-balance2} C. D. Hoyle,  U. Schmidt, B. R. Heckel, E. G. Adelberger, J. H. Gundlach, D. J. Kapner, H. E. Swanson,
%Submillimeter test of the gravitational inverse-square law: A search for ?Large?  extra dimensions, 
Phys. Rev Lett. {\bf 86}, 1418 (2001). 
%
\bibitem{torsion-balance3} D. J. Kapner, T. S. Cook, E. G. Adelberger,  J. H. Gundlach, B. R. Heckel, C. D. Hoyle, H. E. Swanson,
%Tests of the gravitational inverse-square law below the dark-energy length scale, 
Phys. Rev.  Lett. {\bf 98}, 021101 (2007).
%
\bibitem{torsion-balance4} E. G. Adelberger, B. R. Heckel, S. Hoedl,   C. D. Hoyle, D. J. Kapner, A. Upadhye,
%Particle-physics implications of a recent test of the gravitational inverse-square law, 
Phys. Rev. Lett. {\bf 98}, 131104 (2007).
%
\bibitem{acharya} B. ~S.~Acharya, ~G. Kane and E. Kufflic {\em{Bounds on Scalar masses on theories of Moduli Stabilization }} IJMPA {\bf 129},
1450073, (2014)
%
%
\bibitem{kim} Y. M. Cho, and J. H. Kim, 
 %Dilatonic dark matter and its experimental detection,
  Phys. Rev. D {\bf 79}, 023504 (2009). 
%
\bibitem{fifthforce1} Y. M. Cho,
 %Unified cosmology, 
 Phys. Rev. D {\bf 41}, 2462 (1990).
%
\bibitem{fifthforce2} Y.M. Cho, and J.H. Yoon, 
%Geometric symmetry  breaking and cosmological potential in Kaluza-Klein theory, 
 Phys. Rev. D {\bf 47}, 3465 (1993).
%
\bibitem{fifthforce3} Y.M. Cho, \emph{ Proceedings of XXth Yamada Conference}, edited by S. Hayakawa and K. Sato (University Academy,Tokyo 1988).
%
\bibitem{salati} For a comprehensive details one can see:  P. Salati,  A primer on primordial nucleosynthesis, pp. 63 in \emph{ ASp Conference series Vol. 126 1997} ed., D. V. Gabaud, M. Hendry, P. Molaro and K. Chamcham.

\bibitem{fuzzy1} Hooman Davoudiasl, and  Christopher W. Murphy,  
%Fuzzy Dark Matter from Infrared Confining Dynamics, 
Phys. Rev. Lett. {\bf 118}, 141801 (2017).

\bibitem{swe1} Paulo Aguiar and Paulo Crawford, 
%Sachs-Wolfe effect in some anisotropic models, 
arXiv:astro-ph/0110412v2.

\bibitem{swe2} Martin White and Wayne Hu,
 %The Sachs-Wolfe effect, 
 Astron. Astrophys. {\bf 321}, 8-9 (1997).

\bibitem{CASPEr} D. Budker, P. W. Graham, M. Ledbetter, S. Rajendran,  and A. O. Sushkov, 
%Cosmic Axion Spin Precession Experiment (CASPer), 
Phys. Rev. X {\bf 4}, 021030 (2014).

\bibitem{ADMX} N. Du, N. Force, R. Khatiwada, E. Lentz, R. Ottens, L. J Rosenberg, et al.
%G. Rybka, G. Carosi, N. Woollett,  D. Bowring, A. S. Chou 
(ADMX collaborarion), 
%Search for Axion Dark Matter with the Axion Dark Matter Experiment, 
Phys. Rev. Lett. {\bf 120}, 151301 (2018).

\bibitem{LSW} F. Hoogeveen, T. Ziegenhagen, 
%Production and detection of light bosons using optical resonators, 
Nucl. Phys. B {\bf 358}, 3-26(1991).

\bibitem{borexino} G. Bellini et.al. Borexino collaboration,
 %Search for Solar Axions Produced in p(d,3He) A Reaction with Borexino Detector,
  Phys. Rev. D {\bf 85}, 092003 (2012).
 
 \bibitem{borexino2} G. Bellini et.al. Borexino collaboration,
  %Search for Solar Axions emitted in M1-transition of 7Li*   with Borexino CTF, 
  EPJC {\bf 54}, 61 (2008).

\bibitem{Lieu1996} R. Lieu et al., 
%Diffuse extreme-ultraviolet emission from the Coma cluster: evidence for rapidly cooling gases at submegakelvin temperatures, 
Science {\bf 274}, 5291 (1996).
%
\bibitem{Bonamente2002}  M.~Bonamente, R.~Lieu, M.~K.~Joy and J.~H.~Nevalainen,  
%The soft X-ray emission in a large sample of galaxy clusters with the ROSAT Position Sensitive Proportional Counter, 
Ap. J. {\bf 576}, 688 (2002). 
%
\bibitem{Bonamente2001} M.~Bonamente, R.~Lieu, J.~Nevalainen and J.~S.~Kaastra, 
%ROSAT and BeppoSAX Evidence of Soft X-Ray Excess Emission in the Shapley Supercluster: A3571, A3558, A3560, and A3562, 
Ap. J. {\bf 552}, L7 (2002).
%
\bibitem{Nevalainen2003} J.~Nevalainen, R.~Lieu, M.~Bonamente, and D.~Lumb, 
%Soft x-ray excess emission in clusters of galaxies observed with XMM-Newton, 
Ap. J. {\bf 584}, 716 (2003). 
%
\bibitem{Henriksen2003} Daniel S. Hudson, Mark J. Henricksen,
 %Diffuse nonthermal X-ray emission: Evidence of cosmic-ray acceleration at the shock front in IC 1262, 
 Ap. J. {\bf 595}, L1-L4 (2003).
%
\bibitem{Bonamente2007} M. Bonamente, J. Nevalainen, and R. Lieu,  
%Soft and hard X-ray excess emission in Abell 3112 observed with Chandra, 
Ap. J. {\bf 668},796 (2007).
%
\bibitem{Lehto2010} T. Lehto, J. Nevalainen, M. Bonamente, N. Ota, and J. Kaastra,  
%Suzaku observations of X-ray excess emission in the cluster of galaxies A3112,  
A \& A {\bf 524}, A70 (2010).
%

\bibitem{0.5-1KeV}  R. C. Hicox, and M. Markevitch,  
%Absolute measurement of the unresolved cosmic X-ray background in the 0.5--8 keV  band with {\it CHANDRA}, 
Ap. J. {\bf 645 }, 95 (2006).  
%
\bibitem{3.55KeV} Esra Bulbul et. al.,  
%Detection of an unidentified emission line in the stacked  X-ray spectrum  of galaxy clusters, 
ApJ   {\bf 789}, 13 (2014).
%
\bibitem{.511KeVa} J.  Knodlseder et. al., 
%Early SPI/INTEGRAL measurements of 511 keV line emission from the 4th quadrant of the Galaxy, Astron. 
Astrophys. {\bf 411}, L 457 (2003).
%
\bibitem{.511KeVaa} P. Jean   et. al., 
%Early SPI/INTEGRAL measurements of 511 keV line emission from positron annihilation, 
Astron. Astrophys. {\bf 407}, L55  (2003).
%
\bibitem{.511KeVb} J. Knodlseder et. al., 
%The all-sky distribution of 511 keV electron-positron annihilation emission,
 Astron. Astrophys. {\bf 441}, 513 (2003).
%
\bibitem{pankaj} S.~ Das, P.~Jain, J.~ P.~Ralston, and R. Saha, 
%Probing light pseudoscalars with light propagation, resonance and spontaneous polarization, 
JCAP {\bf 0506}, 002 (2005).
%
\bibitem{Craig} N. J. Craig, and S. Raby,   Modulino dark matter and teh INTEGRAL 511 keV Line 
%\emph{Phys. Lett. B.}  {\bf 735}, 338, (2014).
arXiv:0908.1842v2.
%
\bibitem{sikivie2021} P. Sikivie, 
%Invisible Axion Search Methods, 
Rev.  Mod. Phys. {\bf 93}, 015004 (2021).

\bibitem{GKP} A. K. Ganguly, S. Konar and P. B. Pal,
 %Faraday effect: a field theoretical point of view,
  Phys. Rev. D {\bf 60}, 105014 (1999). 
%
\bibitem{GJM} A.~K.~Ganguly, P.~Jain  and S.~Mandal, 
%Photon and axion oscillation in a magnetized medium: A general treatment,
 Phys. Rev. D {\bf 79},  115014 (2009).

\bibitem{CJG} Ankur Chaubey, Manoj K. Jaiswal, Avijit K. Ganguly, 
%Exploring scalar-photon interactions in energetic astrophysical events,  
Phys. Rev. D {\bf 102}, 123029 (2020).

\bibitem{pal-jose} J. F. Nieves, and P.B. Pal,
 %{ {\bf P} and {\bf CP}-odd terms in the photon self-energy within a medium},
  Phys. Rev. D {\bf 39}, 652 (1989).
%
\bibitem{adas}  A. Das, Finite temperature Field theory, (World Scientific, 1997).
%
\bibitem{kapusta}  J. Kapusta,\emph{ Finite Temperature Field Theory}, Cambridge University Press,(1989); M. Le Bellac, \emph{Thermal Field Theory}, Cambridge University Press,(1996).
%
\bibitem{Perez}  H. Perez Rojas, A. E. Shabad,
 %Polarization of relativistic electron and positron gas in a strong magnetic field.Propagation of electromagnetic waves,
  Ann.  Phys.  {\bf121},  432-455 (1964).
%
\bibitem{Braaten-Segel} E. Braaten, and D. Segel, 
%Neutrino energy loss from plasma process at all temperatures and densities, 
Phys. Rev. D {\bf 48}, 1478 (1993).

\bibitem{palash2020} Palash B. Pal, 
% Photon propagation in non-trivial backgrounds, 
Phys. Rev. D {\bf 102}, 036004 (2020). 


\bibitem{GJ} A.~K.~Ganguly, M.~K.~Jaiswal, 
%Lorentz symmetry violating low energy dispersion relations from a dimension-five photon scalar mixing operator,  
Phys. Rev. D {\bf 90},  026002 (2014).
%
\bibitem{Sngupta} S. N. Gupta, Theory of Longitudinal Photons in Quantum Electrodynamics, \emph{The proceedings of the Physical society, Section A}  {\bf63}, part(7), 46 (1950).
\bibitem{elmfors}Per Elmfors , David Peersson and Bo-Sture Skagerstam, 
{\emph{Thermal Fermionic Dispersion Relations in a Magnetic Field}},
Nucl. Phys. {\bf B422},521 (1994 ).
%
\bibitem{Sch} J.Schwinger, 
%On gauge invariance and vacuum polarization, 
Phys.Rev. {\bf 82},664(1951).
%
\bibitem{peccei} R.~D.~Peccei and H.~R.~Quinn, 
%CP conservation in presence of pseudoparticles,   
Phys. Rev. Lett.  {\bf 38}, 1440 (1977).


\bibitem{Beskin_pulsar} A.G. Mikhaylenko, V. S. Beskin, and Ya. N. Istomin,
 %On the thermal effects on radio waves propagating in the pulsar magnetosphere, 
 arXiv: 1912.03731v1 [astro-ph.HE] (2019).

\bibitem{ONS} J. C. DOlivo, J. F. Nieves and S. Sahu,
 %Field theory of the photon self energy in a medium with a magnetic field and the faraday effect, 
  Phys. Rev D {\bf 67}, 025018 (2003).
%
%
\bibitem{shahbad}  H. Perez Rojas, A. E. Shabad,
 %Polarization of relativistic electron and positron gas in a strong magnetic field.Propagation of electromagnetic waves,
  Ann.  Phys.  {\bf121},  432-455 (1964).
%


\bibitem{shore} G.M. Shore , 
%Faster than light photons in gravitational fields II.: 
%Dispersion and vacuum polarisation
Nucl. Phys. {\bf B633}, 271 ( 2002)

\bibitem{avramidi} I. G. Avramidi,
% A covariant technique for the calculation of the effective one loop 
%effective action 
Nucl. Phys. {\bf B355}, 712 (1991). 

\bibitem{Rempel}Laurent Freidel, Trevor Rempel
%Scalar Field Theory in Curved Momentum Space
arXiv(1312.3674 )


\bibitem{Drummond} I. T. Drummond and S. Hathrell,~ Phys. Rev. {\bf D22}, 343 
% superluminal velocity of Light.
(1980).

\bibitem{Richard-Conn-Henry} Richard Conn Henr, 
%Kretcchmann Scalar For a Kerr Neuman Black Hole
ApJ {\bf 535}, 350 (2000).
%

\bibitem{mohanty-nieves-pal} S. Mohanty, J.  F. Nieves, and Palash B. Pal, 
%Optical activity of a neutrino gas,
 Phys. Rev. D {\bf 58}, 093007 (1998).
%

\bibitem{Felix} F. Karbstein, Phys. Rev. D {\bf 88}, 085033 (2013).
\bibitem{Urrutia} L. F. Urrutia, Phys. Rev. D {\bf 17}, 8 (1978).

\bibitem{rs75} M. A. Ruderman, \& P. Sutherland,
 %Theory of pulsars - Polar caps, sparks, and coherent microwave radiation,
  ApJ {\bf 196}, 51 (1975).
%
\bibitem{as79} J. Arons, \& E. T. Scharlemann,
 %Pair formation above pulsar polar caps - Structure of the low altitude acceleration zone, 
 ApJ {\bf 231}, 854 (1979).
%
\bibitem{dh96} J. K. Daugherty, \& A. K. Harding,
% Gamma-Ray Pulsars: Emission from Extended Polar CAP Cascades,
 ApJ {\bf 458}, 278 (1996).
%
\bibitem{beskin-book} V. S. Beskin, \emph{ MHD Flows in Compact Astrophysical  Objects, Accretion, Winds and Jets}, Springer-Verlag Berlin Heidelberg.
%

\bibitem{meszaros} P. Meszaros, \emph{ High-Energy Radiation from Magnetized Neutron Stars}, University of Chicago Press (1992).
  
\bibitem{Radhakrishnan} V. Radhakrishnan, D. J. Cooke, 
%Magnetic poles and the Polarization Structure of Pulsar Radiation, 
ApL {\bf 3}, 225 (1969).
  
\bibitem{usov93} V. V. Usov,  \emph{High frequency emission of X ray Pulsar 1E 2259+586 }, \emph{ApJ} {\bf 410}, 761 (1993).
%
\bibitem{cardano} G.~Cardano,  \emph{Ars Magna } (1545).
%
\bibitem{KOPP} J. Kopp, 
%Efficient numerical diagonalization of hermitian $3 \times 3$ matrices, 
Int. J. Mod. Phys. C {\bf 19}  523-548 (2008).
%
\bibitem{mybook} A. K. Ganguly, \emph{Introduction to Axion Photon Interaction in Particle Physics and Photon Dispersion in Magnetized Media},  Particle Physics, Eugene Kennedy (Ed.), ISBN: 978-953-51-0481-0, InTech, (2012). 
%
\bibitem{VACLAV} Vaclav O. Kostroun, 
%Simple numerical evaluation of modified bessel functions $K_{\nu}(x)$ of fractional order and the integral $\int^{\infty}_{x} K_{\nu}(\eta)$ d$\eta$, 
Nuclear Instrumentation methods {\bf 172}, 371(1980).  

\bibitem{usov88} V. Usov, 
 %Generation of Gamma Rays by rotating white dwarf, Institute of space research,  
 USSR, moscow (1988).


\bibitem{Lesch} H. Lesh, A. Jessner, M. Kramer, and T. Kunzi,
 %On the possibility of curvature radiation from radio pulsars, 
 Astron. and Astrophys. {\bf 332},L21-L24  (1998).
%\bibitem{meszaros} P. Meszaros, \emph{ High-Energy Radiation from Magnetized Neutron Stars}, University of Chicago Press (1992).

\bibitem{Mirizzi-raffelt1} A. Mirizzi, G. G. Raffelt, and Pasquale D. Serpico, 
%Photon-axion conversion as a mechanism for supernova dimming: Limits from  CMB spectral distortion, 
Phys. Rev. D {\bf 72}, 023501 (2005).

\bibitem{Mirrizzi-raffelt2} A. Mirizzi, G. G. Raffelt, and Pasquale D. Serpico,
% Photon-axion conversion in intergalatic magnetic fields and cosmological consequences, 
 arXiv:astro-ph/0607415v1 (2006).

\bibitem{Tiwari} P. Tiwari, 
%Constraining axionlike particles using the distance-duality relation,
 Phys. Rev. D {\bf 95}, 023005 (2017).
%\emph{arXiv:1610.06583v2} (2006).

\bibitem{FERMI_LAT} M. Meyer, M. Giannotti, A. Mirizzi, J. Conrad, and M.A. Sanchez-Conde, 
 %Fermi Large Area Telescope as a Galatic Supernovae Axionscope, 
 Phys. Rev. Lett {\bf 118}, 011103 (2017).

\bibitem{Soda} E. Masaki, A. Aoki, and J. Soda,
% Photon-axion conversion, Magnetic Field Configuration and Polarization of Photons, 
Phys. Rev. D {\bf 96}, 043519 (2017).
%\emph{arXiv:1702.08843v2} (2017).

\bibitem{Grossman} Y. Grossman, S. Roy, and J. Zupan, 
%Effects of initial axion production and photon-axion oscillation on type Ia supernova dimming, 
 Phys. Lett. B {\bf 543}, 1-2 (2002).
%\emph{arXiv:hep-ph/0204216} {\bf 72} (2002).


\bibitem{Bassett} B. A. Bassett, and M. Kunz, 
%Cosmic Acceleration vs Axion Photon Mixing
 ApJ {\bf 607}, 661 (2004).
%\emph{arXiv:astro-ph/0311495v2} (2004).


\bibitem{Jackson}  J. C. Jackson, 
%Tight cosmological constraints from the angular size/redshift relation for ultra-compact radio sources, 
JCAP {\bf 11} ,007 (2004).


\bibitem{ckt1} C. Csaki, N. Kaloper, and J. Terning 
%(CKT I), 
% Dimming supernovae without cosmic acceleration,
 Phys. Rev. Lett. {\bf 88}, 161302 (2002).
\bibitem{ckt2}  C. Csaki, N. Kaloper, and J. Terning, 
% Exorcising $w<-1$, Ann.
 Phys. {\bf 317}, 410 (2005).

%
%\bibitem{Braaten-Segel} E. Braaten and D. Segel, \emph{Neutrino energy loss from plasma process at all temperatures and densities}, \emph{Phys. Rev. D} {\bf 48}, 4 (1993).




\bibitem{Pshirkov:2007st}
M.~S. Pshirkov and S.~B. Popov,
% \emph{{Conversion of Dark Matter Axions to
 % Photons in Magnetospheres of Neutron Stars}},
 % \href{https://doi.org/10.1134/S1063776109030030}{\emph
 J. Exp. Theor. Phys.
  {\bf 108}, 384 (2009).
  %[\href{https://arxiv.org/abs/0711.1264}{{\ttfamily 0711.1264}}].

\bibitem{Hook:2018iia}
A.~Hook, Y.~Kahn, B.~R. Safdi and Z.~Sun, 
%\emph{{Radio Signals from Axion Dark
  %Matter Conversion in Neutron Star Magnetospheres}},
  %\href{https://doi.org/10.1103/PhysRevLett.121.241102}
  Phys. Rev. Lett.
  %{\bfseries 
  {\bf121},  241102 (2018).
  %[\href{https://arxiv.org/abs/1804.03145}{{\ttfamily 1804.03145}}].

\bibitem{Lai:2006af}
D.~Lai and J.~Heyl,
% \emph{{Probing Axions with Radiation from Magnetic Stars}},
 %\href{https://doi.org/10.1103/PhysRevD.74.123003}{\emph
 Phys. Rev. D
 % {\bfseries 
 {\bf 74}, 123003 (2006).
  %[\href{https://arxiv.org/abs/astro-ph/0609775}{{\ttfamily
  %astro-ph/0609775}}].

\bibitem{Huang:2018lxq}
F.~P. Huang, K.~Kadota, T.~Sekiguchi and H.~Tashiro, 
%\emph{{Radio Telescope
  %Search for the Resonant Conversion of Cold Dark Matter Axions from the
 % Magnetized Astrophysical Sources}},
 % \href{https://doi.org/10.1103/PhysRevD.97.123001}
 Phys. Rev. D
  %{\bfseries 
  {\bf 97},  123001 (2018).
  %[\href{https://arxiv.org/abs/1803.08230}{{\ttfamily 1803.08230}}].

\bibitem{Safdi:2018oeu}
B.~R. Safdi, Z.~Sun and A.~Y. Chen, 
%\emph{{Detecting Axion Dark Matter with
  %Radio Lines from Neutron Star Populations}},
  %\href{https://doi.org/10.1103/PhysRevD.99.123021}
  Phys. Rev. D
  %{\bfseries 
  {\bf 99},  123021 (2019).
  %[\href{https://arxiv.org/abs/1811.01020}{{\ttfamily 1811.01020}}].


\bibitem{Leroy:2019ghm}
M.~Leroy, M.~Chianese, T.~D.~P. Edwards and C.~Weniger,
% \emph{{Radio Signal of
 % Axion-Photon Conversion in Neutron Stars: A Ray Tracing Analysis}},
  %\href{https://doi.org/10.1103/PhysRevD.101.123003}
  Phys. Rev. D
  %{\bfseries
  {\bf 101}, 123003(2020).
  %\href{https://arxiv.org/abs/1912.08815}{{\ttfamily 1912.08815}}].

\bibitem{Witte:2021arp}
S.~J. Witte, D.~Noordhuis, T.~D.~P. Edwards and C.~Weniger, 
%\emph{{Axion-Photon
  %Conversion in Neutron Star Magnetospheres: The Role of the Plasma in the
 % Goldreich-Julian Model}}, 
  % \href{https://
  arxiv:2104.07670
  %2104.07670}}.


\bibitem{Battye:2021xvt}
R.~A. Battye, B.~Garbrecht, J.~I. Mcdonald and S.~Srinivasan, 
%\emph{{Radio Line
 % Properties of Axion Dark Matter Conversion in Neutron Stars}},
  %\href{https://
  arxiv:2104.08290.

\bibitem{Foster:2020pgt}
J.~W. Foster, Y.~Kahn, O.~Macias, Z.~Sun, R.~P. Eatough, V.~I. Kondratiev,
  W.~M. Peters, C.~Weniger and B.~R. Safdi, 
  %\emph{{Green Bank and Effelsberg
  %Radio Telescope Searches for Axion Dark Matter Conversion in Neutron Star
  %Magnetospheres}},
  %\href{https://doi.org/10.1103/PhysRevLett.125.171301}{\emph
  Phys. Rev. Lett.
  %{\bfseries 
  {\bf125},  171301 (2020).
  %[\href{https://arxiv.org/abs/2004.00011}{{\ttfamily 2004.00011}}].

\bibitem{Darling:2020uyo}
J.~Darling, 
%\emph{{New Limits on Axionic Dark Matter from the Magnetar PSR
  %J1745-2900}},
 % \href{https://doi.org/10.3847/2041-8213/abb23f}{\emph
 Astrophys. J. Lett.
  %{\bfseries
  {\bf  900}, L28 (2020).
  %[\href{https://arxiv.org/abs/2008.11188}{{\ttfamily 2008.11188}}].

\bibitem{Darling:2020plz}
J.~Darling, 
%\emph{{Search for Axionic Dark Matter Using the Magnetar PSR
  %J1745-2900}},
 % \href{https://doi.org/10.1103/PhysRevLett.125.121103}{\emph{
 Phys. Rev. Lett.
  %{\bfseries 
  {\bf 125},  121103 (2020).
  %[\href{https://arxiv.org/abs/2008.01877}{{\ttfamily 2008.01877}}].

\bibitem{Battye:2021yue}
R.~A. Battye, J.~Darling, J.~McDonald and S.~Srinivasan, 
%\emph{{Towards Robust
 % Constraints on Axion Dark Matter Using PSR J1745-2900}},
  %\href{https://
  arxiv:2107.01225.

\bibitem{Millar2021} Alexander J. Millar, Sebastian Baum, Matthew Lawson, David M. C. Marsh, 
%Axion-photon conversion in strongly magnetised plasmas, 
arxiv: 2107.07399v1 (2021).

\bibitem{Diego} D. F. Torres, Nature Astronomy {\bf 2}, (2018).
\bibitem{Arons} E.T. Scharlemann, J. Arons, and W. M. Fawley, Apj {\bf 222} (1978)
%\bibitem{sikivie2021} P. Sikivie, Invisible Axion Search Methods, Review of Modern Physics {\bf 93}, (2021).
\bibitem{Jackson_book} J. D Jackson,  Third edition, Classical Electrodynamics, Wiley, (1998).

\end{thebibliography}
\end{document}